\documentstyle[pre,aps,epsf]{revtex}

\begin{document}

\draft


\title
{Soluble two-species diffusion-limited Models
            in arbitrary dimensions\footnote{To appear in Physical Review E}}

\author
{M. Mobilia and P.-A. Bares }

\address
{Institute of Theoretical Physics, Swiss Federal Institute of Technology of  Lausanne, CH-1015 Lausanne  EPFL, Switzerland }

\date{\today}

\maketitle

\begin{abstract}
A class of two-species ({\it three-states}) bimolecular diffusion-limited models of
classical particles with hard-core reacting and diffusing in a hypercubic 
lattice of arbitrary
dimension is investigated. The manifolds on which the equations 
of motion of the correlation functions close, are
determined explicitly.
This property allows to solve for the density and the two-point 
(two-time)
correlation functions in arbitrary dimension for both, a translation invariant
class
and another one where translation invariance is broken. Systems with correlated 
as well as uncorrelated, yet random initial states can also be treated 
exactly by this
approach. We discuss the asymptotic behavior of density and correlation functions
in the various cases. The dynamics studied is very rich. 
   
\end{abstract}
\section{Introduction}

Non-equilibrium statistical mechanics has wittnessed recently a resurgence of
interest. Though over 50 years old, the field is still in its infancy. 
Powerful concepts and tools are being developped, and yet much progress remains
to be done.
The understanding of classical stochastic many-body systems
is of relevance to a wide class of phenomena in physics and beyond. 
In this context, a class of models describing diffusion-limited reactions plays an important role \cite{Privman}. 
The natural language to describe the stochastic dynamics of $N$ classical bodies 
is that of 
the Master equation. Formally, the dynamics can be coded in an 
imaginary time Schr\"odinger equation, where the Markov generator plays the role
of the Hamiltonian (see e.g. \cite{Privman,Schutz0} and references therein). In the past,
various representations in terms of spins, fermions or bosons have been used
depending on the physics being emphasized. A powerful method that relies on 
bosonic field theory and
the renomalization group has been applied by Cardy and collaborators to deal
with low density systems in arbitrary dimensions (see e.g. \cite{Cardy}). In one spatial dimension
alternative approaches have been proposed which take into account the hard core
of the classical particles  (see e.g. \cite{Privman,Schutz0} and references therein).
Since the pioneering work of Glauber on the stochastic Ising model \cite{Glauber} various
generalizations and extensions have appeared (e.g. \cite{Racsz,Lushnikov}) .
A general approach has been proposed by Sch\"utz \cite{Schutz}.
The latter investigates the most general class of single-species models of 
bimolecular diffusion-limited reactions that can be solved exactly. Upon imposing
constraints on the available manifold, the equations of motion of all correlation
functions close and, in that sense, the
dynamics is completely soluble. Via a duality transformation, Sch\"utz further
shows that on the 10-parametric soluble manifold, the spectrum of the stochastic
Hamiltonian coincides with that of the $XXZ-$Heisenberg model 
(see also \cite{Henkel}).
In Reference \cite{Fujii}, Fujii and Wadati extend Sch\"utz's ideas to
the $s-$species models of 
bimolecular diffusion-limited reaction processes. These authors derive the 
general constraints that allow for the equations of motion of correlation 
functions to close and, similarly to the single species case, introduce a 
dual Hamiltonian with identical spectrum. 
They further note that in the general multi-species case, the constraints of 
solubility (in the sense given above) do not seem to imply a simple 
relationship to integrable quantum Hamiltonians.

The general $s-$species bimolecular reaction-diffusion processes are 
characterized by $(s+1)^{4}-(s+1)^{2}$ independent parameters (reaction-rates) 
and we have to impose $2s^{3}$ constraints to close the hierarchy of the 
equations of motion of correlation functions. 
As few exact and explicit  
results for the {\it dynamics} of multi-species 
processes are available (in particular in dimensions $d>1$, see the discussion at the end of section III) , 
we decided 
to investigate in some details and generality the soluble two-species bimolecular 
diffusion-limited reaction systems. 
In this paper, 
we focus on the two-species 
problem ($s=2$) and obtain, in 
arbitrary dimension, exact results. A particular physical application of this work to a {\it three-states} growth model will be presented elsewhere \cite{Mobar2}
\\
The paper is organized as follows: the remaining of this section will be devoted
to definitions and notations. 
In section II, the equations of motion of the density and 
two-point correlation functions are derived. 
The constraints that ensure the solubility of the problem are explicitly identified. We 
classify the soluble manifolds which will be investigated in the sequel. 
In the 
first part of section III, we study the Fourier-Laplace transform 
of the density in the soluble case (on a $56-$parameters manifold). 
In the second part, we 
compute on two manifolds the exact expression of the density in 
arbitrary dimensions. 
We provide the asymptotic behavior of the latter for three 
different initial conditions. At the end of section III we discuss the relationship between our results and the solution of some models solved exactly in dimension $d\geq 1$ \cite{Clement,Fichthorn,Toussaint,Bramson,Belitsky,Bennaim}.
In the first part of section IV, we give the exact  
dynamic form factor for
an homogeneous and uncorrelated (yet random) initial state. 
In the second part, we 
compute, in arbitrary dimension, the exact two-time two-point 
correlation functions for random (homogeneous) uncorrelated as well as 
correlated initial states (we discuss the sensitiveness of the system to
the presence of initial correlations) . Section V is devoted to the study of the 
instantaneous two-points correlation functions on a manifold of 
translationally-invariant models. We first deal with the 
one-dimensional case, which is investigated for random uncorrelated 
as well as correlated initial states. Further, we consider
the higher dimensional case with random (yet homogeneous) and uncorrelated
initial states.  The last section, is devoted to the conclusion. 

For clarity and brevity's sake, some definitions, as well as some technical details,  are given in the appendices.
\\
\\
Consider an hypercubic lattice of dimension $d$ with periodic boundary conditions and $N$ sites ($N=L^d$), where 
$L$ represents the linear dimension of the hypercube.
Further assume that local bimolecular reactions between species $A$ and $B$ 
take place. Each site is either empty (denoted by the symbol $0$)  
or occupied at most by one particle of type $A$ (respectively $B$) 
denoted in the following by the
index $1$ (respectively $2$). 
The dynamics is parametrized by the transition rates
$\Gamma_{\beta_{1} \beta_{2}}^{\beta_{3} \beta_{4}}$, where $\beta_{1},\beta_{2},\beta_{3},
\beta_{4}=0,1,2$:
\begin{eqnarray}
\label{eq.0.1}
\forall (\beta_{1}, \beta_{2})\neq (\beta_{3}, \beta_{4})\;\;,\;\;  
\Gamma_{\beta_{1} \beta_{2}}^{\beta_{3} \beta_{4}}\;: \;\beta_{1} + \beta_{2} \longrightarrow \beta_{3} + \beta_{4}
\end{eqnarray}
Probability conservation implies
\begin{eqnarray}
\label{eq.0.2}
\Gamma_{\beta_{1} \beta_{2}}^{\alpha_{1} \beta_{2}}=-\sum_{(\beta_{1},\beta_{2})\neq(\beta_{1}',\beta_{2}')}
\Gamma_{\beta_{1} \beta_{2}}^{\beta_{1}' \beta_{2}'}, 
\end{eqnarray}
with
\begin{eqnarray}
\label{eq.0.2.1}
\Gamma_{\beta_{1} \beta_{2}}^{\beta_{3} \beta_{4}}\geq 0, \;\;\forall (\beta_{1},\beta_{2})\neq(\beta_{3},\beta_{4})
\end{eqnarray}

For example the rate $\Gamma_{1 1}^{1 2}$ corresponds to the process
$A + A \longrightarrow A + B$, while conservation of probability leads to 
$\Gamma_{1 1}^{1 1}=-(\Gamma^{1 0}_{1 1}+\Gamma^{0 1}_{1 1}+
\Gamma^{0 0}_{1 1}+\Gamma^{0 2}_{1 1}+
\Gamma^{2 0}_{1 1}+\Gamma^{2 1}_{1 1}+ \Gamma^{1 2}_{1 1}+ 
\Gamma^{2 2}_{1 1})$

The state of the system is determined by specifying the probability for the
occurence of
configuration $\{ n\}$ at time $t$. It is represented by the ket
\begin{eqnarray}
\label{eq.0.3}
|P(t)\rangle=\sum_{\{n\}}P(\{n\},t)| \{n\} \rangle
\end{eqnarray}
where the sum runs over the $3^N$ configurations ($N=L^d$).
At site $i$ the local state is denoted by the ket 
$|n_i\rangle =(1 \; 0 \; 0)^{T}$ if the site 
$i$ is empty,  $|n_i\rangle =(0 \; 1 \; 0)^{T}$ if the site $i$ is occupied by 
a particle of type $A$ 
($1$) and  $|n_i\rangle =(0 \; 0 \; 1)^{T}$ 
otherwise.
It is by now well known that the Master equation governing the dynamics of the systems can be rewritten as an imaginary-time Schr\"odinger equation:
\begin{eqnarray}
\label{eq.0.3.1}
\frac{\partial}{\partial t}|P(t)\rangle=-H |P(t)\rangle,
\end{eqnarray}
where $H$ is the Markov generator, also called {\it stochastic Hamiltonian}, which in general is neither hermitian nor normal. The construction of the {\it stochastic Hamiltonian} $H$ from the master equation follows a known procedure (see, e.g. \cite{Privman,Schutz0,Schutz,Fujii}).
We  define \cite{Privman,Schutz0,Schutz,Fujii} the {\it left vacuum} $\langle \widetilde \chi|$ :
\begin{eqnarray}
\label{eq.0.4}
\langle \widetilde \chi|\equiv \sum_{\{n\}} \langle \{n\} |
\end{eqnarray}
Probability conservation yields the local equation (stochasticity of $H$)
\begin{eqnarray}
\label{eq.0.5}
\langle \widetilde \chi|H = \sum_{\alpha=1,\dots, d}\sum_{m}\langle \widetilde 
\chi|H_{m, m+e^{\alpha}}=0 \Longrightarrow \langle \widetilde 
\chi|H_{m, m+e^{\alpha}}=0, 
\end{eqnarray}
where $e^{\alpha}, 1\leq \alpha \leq d, $ designates, in cartesian coordinates, the unit vector along the $\alpha$-direction.

The two-species local Markov generator acts on two adjacent sites, i.e.,  
\begin{eqnarray}
\label{eq.0.5.1}
-H_{m,m+e^{\alpha}} =
\left(
 \begin{array}{c c c c c c c c c}
 \Gamma_{0 0}^{0 0} & \Gamma_{0 1}^{0 0} &\Gamma_{0 2}^{0 0}&\Gamma_{1 0}^{0 0} &\Gamma_{1 1}^{0 0} & \Gamma_{1 2}^{0 0} &  \Gamma_{2 0}^{0 0} & \Gamma_{2 1}^{0 0} &\Gamma_{2 2}^{0 0} \\
 \Gamma_{0 0}^{0 1} & \Gamma_{0 1}^{0 1} &\Gamma_{0 2}^{0 1}&\Gamma_{1 0}^{0 1} &\Gamma_{1 1}^{0 1} & \Gamma_{1 2}^{0 1} &  \Gamma_{2 0}^{0 1} & \Gamma_{2 1}^{0 1} &\Gamma_{2 2}^{0 1} \\
 \Gamma_{0 0}^{0 2} & \Gamma_{0 1}^{0 2} &\Gamma_{0 2}^{0 2}&\Gamma_{1 0}^{0 2} &\Gamma_{1 1}^{0 2} & \Gamma_{1 2}^{0 2} &  \Gamma_{2 0}^{0 2} & \Gamma_{2 1}^{0 2} &\Gamma_{2 2}^{0 2} \\
 \Gamma_{0 0}^{1 0} & \Gamma_{0 1}^{1 0} &\Gamma_{0 2}^{1 0}&\Gamma_{1 0}^{1 0} &\Gamma_{1 1}^{1 0} & \Gamma_{1 2}^{1 0} &  \Gamma_{2 0}^{1 0} & \Gamma_{2 1}^{1 0} &\Gamma_{2 2}^{1 0} \\
 \Gamma_{0 0}^{1 1} & \Gamma_{0 1}^{1 1} &\Gamma_{0 2}^{1 1}&\Gamma_{1 0}^{1 1} &\Gamma_{1 1}^{1 1} & \Gamma_{1 2}^{1 1} &  \Gamma_{2 0}^{1 1} & \Gamma_{2 1}^{1 1} &\Gamma_{2 2}^{1 1} \\
 \Gamma_{0 0}^{1 2} & \Gamma_{0 1}^{1 2} &\Gamma_{0 2}^{1 2}&\Gamma_{1 0}^{1 2} &\Gamma_{1 1}^{1 2} & \Gamma_{1 2}^{1 2} &  \Gamma_{2 0}^{1 2} & \Gamma_{2 1}^{1 2} &\Gamma_{2 2}^{1 2} \\
\Gamma_{0 0}^{2 0} & \Gamma_{0 1}^{2 0} &\Gamma_{0 2}^{2 0}&\Gamma_{1 0}^{2 0} &\Gamma_{1 1}^{2 0} & \Gamma_{1 2}^{2 0} &  \Gamma_{2 0}^{2 0} & \Gamma_{2 1}^{2 0} &\Gamma_{2 2}^{2 0} \\
 \Gamma_{0 0}^{2 1} & \Gamma_{0 1}^{2 1} &\Gamma_{0 2}^{2 1}&\Gamma_{1 0}^{2 1} &\Gamma_{1 1}^{2 1} & \Gamma_{1 2}^{2 1} &  \Gamma_{2 0}^{2 1} & \Gamma_{2 1}^{2 1} &\Gamma_{2 2}^{2 1} \\
 \Gamma_{0 0}^{2 2} & \Gamma_{0 1}^{2 2} &\Gamma_{0 2}^{2 2}&\Gamma_{1 0}^{2 2} &\Gamma_{1 1}^{2 2} & \Gamma_{1 2}^{2 2} &  \Gamma_{2 0}^{2 2} & \Gamma_{2 1}^{2 2} &\Gamma_{2 2}^{2 2} \\
 \end{array}\right)\
\end{eqnarray}
where the same notations as in reference \cite{Fujii} have been used.
Probability conservation implies that each column in the above representation
sums
up to zero. 
Locally, the left vacuum $\langle \widetilde \chi |$ has the representation
\begin{eqnarray}
\label{eq.0.5.2}
\langle \widetilde \chi |=(1\,  1\,  1)\otimes (1 \, 1 \, 1)= 
(1\, 1\,  1\,  1\,  1\,  1\,  1\,  1\,  1).
\end{eqnarray}
The action of any operator on the left-vacuum has then a simple summation 
interpretation. This observation will be crucial in the following computations.
Below we shall assume an initial state $|P(0)\rangle$ and investigate the
expectation value of an operator $O$ (observables such as density, etc.) 
\begin{eqnarray}
\label{eq.0.6}
\langle O \rangle(t)\equiv \langle \widetilde \chi|O e^{-Ht}|P(0)\rangle
\end{eqnarray}
\section{Dynamical Equations of motion}
Exploiting the properties of the left vacuum $\langle \widetilde \chi|$
and denoting by $n_{i}^{\beta}, \beta\in \{0,1,2\}$ 
the occupation of site $i$ by a particle of type $\beta$, 
we derive below the equations of motion of the density and two-point correlation
function.  
For $\beta=0$, $n_i^{0}$ denotes the empty state at site $i$, i.e.,
\begin{eqnarray}
\label{eq.1.1}
n_i^{0}=\openone-n_i^{A}-n_i^{B}
\end{eqnarray}
We evaluate the action of $H$ on the 
operators $n_{i}^{A}, \,n_{i}^{B}$, taking into account the local nature of the
Markov generator: 
\begin{eqnarray}
\label{eq.1.2}
-\langle \widetilde \chi|n_{m}^{A}H_{m,m\pm e^{\alpha}} \; ;\;
-\langle \widetilde \chi|n_{m}^{B}H_{m,m\pm e^{\alpha}} \nonumber\\
\end{eqnarray}
As an example, the first term in the above yields
\begin{eqnarray}
\label{eq.1.3}
&-&\langle \widetilde \chi|n_{m}^{A}H_{m,m+e^{\alpha}}=\sum_{\gamma,\delta=0,1,2}\left\{(\Gamma_{\gamma \delta}^{1 0}+\Gamma_{\gamma \delta}^{1 1} + \Gamma_{\gamma \delta}^{1 2})\langle \widetilde \chi| n_{m}^{\gamma} n_{m+e^{\alpha}}^{\delta}\right\}=\nonumber\\
&& \langle \widetilde \chi| (\Gamma^{1 0}_{0 0}+\Gamma^{1 1}_{0 0}+\Gamma^{1 2}_{0 0})+\left((\Gamma^{1 0}_{1 0}+\Gamma^{1 1}_{1 0}+\Gamma^{1 2}_{1 0})-(\Gamma^{1 0}_{0 0}+\Gamma^{1 1}_{0 0}+\Gamma^{1 2}_{0 0} )\right)\langle \widetilde \chi| n_m^{A} \nonumber\\ &+&\left((\Gamma^{1 0}_{0 1}+\Gamma^{1 1}_{0 1}+\Gamma^{1 2}_{0 1})-(\Gamma^{1 0}_{0 0}+\Gamma^{1 1}_{0 0}+\Gamma^{1 2}_{0 0} )\right)\langle \widetilde \chi| n_{m+e^{\alpha}}^{A}
\nonumber\\&+&\left((\Gamma^{1 0}_{2 0}+\Gamma^{1 1}_{2 0}+\Gamma^{1 2}_{2 0})-(\Gamma^{1 0}_{0 0}+\Gamma^{1 1}_{0 0}+\Gamma^{1 2}_{0 0} )\right)\langle \widetilde \chi| n_{m}^{B}
+
 \left((\Gamma^{1 0}_{0 2}+\Gamma^{1 1}_{0 2}+\Gamma^{1 2}_{0 2})-(\Gamma^{1 0}_{0 0}+\Gamma^{1 1}_{0 0}+\Gamma^{1 2}_{0 0} )\right)\langle \widetilde \chi| n_{m+e^{\alpha}}^{B}
\nonumber\\
&+&\left((\Gamma^{1 0}_{0 0}+\Gamma^{1 1}_{0 0}+\Gamma^{1 2}_{0 0})-(\Gamma^{1 0}_{0 1}+\Gamma^{1 1}_{0 1}+\Gamma^{1 2}_{0 1})+(\Gamma^{1 0}_{1 1}+\Gamma^{1 1}_{1 1}+\Gamma^{1 2}_{1 1})-(\Gamma^{1 0}_{1 0}+\Gamma^{1 1}_{1 0}+\Gamma^{1 2}_{1 0})\right)\langle \widetilde \chi| n_m^{A}n_{m+e^{\alpha}}^{A} \nonumber\\
&+&\left((\Gamma^{1 0}_{0 0}+\Gamma^{1 1}_{0 0}+\Gamma^{1 2}_{0 0})-(\Gamma^{1 0}_{0 2}+\Gamma^{1 1}_{0 2}+\Gamma^{1 2}_{0 2})+(\Gamma^{1 0}_{2 2}+\Gamma^{1 1}_{2 2}+\Gamma^{1 2}_{2 2})  -(\Gamma^{1 0}_{2 0}+\Gamma^{1 1}_{2 0}+\Gamma^{1 2}_{2 0})\right)\langle \widetilde \chi| n_m^{B}n_{m+e^{\alpha}}^{B} \nonumber\\
&+&\left((\Gamma^{1 0}_{0 0}+\Gamma^{1 1}_{0 0}+\Gamma^{1 2}_{0 0})-(\Gamma^{1 0}_{0 2}+\Gamma^{1 1}_{0 2}+\Gamma^{1 2}_{0 2})+(\Gamma^{1 0}_{1 2}+\Gamma^{1 1}_{1 2}+\Gamma^{1 2}_{1 2})-(\Gamma^{1 0}_{1 0}+\Gamma^{1 1}_{1 0}+\Gamma^{1 2}_{1 0})\right)\langle \widetilde \chi| n_m^{A}n_{m+e^{\alpha}}^{B} \nonumber\\
&+&\left((\Gamma^{1 0}_{0 0}+\Gamma^{1 1}_{0 0}+\Gamma^{1 2}_{0 0})-(\Gamma^{1 0}_{0 1}+\Gamma^{1 1}_{0 1}+\Gamma^{1 2}_{0 1})+(\Gamma^{1 0}_{2 1}+\Gamma^{1 1}_{2 1}+\Gamma^{1 2}_{2 1})-(\Gamma^{1 0}_{2 0}+\Gamma^{1 1}_{2 0}+\Gamma^{1 2}_{2 0})\right)\langle \widetilde \chi| n_m^{B}n_{m+e^{\alpha}}^{A} \nonumber\\
\end{eqnarray}
where the use of equation (\ref{eq.0.2}) is required to substitute for
$\Gamma_{1 1}^{1 1},\Gamma_{1 0}^{1 0}, \Gamma_{1 2}^{1 2}$.  
As expected, the stochastic Hamiltonian connects the one-body 
initial operator to a two-body expression.

The equation of motion for the density becomes (at site $m$)
\begin{eqnarray}
\label{eq.1.4}
\frac{d}{dt}\langle n_m^{A,B}\rangle (t)\equiv
\frac{d}{dt}\langle \widetilde \chi| n_m^{A,B}e^{-Ht}|P(0)\rangle=
-\sum_{e^{\alpha}}\langle n_{m}^{A,B}(H_{m,m+e^{\alpha}}+ H_{m-e^{\alpha},m})\rangle(t)
\end{eqnarray}
In order to determine the second moments, we also need to evaluate the following
terms
\begin{eqnarray}
\label{eq.1.5}
-\langle \widetilde \chi|n_{m}^{A}n_{m+e^{\alpha}}^{A} H_{m,m+e^{\alpha}} \; ; \;
-\langle \widetilde \chi|n_{m}^{A}n_{m+e^{\alpha}}^{B}H_{m,m+e^{\alpha}} \; ; \;
-\langle \widetilde \chi|n_{m}^{B}n_{m+e^{\alpha}}^{B} H_{m,m+e^{\alpha}} \; ; \;
-\langle \widetilde \chi|n_{m}^{A}n_{m+e^{\alpha}}^{B}H_{m,m+e^{\alpha}} \nonumber\\
\end{eqnarray}
For the sake of illustration, the first term above yields
\begin{eqnarray}
\label{eq.1.6}
&-&\langle \widetilde \chi|n_{m}^{A}n_{m+e^{\alpha}}^{A} H_{m,m+e^{\alpha}}= \nonumber\\
&&\Gamma_{0 0}^{1 1} \langle \widetilde \chi| +(\Gamma_{1 0}^{1 1}-\Gamma_{0 0}^{1 1})\langle \widetilde\chi|n_{m}^{A}+(\Gamma_{0 1}^{1 1}-\Gamma_{0 0}^{1 1})\langle \widetilde \chi|n_{m+e^{\alpha}}^{A}
+(\Gamma_{2 0}^{1 1}-\Gamma_{0 0}^{1 1})\langle \widetilde\chi|n_{m}^{B}+
(\Gamma_{0 2}^{1 1}-\Gamma_{0 0}^{1 1})\langle \widetilde\chi|n_{m+e^{\alpha}}^{B}\nonumber\\&+& (\Gamma_{0 0}^{1 1}+\Gamma_{1 1}^{1 1}-\Gamma_{1 1}^{0 1}-\Gamma_{1 1}^{1 0})\langle \widetilde\chi|n_{m}^{A}n_{m+e^{\alpha}}^{A} +
 (\Gamma_{0 0}^{1 1}+\Gamma_{2 2}^{1 1}-\Gamma_{0 2}^{1 1}-\Gamma_{2 0}^{1 1})\langle \widetilde \chi|n_{m}^{B}n_{m+e^{\alpha}}^{B}  \nonumber\\
&+& (\Gamma_{0 0}^{1 1}+\Gamma_{1 2}^{1 1}-\Gamma_{0 2}^{1 1}-\Gamma_{1 0}^{1 1})\langle \widetilde\chi|n_{m}^{A}n_{m+e^{\alpha}}^{B}+ (\Gamma_{0 0}^{1 1}+\Gamma_{2 1}^{1 1}-\Gamma_{2 0}^{1 1}-\Gamma_{0 1}^{1 1})\langle \widetilde\chi|n_{m}^{B}n_{m+e^{\alpha}}^{A}
\end{eqnarray}
Notice that the evolution operator connects a two-body operator to a 
two-body expression.

To compute the two-point correlation functions, we have to distinguish
the sites that are nearest neighbors from those that are not.
If the sites $m$ and$n$ are not nearest neighbors $(dist(m,n)>1)$, 
the equation of motion reads
\begin{eqnarray}
\label{eq.1.7}
-\frac{d}{dt}\langle n_{m}^{A,B}n_{n}^{A,B}\rangle(t)&=&\sum_{\alpha}\left(\langle(n_{m}^{A,B}H_{m-e^{\alpha},m}) n_{n}^{A,B}\rangle(t)+\langle n_{m}^{A,B}(n_{n}^{A,B}H_{n-e^{\alpha},n}) \rangle(t)\right)\nonumber\\ &+& \sum_{\alpha}\left( \langle (n_{m}^{A,B}H_{m,m+e^{\alpha}}) n_{n}^{A,B} \rangle (t) +\langle n_{m}^{A,B}(n_{n}^{A,B} H_{n,n+e^{\alpha}}) \rangle (t) \right)
\end{eqnarray}

while if the sites are nearest neighbors, we have
\begin{eqnarray}
\label{eq.1.8}
-\frac{d}{dt}\langle n_{m}^{A,B}n_{m+e^{\alpha}}^{A,B}\rangle(t)&=&
\langle n_{m}^{A,B} n_{m+e^{\alpha}}^{A,B} H_{m,m+e^{\alpha}}\rangle (t) \nonumber\\&+&
\sum_{\alpha'}\left(\langle(n_m^{A,B}H_{m-e^{\alpha'},m}) n_{m+e^{\alpha}}^{A,B}\rangle(t) +
 \langle n_{m}^{A,B}(n_{m+e^{\alpha}}^{A,B}H_{m+e^{\alpha},m+e^{\alpha}+e^{\alpha'}})\rangle (t) \right) \nonumber\\
&+&\sum_{\alpha'\neq \alpha}\left(\langle(n_m^{A,B} H_{m,m+e^{\alpha'}}) n_{m+e^{\alpha}}^{A,B}\rangle(t) +
 \langle n_{m}^{A,B}(n_{m+e^{\alpha}}^{A,B} H_{m+e^{\alpha}-e^{\alpha'}, m+e^{\alpha}}) \rangle (t) \right)
\end{eqnarray}

The equations of motion of $n-$points correlation functions are obtained 
in a similar way,  with help of  (\ref{eq.1.2}, 
\ref{eq.1.3}) and (\ref{eq.1.5}, \ref{eq.1.6}) .
As is well known, the equations of motion of classical (or quantum)
correlation functions
constitute an open hierarchy which is not soluble in general.
However, if we impose on the $3^4-3^2=72$ bimolecular transition rates
involving two adjacent sites,  
the following $16$ contraints (see appendix A for the definitions (\ref{eq.A.1},\ref{eq.A.2})).
\begin{eqnarray}
\label{eq.1.9}
&1&) A_{2}^{a}+A_{1}^{a}+A_{0}^{a}=\Gamma_{1 1}^{1 0}+ \Gamma_{1 1}^{1 1}+\Gamma_{1 1}^{1 2} \nonumber\\
&2&) B_{2}^{a}+B_{1}^{a}+A_{0}^{a}=\Gamma^{1 0}_{2 2}+ \Gamma^{1 1}_{2 2}+\Gamma^{1 2}_{2 2} \nonumber\\
&3&) B_{2}^{a}+A_{1}^{a}+A_{0}^{a}=\Gamma^{1 0}_{1 2}+ \Gamma^{1 1}_{1 2}+\Gamma^{1 2}_{1 2} \nonumber\\
&4&) A_{2}^{a}+B_{1}^{a}+A_{0}^{a}=\Gamma^{1 0}_{2 1}+ \Gamma^{1 1}_{2 1}+\Gamma^{1 2}_{2 1} \nonumber\\
&5&) C_{2}^{a}+C_{1}^{a}+C_{0}^{a}=\Gamma^{0 1}_{1 1}+ \Gamma^{1 1}_{1 1}+\Gamma^{2 1}_{1 1} \nonumber\\
&6&) D_{2}^{a}+D_{1}^{a}+C_{0}^{a}=\Gamma^{0 1}_{2 2}+ \Gamma^{1 1}_{2 2}+\Gamma^{2 1}_{2 2} \nonumber\\
&7&) D_{2}^{a}+C_{1}^{a}+C_{0}^{a}=\Gamma^{0 1}_{1 2}+ \Gamma^{1 1}_{1 2}+\Gamma^{2 1}_{1 2} \nonumber\\
&8&) C_{2}^{a}+D_{1}^{a}+C_{0}^{a}=\Gamma_{2 1}^{0 1}+ \Gamma_{2 1}^{1 1}+\Gamma_{2 1}^{2 1} \nonumber\\
&9&) A_{2}^{b}+A_{1}^{b}+A_{0}^{b}=\Gamma^{2 0}_{1 1}+ \Gamma^{2 1}_{1 1}+\Gamma^{2 2}_{1 1} \nonumber\\
&10&) B_{2}^{b}+B_{1}^{b}+A_{0}^{b}=\Gamma^{2 0}_{2 2}+ \Gamma^{2 1}_{2 2}+\Gamma^{2 2}_{2 2} \nonumber\\
&11&) B_{2}^{b}+A_{1}^{b}+A_{0}^{b}=\Gamma^{2 0}_{1 2}+ \Gamma^{2 1}_{1 2}+\Gamma^{2 2}_{1 2} \nonumber\\
&12&) A_{2}^{b}+A_{0}^{b}+B_{1}^{b}=\Gamma^{2 0}_{2 1}+ \Gamma^{2 1}_{2 1}+\Gamma^{2 2}_{2 1} \nonumber\\
&13&) C_{0}^{b}+C_{1}^{b}+C_{2}^{b}=\Gamma^{0 2}_{1 1}+ \Gamma^{1 2}_{1 1}+\Gamma^{2 2}_{1 1} \nonumber\\
&14&) C_{0}^{b}+D_{1}^{b}+D_{2}^{b}=\Gamma^{0 2}_{2 2}+ \Gamma^{1 2}_{2 2}+\Gamma^{2 2}_{2 2} \nonumber\\
&15&) C_{0}^{b}+C_{1}^{b}+D_{2}^{b}=\Gamma^{0 2}_{1 2}+ \Gamma^{1 2}_{1 2}+\Gamma^{2 2}_{1 2} \nonumber\\
&16&) C_{0}^{b}+C_{2}^{b}+D_{1}^{b}=\Gamma^{0 2}_{2 1}+ \Gamma^{1 2}_{2 1}+\Gamma^{2 2}_{2 1} \nonumber\\
\end{eqnarray}
the equations of motion of the density and
two-point correlation functions (and all multi-points correlation functions)  
become closed. It is worth emphasizing that when the hierarchy closes at the
lowest level, i.e., at the level of the density, the equations of motion of 
{\it all} higher correlation functions also close. This is a remarkable
property.  

In the expressions (\ref{eq.1.8}) and (\ref{eq.1.9}), 
the rates
$\Gamma_{\alpha \beta}^{\alpha \beta}$ have not been made explicit for brevity.\\
A general diffusion-limited two-species reaction model
is defined on the manifold, 
$V_{par}=\Big\{\Gamma_{\alpha \beta}^{\gamma \delta} 
-\{\Gamma_{\alpha \beta}^{\alpha \beta}\}|\alpha,\beta \in (0, 1, 2)\Big\}$,
which has here $3^4-9=72$ independent parameters.
Let us denote by $V_{sol}$ the restriction of $V_{par}$ on the ($72-16=56$
parameters) 
manifold defined
by the additional constraints (\ref{eq.1.9}):
$V_{sol}\equiv V_{par}\cap(\ref{eq.1.9}) $. The latter represents the manifold
on which the equations of motion of the correlation functions are closed, i.e.,
the soluble manifold.   
We can further require translation invariance, i.e., 
$\langle n_{m}^{i}n_{m+|r|}^{j}\rangle(t) 
= \langle n_{0}^{i}n_{|r|}^{j}\rangle (t)\equiv{\cal G}_{|r|}^{ij}(t) , \, 
\forall r, t$ ($i,j \in(A, B)$) and in particular
$\langle n_{m}^{A}n_{n}^{B}\rangle(t)= \langle n_{m}^{B}n_{n}^{A}\rangle(t)
\equiv {\cal G}_{|n-m|}^{AB}(t)$. 
Imposing the above conditions in equations (\ref{eq.1.7}) and(\ref{eq.1.8}) 
and taking into account the conditions of solubility
(\ref{eq.1.9}), we arrive at the manifold $V_{transl-invar}$, 
the restriction of $V_{sol}$ on the translation invariant soluble dynamics.
Notice that $V_{transl-invar}(d)=V_{sol}\cap V'(d)$,  where $V'(d)=\Big\{
E_{0}^{ab}=E_{0}^{ba}; \; F_{1}^{ab}+F_{2}^{ab}+A_{0}^{b}d +C_{0}^{b}(d-1)=
F_{1}^{ba}+F_{2}^{ba}+C_{0}^{b}d +A_{0}^{b}(d-1); \;
 F_{3}^{ab}+F_{4}^{ab}+A_{0}^{a}(d-1) +C_{0}^{a}d=
F_{3}^{ba}+F_{4}^{ba}+C_{0}^{a}(d-1) +A_{0}^{a}d; \;
H_{1}^{ab}+H_{2}^{ab}+(C_{2}^{a}+B_{1}^{b})d + (A_{1}^{a}+D_{2}^{b})(d-1)=
H_{1}^{ba}+H_{2}^{ba}+(A_{1}^{a}+D_{2}^{b})d + (B_{1}^{b}+C_{2}^{a})(d-1); \;
A_{2}^{a}+D_{1}^{b}=B_{2}^{b}+C_{1}^{a}; \;
G_{1}^{ba}+C_{2}^{b}d + A_{1}^{b}(d-1)=G_{1}^{ab}+C_{2}^{b}(d-1) + A_{1}^{b}d; \;
G_{2}^{ab}+B_{1}^{a}(d-1)+D_{2}^{a}d =G_{2}^{ba}+B_{1}^{a}d+D_{2}^{a}(d-1); \;
B_{2}^{a}=D_{1}^{a}; \; A_{2}^{b}=C_{1}^{b}
 \Big\}$. 
Therefore this manifold has $72-16-9=47$ independent parameters.
In practice, however further constraints may be required for the computations to
be accessible. With this remark in mind, we define the manifolds
$V''=\Big\{ A_{1}^{b}+C_{2}^{b}=A_{2}^{b}= C_{1}^{b}=B_{1}^{a}+D_{2}^{a}
 =B_{2}^{a}=D_{1}^{a}=0 \Big\}$ and
$V'''=\Big\{A_{n}^{b}=B_{n}^{a}=C_{n}^{b}=D_{n}^{a}=G_{2}^{a}=G_{1}^{b}=
G_{1}^{ab}=G_{2}^{ab}=H_{n}^{a,b}=0\;   |n=1,2\Big\} $.

Summarizing the cases that we will discuss
in this paper: \\

i) For the case where translation invariance is broken, we shall compute the
exact density on the manifold $V_{1}$, which has $72-16-6=50$ independent
parameters,
\begin{eqnarray}
\label{eq.1.12}
V_{1}\equiv \cap V_{sol}\cap V''
\end{eqnarray}

ii) For the translation invariant case, we shall evaluate both, the density and
two-point correlation function exactly on the manifold $V_{2}(d)$, 
\begin{eqnarray}
\label{eq.1.13}
V_{2}(d)\equiv V_{sol}\cap V'(d)\cap V'''\equiv V_{transl-invar}(d)\cap V'''
\end{eqnarray}
which has $47-16=31$ independent parameters.

To conclude this section, it is worth noting 
that there are few cases in which the open hierarchy of equations of motion
can be solved 
analytically. 
The class of single-species one-dimensional models for which 
the evolution operator can 
be cast into a {\it free fermionic} form is an important example. 
However, the procedure of free "fermionization" cannot be 
applied to higher dimensions and/or multispecies 
problems, contrary to the method followed here.

\section{The density: general discussion}
In the first part of this section we compute exactly the Fourier-Laplace transform 
of the density on the $56-$parameters manifold $V_{sol}$, which is, as are
correlation functions, 
directly related to light scattering measurements in 
{\it real} reaction-diffusion systems \cite{Kroon,Schutz2,Grynberg2}.
The computation on the most general soluble manifold is here manageable because the linear differential difference equations 
governing the dynamics give rise to a $2\times 2$ matrix, the properties 
of which can be studied analytically. For higher order correlation functions
and/or for the s-species case, with $s>2$, the 
problem is however technically much harder (we shall come back to the general case in a future work).
In the second part of this section we provide the density of 
species $A$ and $B$ in space and time,
both in the translation invariant case and in a situation where
translation
invariance is broken.
On the manifold $(V_{par}\cap V_{sol})\supset 
(V_{transl-invar}\cap V_{sol}) $, we have
\begin{eqnarray}
\label{eq.2.2}
\frac{d}{dt}\langle n_m^{A}\rangle=(A_0^{a}+C_0^{a})d+
\langle n_m^{A}\rangle(t)(A_{1}^{a} + C_{2}^{a})d+\sum_{\alpha}\left(A_2^{a}\langle n_{m+e^{\alpha}}^{A}\rangle(t) + C_1^{a}\langle n_{m-e^{\alpha}}^{A}\rangle(t)\right) \nonumber\\ +\langle n_m^{B}\rangle(t) (B_{1}^{a} + D_{2}^{a})d +\sum_{\alpha}\left(B_2^{a}\langle n_{m+e^{\alpha}}^{B}\rangle(t) + D_1^{a}\langle n_{m-e^{\alpha}}^{B}\rangle(t)\right)
\end{eqnarray}
\begin{eqnarray}
\label{eq.2.3}
\frac{d}{dt}\langle n_m^{B}\rangle=(A_0^{b}+C_0^{b})d+
\langle n_m^{B}\rangle(t)(B_{1}^{b} + D_{2}^{b})d +\sum_{\alpha}\left(B_2^{b}\langle n_{m+e^{\alpha}}^{B}\rangle(t) + D_1^{b}\langle n_{m-e^{\alpha}}^{B}\rangle(t)\right) \nonumber\\ +\langle n_m^{A}\rangle(t) (A_{1}^{b} + C_{2}^{b})d +\sum_{\alpha}\left(A_2^{b}\langle n_{m+e^{\alpha}}^{A}\rangle(t) + C_1^{b}\langle n_{m-e^{\alpha}}^{A}\rangle(t)\right)
\end{eqnarray}
Let us first consider the most general soluble case which is characterized by 
the set of equations (\ref{eq.2.2},\ref{eq.2.3}). 
The solution of (\ref{eq.2.2},\ref{eq.2.3}) is split into the solution of the
homogeneous system
$\langle n_{m}^{A}\rangle_{h}(t)$  
($\langle n_{m}^{B}\rangle_{h}(t)$) and a function $f_{A}(t)$ ($f_{B}(t)$) that
takes into account the inhomogeneity, i.e., 
\begin{eqnarray}
\label{eq.2.2.1}
\frac{d}{dt} f_{A}(t)=(A_{0}^{a}+ C_{0}^{a})d + f_{A}(t) (A_{1}^{a}+ A_{2}^{a} + C_{1}^{a}+ C_{2}^{a})d  +  f_{B}(t) (B_{1}^{a}+ B_{2}^{a} + D_{1}^{a}+ D_{2}^{a})d,
\end{eqnarray}
\begin{eqnarray}
\label{eq.2.2.2}
\frac{d}{dt} f_{B}(t)= (A_{0}^{b}+ C_{0}^{b})d +  f_{B}(t) (B_{1}^{b}+ B_{2}^{b} + D_{1}^{b}+ D_{2}^{b})d +  f_{A}(t) (A_{1}^{b}+ A_{2}^{b} + C_{1}^{b}+ C_{2}^{b}) d,
\end{eqnarray}
We introduce the Fourier transforms of $\langle n_m^{A,B}\rangle_{h}(t)$, i.e., 
\begin{eqnarray}
\label{eq.2.2.3}
\langle n_m^{A,B}\rangle_{h}(t)=\sum_{\vec p \in 1.B.Z.} 
\langle \hat{n}_{\vec{p}}^{A,B}\rangle(t) e^{i\vec{p}.m} \Longleftrightarrow
\langle \hat{n}_{\vec{p}}^{A,B}\rangle(t) =\frac{1}{L^d}\sum_{m} 
\langle n_m^{A,B}\rangle_{h}(t)   e^{-i\vec{p}.m}
\end{eqnarray}
where the sum on $\vec{p}$ runs over the first Brillouin zone ($ 1.B.Z.$).
The solution of the  homogeneous problem in Fourier space reads

\begin{eqnarray}
\label{eq.2.2.6}
\left(
 \begin{array}{c c}
\langle \hat{n}_{\vec{p}}^{A}\rangle(t)   \\
\langle \hat{n}_{\vec{p}}^{B}\rangle(t)   \\
\end{array}\right)= e^{{\cal M}(p) t}
\left(
 \begin{array}{c c }
 \langle \hat{n}_{\vec{p}}^{A}\rangle(t=0)   \\
 \langle \hat{n}_{\vec{p}}^{B}\rangle(t=0)   \\
\end{array}\right)
\end{eqnarray}
where ${\cal M}_{i,j}(p), (i,j)\in (1,2)$ is a  $s\times s=2 \times 2$ matrix 
with the entries
\begin{eqnarray}
\label{eq.2.2.7}
{\cal M}_{1,1}(p)&=&(A_{1}^{a}+C_{2}^{a})d + \sum_{\alpha}\left(A_{2}^{a}e^{i\vec{p}.e_{\alpha}} +C_{1}^{a}e^{-i\vec{p}.e_{\alpha}}  \right) \nonumber\\
{\cal M}_{1,2}(p)&=&  (B_{1}^{a}+D_{2}^{a})d + \sum_{\alpha}\left(B_{2}^{a}e^{i\vec{p}.e_{\alpha}} +D_{1}^{a}e^{-i\vec{p}.e_{\alpha}}  \right) \nonumber\\
{\cal M}_{2,1}(p)&=& (A_{1}^{b}+C_{2}^{b})d + \sum_{\alpha}\left(A_{2}^{b}e^{i\vec{p}.e_{\alpha}} +C_{1}^{b}e^{-i\vec{p}.e_{\alpha}}  \right)  \nonumber\\
{\cal M}_{2,2}(p)&=& (B_{1}^{b}+D_{2}^{b})d + \sum_{\alpha}\left(B_{2}^{b}e^{i\vec{p}.e_{\alpha}} +D_{1}^{b}e^{-i\vec{p}.e_{\alpha}}  \right)
\end{eqnarray}
The eigenvalues of the  matrix ${\cal M}$ , which represent 
the inverse relaxation times of the system, control the 
asymptotic behavior of the density, 
\begin{eqnarray}
\label{eq.2.2.8}
\lambda_{\pm}(p)=\frac{{\cal M}_{1,1}(p)+{\cal M}_{2,2}(p)}{2}
\pm\frac{\sqrt{({\cal M}_{1,1}(p) -{\cal M}_{2,2}(p))^2 + 4{\cal M}_{1,2}(p)
{\cal M}_{2,1}(p) }}{2}
\end{eqnarray}
It has been shown in considering the one-dimensional alternating-bonds Ising 
model 
obeying Glauber's dynamics \cite{Droz} that the relaxational eigenvalues of the 
analog of the matrix ${\cal M}$ allow to identify the critical 
(but non-universal) behavior: it is determined by the long wave-length $p$ 
modes of the 
analog of the acoustic $\lambda_{-}$ branch. 
One-dimensional alternating-bonds ($J_{1}>J_{2}>0$) Ising model, with 
Glauber's dynamics, exhibits a non-universal critical dynamical
exponent $z=1+\frac{J_{1}}
{J_{2}}$ \cite{Droz}.
\\
\\
In the sequel we shall need the zero-momentum 
$2 \times 2$ matrix ${\cal M}(p=0)\equiv{\cal M}(0) $ 
\begin{eqnarray*}
\label{eq.2.2.9}
{\cal M}_{1,1}(0)&=&(A_{1}^{a}+A_{2}^{a} +C_{1}^{a} + C_{2}^{a})d \nonumber\\
{\cal M}_{1,2}(0)&=&(B_{1}^{a}+B_{2}^{a} +D_{1}^{a}+D_{2}^{a})d   \nonumber\\
{\cal M}_{2,1}(0)&=& (A_{1}^{b} +A_{2}^{b}+C_{1}^{b}+ C_{2}^{b})d   \nonumber\\
{\cal M}_{2,2}(0)&=& (B_{1}^{b}+B_{2}^{b} +D_{1}^{b}+D_{2}^{b})d
\end{eqnarray*}
whose eigenvalues we shall denote, for short,
\begin{eqnarray*}
\label{eq.2.2.10}
\gamma_{\pm}=\lambda_{\pm}(p=0)
\end{eqnarray*}

Notice that at $p=0$, $Tr {\cal M}(0)<0$ and $det{\cal M}(0)\geq 0 $

We are now in a position to compute the Fourier-Laplace transform 
${\cal S}_{0}^{A,B}(\vec{p},\omega)$ of the density : ${\cal S}_{0}^{A,B}(\vec{p},\omega) \equiv \frac{1}{L^d}\sum_{m}\int_{0}^{\infty} dt e^{-i\vec{p}.m -\omega t} \langle n_{m}^{A,B}\rangle (t)$.

We consider initial states 
$\langle\hat{n}_{p}^{A,B}\rangle (0)= \frac{1}{L^d}\sum_{m} 
\langle {n}_{m}^{A,B}\rangle(0) e^{-i\vec{p}.m} (\leftrightarrow
\langle {n}_{m}^{A,B}\rangle(0)=\sum_{\vec{p} \in 1B.Z.} \langle \hat{n}_{p}^{A,B} \rangle(0) e^{i\vec{p}.m} )$ and assume that the 
matrix ${\cal M}(p)$ is regular (i.e. $det {\cal M}(p)\neq 0$).
We shall distinguish four cases. 

1) We assume that the matrix  ${\cal M}(p)$ is diagonalizable, i.e.,
 $\lambda_{+}(p)\neq\lambda_{-}(p)$, but not triangular, ${\cal M}_{1,2}(p)\neq 0 $, and obtain
\begin{eqnarray}
\label{eq.2.2.13}
&{\cal S}_{0}^{A}(\vec{p},\omega)&=
\frac{1}{\lambda_{-}(p)-\lambda_{+}(p)}\left[\frac{\lambda_{-}(p)-
{\cal M}_{1,1}(p)}{\omega -\lambda_{+}(p)} -\frac{\lambda_{+}(p)-
{\cal M}_{1,1}(p)}{\omega -\lambda_{-}(p)}  \right]\langle \hat{n}_{p}^{A}(0) 
\rangle 
 + \frac{{\cal M}_{1,2}(p) \langle \hat{n}_{p}^{B}(0) \rangle  }
 {(\omega -\lambda_{+}(p))(\omega -\lambda_{-}(p))} \nonumber\\
&+& \frac{\delta_{p,0}}{\gamma_{-}-\gamma_{+}}\left[(\gamma_{-}-
{\cal M}_{1,1}(0))(A_{0}^{a}+ C_{0}^{a})d -{\cal M}_{1,2}(0)
(A_{0}^{b}+ C_{0}^{b})d\right] \frac{1}{\omega(\omega -\gamma_{+})} 
\nonumber\\
&-& \frac{\delta_{p,0}}{\gamma_{-}-\gamma_{+}}\left[(\gamma_{+}-
{\cal M}_{1,1}(0))(A_{0}^{a}+ C_{0}^{a})d -{\cal M}_{1,2}(0)(A_{0}^{b}
+ C_{0}^{b})d\right] \frac{1}{\omega(\omega -\gamma_{-})}
\end{eqnarray}
and
\begin{eqnarray}
\label{eq.2.2.14}
&{\cal S}_{0}^{B}(\vec{p},\omega)&=
\langle\hat{n}_{p}^{B}(0)\rangle\left[\frac{\lambda_{-}(p) - 
{\cal M}_{1,1}(p)}{ \omega-\lambda_{-}(p)}-\frac{\lambda_{+}(p) - 
{\cal M}_{1,1}(p)}{\omega-\lambda_{+}(p)}\right]-
\left[\frac{(\lambda_{+}(p) - {\cal M}_{1,1}(p) )(\lambda_{-}(p) - 
{\cal M}_{1,1}(p) )}{{\cal M}_{1,2} (p) (\omega-\lambda_{+}(p))
(\omega-\lambda_{-}(p))}\right]\langle\hat{n}_{p}^{A}(0)\rangle 
\nonumber\\
&+& \frac{\delta_{p,0}}{\gamma_{-}-\gamma_{+}}\left[\frac{(\gamma_{-}-
{\cal M}_{1,1}(0))}{{\cal M}_{1,2}(0)}(A_{0}^{a}+ C_{0}^{a})d -
(A_{0}^{b}+ C_{0}^{b})d\right] \frac{\gamma_{+}-{\cal M}_{1,1}(0) }
{\omega(\omega -\gamma_{+})} \nonumber\\
&-& \frac{\delta_{p,0}}{\gamma_{-}-\gamma_{+}}\left[\frac{(\gamma_{+}-
{\cal M}_{1,1}(0))}{{\cal M}_{1,2}(0)}(A_{0}^{a}+ C_{0}^{a})d -
(A_{0}^{b}+ C_{0}^{b})d\right] \frac{\gamma_{-}-{\cal M}_{1,1}(0)}
{\omega(\omega -\gamma_{-})}
\end{eqnarray}
Notice that the inhomogeneous part of the equations of motion give rise 
to a zero-momentum contribution which we shall omit hereafter. As expected the
poles in the $\omega$-plane occur at the relaxational eigenvalues.

2) Next consider the case where the matrix ${\cal M} (p)$ is non-diagonalizable 
and non-triangular: $\lambda(p)=\lambda_{+}(p)=\lambda_{-}(p)=
\frac{{\cal M}_{1,1}(p)+{\cal M}_{2,2}(p)}{2}$  and  ${\cal M}_{1,1}(p)\neq 
{\cal M}_{2,2}(p)$ .
We can compute $e^{{\cal M}(p)t}$ with the help of a Jordan decomposition of the 
matrix ${\cal M}(p)$, namely,
\begin{eqnarray}
\label{eq.2.2.15}
e^{{\cal M}(p)t}=P(p) e^{{\cal M}'(p) t}P^{-1}(p)
\end{eqnarray}
where $P$ is a regular $2 \times 2$ matrix and ${\cal M}'(p)={\cal M}_{1}(p)+ 
{\cal M}_{2}(p)$
is the sum of a diagonal matrix ${\cal M}_{1}$ and a Jordan-block matrix. 
${\cal M}_{2}(p) $ is chosen such that $[{\cal M}_{1}(p), {\cal M}_{2}(p)] = 0$. 
${\cal M}_{2} (p)$ is nilpotent ($({\cal M}_{2}(p))^2 =0$).
Thus,
\begin{eqnarray}
\label{eq.2.2.15}
e^{{\cal M}(p) t}=P(p) e^{{\cal M}'(p) t}P(p)^{-1}=
\left(
 \begin{array}{c c}
 \alpha(p) & \epsilon(p) \\
 \beta(p) & \delta(p) 
 \end{array}\right)\ 
\left(
 \begin{array}{c c}
 e^{\lambda(p)t} &  \lambda (p) t  e^{\lambda(p)t}  \\
 0 &  e^{\lambda(p)t}
 \end{array}\right)\
\left(
 \begin{array}{c c}
 \delta(p) & -\epsilon(p) \\
 -\beta(p) & \alpha(p) 
 \end{array}\right)\frac{1}{(\alpha(p) \delta(p) - \beta(p) \epsilon(p))}\
\end{eqnarray}
With
\begin{eqnarray}
\label{eq.2.2.15.0}
P(p) =
\left(
 \begin{array}{c c}
 \alpha(p) & \epsilon(p) \\
 \beta(p) & \delta(p) 
 \end{array}\right)\ 
\end{eqnarray}
which entries are:
\begin{eqnarray}
\label{eq.2.2.16}
\alpha(p)&\equiv& \frac{\lambda(p)}{{\cal M}_{1,1}(p)} 
\left(1+\frac{{\cal M}_{1,2}(p){\cal M}_{2,1}(p)}{det {\cal M}(p)}\right) 
\nonumber\\
\beta(p)&\equiv& -\frac{\lambda(p){\cal M}_{2,1}(p)}{det{\cal M}(p)} 
\nonumber\\
\epsilon(p)&\equiv&  \frac{\lambda(p)({\cal M}_{2,2}(p) - {\cal M}_{1,2}(p))}
{det{\cal M}(p)} \nonumber\\
\delta(p)&\equiv& \frac{\lambda(p)}{{\cal M}_{1,2}(p)} 
\left(1-\frac{{\cal M}_{1,1}(p)( {\cal M}_{2,2}(p) -{\cal M}_{1,2}(p))}
{det {\cal M}(p)}\right)
\end{eqnarray}
The matrix $P(p)$ is regular ($det P(p) \neq 0$) if ${\cal M}(p)$ is regular.
This decomposition leads to the form factors ($\vec{p}\neq 0$)
\begin{eqnarray}
\label{eq.2.2.17}
{\cal S}_{0}^{A}(\vec{p},\omega)=\frac{1}{\omega-\lambda(p)}\left[
\langle \hat{n}_{p}^{A}\rangle (0) - \frac{\lambda(p)}{(\omega-\lambda(p)) 
det P(p)}\left( \alpha^2(p) \langle\hat{n}_{p}^{B}\rangle (0) -\alpha(p)\beta(p)  
\langle\hat{n}_{p}^{A}\rangle (0) \right)\right]
\end{eqnarray}
\begin{eqnarray}
\label{eq.2.2.18}
{\cal S}_{0}^{B}(\vec{p},\omega)=\frac{1}{\omega-\lambda(p)}\left[
\langle \hat{n}_{p}^{B}\rangle (0) - \frac{\lambda(p)}{(\omega-\lambda(p)) 
det P(p)}\left( \alpha(p)\beta(p) \langle\hat{n}_{p}^{B}\rangle (0) -\beta^2(p)  
\langle\hat{n}_{p}^{A}\rangle (0) \right)\right]
\end{eqnarray}

Notice that $\lambda_{+}(p)=\lambda_{-}(p)=
\lambda(p)=\frac{{\cal M}_{1,1}(p)+{\cal M}_{2,2}(p)}{2}$, imply the 
following relations 
on the reaction-rates:
\begin{eqnarray}
\label{eq.2.2.18.1}
&&\sum_{\beta,\beta'=0,1,2}\left(\Gamma_{1 0}^{\beta 2}\Gamma_{2 0}^{\beta' 1}
 + \Gamma_{0 0}^{\beta 1}\Gamma_{0 0}^{\beta' 2} \right) \leq
\sum_{\beta,\beta'=0,1,2}\left(\Gamma_{1 0}^{\beta 2}\Gamma_{0 0}^{\beta' 1}
 + \Gamma_{0 0}^{\beta 2}\Gamma_{2 0}^{\beta' 1} \right) \nonumber\\
\nonumber \\
&&\sum_{\beta,\beta'=0,1,2}\left(\Gamma_{0 1}^{2 \beta}\Gamma_{0 2}^{1 \beta'}
 + \Gamma_{0 0}^{1 \beta}\Gamma_{0 0}^{2 \beta'} \right) \leq 
\sum_{\beta,\beta'=0,1,2}\left(\Gamma_{0 0}^{2 \beta}\Gamma_{0 2}^{1 \beta'}
 + \Gamma_{0 0}^{1 \beta}\Gamma_{0 1}^{2 \beta'} \right) \nonumber\\
\nonumber\\
&&\sum_{\beta,\beta'=0,1,2}\left(\Gamma_{2 0}^{1 \beta}\Gamma_{1 0}^{2 \beta'}
 + \Gamma_{2 0}^{1 \beta}\Gamma_{0 1}^{\beta' 2} +
 \Gamma_{0 2}^{\beta 1}\Gamma_{0 1}^{\beta' 2} +\Gamma_{0 2}^{\beta 1}\Gamma_{1 0}^{2 \beta'} + (\Gamma_{0 0}^{1 \beta}+
\Gamma_{0 0}^{\beta 1})(\Gamma_{0 0}^{2 \beta'}+
\Gamma_{0 0}^{\beta' 2} )\right)  \nonumber\\
 &\leq& 
\sum_{\beta,\beta'=0,1,2}\left((\Gamma_{0 0}^{1 \beta}+
\Gamma_{0 0}^{\beta 1})(\Gamma_{0 0}^{2 \beta'}+
\Gamma_{0 0}^{\beta' 2} )+ (\Gamma_{2 0}^{1 \beta}+
\Gamma_{0 2}^{\beta 1})(\Gamma_{0 0}^{2 \beta'}+
\Gamma_{0 0}^{\beta' 2} )  \right) \nonumber\\
\nonumber\\
&& \sum_{\beta,\beta'=0,1,2}\left(\Gamma_{0 2}^{1 \beta}
\Gamma_{1 0}^{\beta' 2}+
 \Gamma_{0 0}^{1 \beta}\Gamma_{0 0}^{\beta' 2} +  \Gamma_{0 0}^{2 \beta}
 \Gamma_{0 0}^{\beta' 1}  +  \Gamma_{0 0}^{\beta 1}\Gamma_{0 1}^{2 \beta'}
 \right)= 
 \sum_{\beta,\beta'=0,1,2}\left(\Gamma_{0 0}^{1 \beta}\Gamma_{1 0}^{\beta' 2}+
  \Gamma_{0 0}^{\beta 2}\Gamma_{0 2}^{1 \beta} +  \Gamma_{0 1}^{2 \beta}
  \Gamma_{2 0}^{\beta' 1}  +  \Gamma_{0 0}^{2 \beta}\Gamma_{0 0}^{\beta' 1}  
  \right) \nonumber\\
\nonumber\\
&&\sum_{\beta,\beta'=0,1,2}\left[\Gamma_{0 1}^{1 \beta}\Gamma_{1 0}^{\beta' 1}+
\Gamma_{0 2}^{2 \beta}\Gamma_{2 0}^{\beta' 2}- 4\Gamma_{0 1}^{2 \beta}\Gamma_{0 2}^{1 \beta'}- 4\Gamma_{0 0}^{1 \beta}\Gamma_{0 0}^{2 \beta'}+  4\Gamma_{0 0}^{2 \beta}\Gamma_{0 2}^{1 \beta'}+  4\Gamma_{0 0}^{1 \beta}\Gamma_{0 1}^{2 \beta'}-(\Gamma_{0 0}^{1 \beta}-
\Gamma_{0 0}^{2 \beta})(\Gamma_{0 0}^{\beta' 1}-
\Gamma_{0 0}^{\beta' 2}) \right]\nonumber\\&=&
\sum_{\beta,\beta'=0,1,2}\left[\Gamma_{0 2}^{2 \beta}\Gamma_{1 0}^{\beta' 1} + 
\Gamma_{0 1}^{1 \beta}\Gamma_{2 0}^{\beta' 2}   -(\Gamma_{0 0}^{1 \beta}-
\Gamma_{0 0}^{2 \beta})(\Gamma_{0 1}^{1 \beta'}-
\Gamma_{0 2}^{2 \beta'})-(\Gamma_{0 0}^{\beta 1}-
\Gamma_{0 0}^{\beta 2})(\Gamma_{1 0}^{\beta' 1}-
\Gamma_{2 0}^{\beta' 2}) \right]
\end{eqnarray}
These are necessary conditions (but not sufficient) for the matrix ${\cal M}(p)$ 
to be non-diagonalizable. This means in turn that it is {\it sufficient} 
(but not {\it necessary}) that one of relations (\ref{eq.2.2.18.1}) be 
violated for the  matrix ${\cal M}(p)$ to be diagonalizable.

If ${\cal M}_{2,1}(p)=0$, the matrix ${\cal M}(p)$ is triangular, i.e.,
\begin{eqnarray}
\label{eq.2.2.18.2}
A_{1}^{b}+C_{2}^{b}=A_{2}^{b}=C_{1}^{b}=0,
\end{eqnarray}
which in terms of reaction-rates imply
\begin{eqnarray}
\label{eq.2.2.18.3}
\sum_{\beta=0,1,2}\left(\Gamma_{1 0}^{2 \beta}+ \Gamma_{0 1}^{\beta 2} - 
\Gamma_{0 0}^{2 \beta} - \Gamma_{0 0}^{\beta 2}  \right)=
\sum_{\beta=0,1,2}\left(\Gamma_{0 1}^{2 \beta}- \Gamma_{0 0}^{2 \beta}\right)
= \sum_{\beta=0,1,2}\left(\Gamma_{1 0}^{\beta 2}- \Gamma_{0 0}^{\beta 2}\right)
\end{eqnarray}
So for example,  if $\Gamma_{0 0}^{2 \beta}=\Gamma_{0 0}^{\beta 2}=
\Gamma_{0 1}^{2 \beta}=\Gamma_{0 1}^{\beta 2}=\Gamma_{1 0}^{\beta 2}=\Gamma_{1 0}^{2 \beta} $, the relations (\ref{eq.2.2.18.3}) are fulfilled.

If the matrix ${\cal M}$ is diagonal, i.e.,  ${\cal M}_{2,1}(p)= {\cal M}_{1,2}(p)=0$,
and in addition to (\ref{eq.2.2.18.2}-\ref{eq.2.2.18.3}), we have 
\begin{eqnarray}
\label{eq.2.2.18.4}
&&B_{1}^{a}+D_{2}^{a}=B_{2}^{a}=D_{1}^{a}=0 \Longrightarrow\nonumber\\
&&\sum_{\beta=0,1,2}\left(\Gamma_{2 0}^{1 \beta}+ \Gamma_{0 2}^{\beta 1} - 
\Gamma_{0 0}^{1 \beta} - \Gamma_{0 0}^{\beta 1}  \right)=
\sum_{\beta=0,1,2}\left(\Gamma_{0 2}^{1 \beta}- \Gamma_{0 0}^{1 \beta}\right)
= \sum_{\beta=0,1,2}\left(\Gamma_{2 0}^{\beta 1}- \Gamma_{0 0}^{\beta 1}\right)
\end{eqnarray}
As an example, relations (\ref{eq.2.2.18.2}-\ref{eq.2.2.18.4}) are fulfilled 
if one has
$\Gamma_{0 0}^{1 \beta}=\Gamma_{0 0}^{\beta 1}=\Gamma_{0 2}^{1 \beta}=\Gamma_{2 0}^{\beta 1}= \Gamma_{2 0}^{1 \beta}=\Gamma_{0 2}^{\beta 1}$ and 
$\Gamma_{0 0}^{2 \beta}=\Gamma_{0 0}^{\beta 2}=\Gamma_{0 1}^{2 \beta}=\Gamma_{1 0}^{\beta 2}= \Gamma_{1 0}^{2 \beta}=\Gamma_{0 1}^{\beta 2}$

It follows from this discussion that when the reaction-rates, in addition 
to the solubility constraints (\ref{eq.1.9}), also violate conditions 
(\ref{eq.2.2.18.2}) and one of the relations (\ref{eq.2.2.18.1}) which are 
{\it sufficient} but not {\it necessary}, the first case applies. When, 
in addition to (\ref{eq.1.9}), the relations  
(\ref{eq.2.2.18.1}), are fulfilled (recall that (\ref{eq.2.2.18.1}) 
are {\it necessary} 
but not {\it sufficient} constraints) and the conditions  (\ref{eq.2.2.18.2}) 
are violated, then the second case applies.
When reaction-rates satisfy (\ref{eq.2.2.18.2}-\ref{eq.2.2.18.3}) in addition 
to the relation (\ref{eq.1.9}), then the third case (see below) applies.

Similarly, when reaction-rates satisfy (\ref{eq.2.2.18.2}-\ref{eq.2.2.18.4}) 
in addition to the relation (\ref{eq.1.9}), then the fourth case (see below) 
applies.
\\

3) If ${\cal M}_{2,1}(p)=0$, the matrix ${\cal M}(p)$ is triangular, 
thus the eigenvalues of ${\cal M}(p)$ are ${\cal M}_{1,1}(p)$ and ${\cal M}_{2,2}(p)$. 
We have already discussed the physical implication of this case 
(see equation (\ref{eq.2.2.18.3}) above), and we have ($\vec{p}\neq 0$)
\begin{eqnarray}
\label{eq.2.2.19}
e^{{\cal M}(p)t}= e^{{\cal M}_{1,1}(p)t}
\left(
 \begin{array}{c c}
 1 & {\cal M}_{1,2}(p)t \\
 0 & 1 
 \end{array}\right)
\end{eqnarray}
which leads to ($\vec{p}\neq 0$)
\begin{eqnarray}
\label{eq.2.2.20}
{\cal S}_{0}^{A}(\vec{p},\omega)=\frac{1}{\omega-{\cal M}_{1,1}(p)}\left[
\langle\hat{n}_{p}^{A}\rangle(0) +\frac{{\cal M}_{1,2}(p)}{\omega -
{\cal M}_{1,1}(p) } \langle\hat{n}_{p}^{B}\rangle \right]\;;\;
{\cal S}_{0}^{B}(\vec{p},\omega)=\frac{\langle\hat{n}_{p}^{B}\rangle(0)}{\omega-{\cal M}_{1,1} (p)}
\end{eqnarray}
%
%

4) If both ${\cal M}_{2,1}(p)={\cal M}_{1,2}(p)=0$, the matrix ${\cal M}(p)$ is already 
diagonal (see (\ref{eq.2.2.18.3})), and the form factors read ($\vec{p}\neq 0$)
\begin{eqnarray}
\label{eq.2.2.22}
&{\cal S}_{0}^{A}(\vec{p},\omega)=\frac{\langle\hat{n}_{p}^{A}\rangle(0) }
{\omega-{\cal M}_{1,1} (p)}\;;\;{\cal S}_{0}^{B}(\vec{p},\omega)=\frac{ \langle\hat{n}_{p}^{B}\rangle(0) }{\omega-{\cal M}_{2,2} (p)}
\end{eqnarray}
%
\\
\\
In the sequel we focus on 
the case where the matrix ${\cal M}(p)$ is diagonal and 
provide explicit expressions in real space and time for the density. 
This is equivalent to imposing the six supplementary conditions characterizing  
$V_1$ (see (\ref{eq.2.2.18.2}-\ref{eq.2.2.18.4})), $B_{1}^{a}+D_{2}^{a}=B_{2}^{a}=D_{1}^{a}=A_{1}^{b}+C_{2}^{b}=A_{2}^{b}=C_{1}^{b}=0 $.
%
%
Let us first compute the density for the case where translation
invariance is broken, i.e., the manifold
$V_{1}\supset V_{2}$ of dimension
$72-(16+6)$. With the above constraints, the densities obey the following 
equations of motion, 
\begin{eqnarray}
\label{eq.2.5}
\frac{d}{dt}\langle n_m^{A}\rangle=(A_0^{a}+C_0^{a})d+
\langle n_m^{A}\rangle(t)(A_{1}^{a} + C_{2}^{a})d +\sum_{\alpha}\left(A_2^{a}
\langle n_{m+e^{\alpha}}^{A}\rangle(t) + C_1^{a}\langle n_{m-e^{\alpha}}^{A}
\rangle(t)\right)
\end{eqnarray}
\begin{eqnarray}
\label{eq.2.6}
\frac{d}{dt}\langle n_m^{B}\rangle=d(A_0^{b}+C_0^{b})+
 \langle n_m^{B}\rangle(t) (B_{1}^{b} + D_{2}^{b})d +\sum_{\alpha}\left(B_2^{b}\langle n_{m+e^{\alpha}}^{B}\rangle(t) + D_1^{b}\langle n_{m-e^{\alpha}}^{B}\rangle(t)\right)
\end{eqnarray}
In order to discuss the solutions of these equations, it is convenient to 
define
the following quantities:
$\mu_{A}\equiv \sqrt{\frac{C_{1}^{a}}{A_{2}^{a}}}, \;
\mu_{B}\equiv \sqrt{\frac{D_{1}^{b}}{B_{2}^{b}}}, \; C_{A}
\equiv\sqrt{A_{2}^{a}C_{1}^{a}}$ et $C_{B}\equiv\sqrt{B_{2}^{b}D_{1}^{b}}$.
Furthermore, we introduce $ B_A \equiv 2(A_{1}^{a}+C_{2}^{a})\leq 0, \; 
B_B \equiv 2(B_{1}^{b}+D_{2}^{b})\leq 0$, $\gamma_A\equiv A_{1}^{a}+A_{2}^{a}
+C_{1}^{a}+C_{2}^{a} \leq 0, \;
\gamma_B\equiv B_{1}^{b}+B_{2}^{b}+D_{1}^{b}+D_{2}^{b} \leq 0$  
and $\epsilon_{j}\equiv \ln{\mu_{j}},\; j=A,B$.
A site on the hypercube is denoted by $m$ 
with $d$ components: $m=(m_1,\dots, m_i, \dots, m_d)$.

Exploiting the well known properties of Bessel functions ($j\in(A,B)$), we arrive
at
\begin{eqnarray}
\label{eq.2.7}
\langle n_m^{j}\rangle (t)=\rho_{j}(\infty) (1-e^{-d|\gamma_{j}| t})+
 e^{-d| \frac{B_{j}}{2}|t}\sum_{m=(m_{1}',\dots, m_{d}')} 
 \langle n_{m'}^{j}\rangle (0)\prod_{i=1\dots d}
 \mu_{j}^{m_{i}-m'_{i}}I_{m_{i}-m'_{i}}(2 C_{j}t)
\end{eqnarray}
The steady densities are respectively,
$\rho_{A}(\infty)=\frac{A_{0}^{a}+C_{0}^{a}}{2|\gamma_{A}|}$, and  
$\rho_{B}(\infty)=\frac{A_{0}^{b}+C_{0}^{b}}{2|\gamma_{B}|}$.
Below, we investigate three different cases

i) The (at $t=0$) initial density (for each species) is given by $\rho_{j}$,
\begin{eqnarray}
\label{eq.2.8}
\langle n_{m'}^{j}\rangle(t=0)=\rho_{j}(0) \delta_{m,m'}
\end{eqnarray}
The asymptotic behavior is then
($m_{i}\gg 1$ and $u_{j}=L^2/{4|C_{j}|t}$)
\begin{eqnarray}
\label{eq.2.9}
\langle n_{m}^{j}\rangle (t)\sim \rho_{j}(\infty)+\psi_{m}^{j}
\frac{e^{-\Theta_{j} t}}{t^{\phi_{j}}}
\end{eqnarray}
where $\psi_{m}^{j}$ are known functions.
Further, we assume $|\gamma_{j}|>0$, since 
for $|\gamma_{j}|=0$, equation (\ref{eq.1.9}) 
tells us that
$\langle n_{m}^{j}\rangle (t)=\rho_{j}(\infty)=\rho_{j}(0)=cste, \forall t$,
where 
\begin{eqnarray}
\label{eq.2.10}
\Theta_{j}=min(d|\gamma_{j}|, d(\frac{|B_{j}|}{2}- |C_{j}|(2+\epsilon^{2}/2)))
\end{eqnarray}
\begin{eqnarray}
\label{eq.2.11}
\phi_{j}=\left\{
\begin{array}{l l}
 0 &\mbox{, if $\Theta_{j}=d|\gamma_{j}|>0$ }\\
 d/2 &\mbox{, if $\Theta_{j}=d(\frac{|B_{j}|}{2}-|C_{j}|(2+\epsilon_{j}^2/2))$ }
\end{array}
\right.
\end{eqnarray}
For $\Theta=0$ and $\gamma \neq 0$, the density decays algebraically, i.e.,
$\langle n_{m}^{j}\rangle (t)\sim \rho_{j}(\infty)+\psi_{m}^{j}t^{-d/2}$.

ii) Initially, the particles (for each species) are confined in some region of
space, i.e., 
\begin{eqnarray}
\label{eq.2.12.0}
\langle n_{m}^{j}\rangle (t=0)=\left\{
\begin{array}{l l}
n_{0}^{j}  &\mbox{, if $0\leq m_{i}\leq L/2$ }\\
0  &\mbox{, otherwise}
\end{array}
\right.
\end{eqnarray}
\begin{eqnarray}
\label{eq.2.12.1}
\langle n_{m}^{j'\neq j}\rangle (t=0)=\left\{
\begin{array}{l l}
n_{0}^{j'}  &\mbox{, if $ L/2<m_{i}\leq L$ }\\
0  &\mbox{, otherwise}
\end{array}
\right.
\end{eqnarray}
We have then
\begin{eqnarray}
\label{eq.2.13}
n_{m}^{j}(t)\sim \rho_{j}(\infty)+ e^{-\Theta_{j} t}\left(\psi_{m,1}^{j}+\frac{\psi_{m,2}^{j}}{t^{\phi_{j}}}\right)
\end{eqnarray}
where $\Theta_{j}$ has been defined in (\ref{eq.2.10}) and 
for $r=m-m', |r|\gg 1, r^2/{\cal C}t < \infty$:
\begin{eqnarray}
\label{eq.2.14}
\phi_{j}=\left\{
\begin{array}{l l}
 0 &\mbox{, if $\Theta_{j}=d|\gamma_{j}|>0$ }\\
 1 &\mbox{, if $\Theta_{j}=d(\frac{|B_{j}|}{2}- C_{j}(2+\epsilon_{j}^2/2))$ }
\end{array}
\right.
\end{eqnarray}
If $\Theta=0$ and $\gamma \neq 0$, 
the density decays as a power law, i.e., 
 $\langle n_{m}^{j}\rangle (t)\sim \rho_{j}(\infty)+
\left(\psi_{m,1}^{j}+\frac{\psi_{m,2}^{j}}{t} \right)$.

iii) The initial distribution of particles is assumed to be non-uniform, i.e., 
\begin{eqnarray}
\label{eq.2.15}
\langle n_{m}^{j} \rangle (t=0)= \left\{
\begin{array}{l l}
n_{0}^{j}   &\mbox{, if $m=0$}\\
n_{0}^{j} \prod_{i=1\dots d}(1-\delta_{m_{i},0})|m_{i}|^{-\alpha_{i}} &\mbox{, if $|m|>0$ } 
\end{array}
\right.
\end{eqnarray}

With (\ref{eq.2.15}), we arrive at ($m_{i}\gg 1$ and$u_{j}=L^2/{4 C_{j}t}$):
\begin{eqnarray}
\label{eq.2.16}
\langle n_{m}^{j}\rangle (t)\sim \rho_{j}(\infty)+\psi_{m}^{j}\frac{e^{-\Theta_{j} t}}{t^{\phi_{j}}}
\end{eqnarray}
When $\Theta_{j}=d(\frac{|B_{j}|}{2}- C_{j}
(2+\frac{\epsilon_{j}^2}{2}))$, decays as (\ref{eq.2.16}), with
\begin{eqnarray}
\label{eq.2.17}
\phi_{j}=\left\{
\begin{array}{l l }
 \sum_{i} \frac{\alpha_{i}}{2} &\mbox{, if $0 \leq\alpha_{i}<1$ }\\
 \frac{d}{2} &\mbox{, if $\alpha_{i}\geq 1$ } 
\end{array} 
\right.
\end{eqnarray}
Again, as $\Theta_{j}=0$, equation
(\ref{eq.2.17}) holds.

Notice the cross-over at 
$\alpha_{i}=\alpha=1$, where the density decays as 
$\langle n_{m}^{j}\rangle (t) \sim \rho_{j}(\infty) + 
\psi_{m}^{j}\left(e^{-d(\frac{|B_{j}|}{2}-|C_{j}|
(2+\frac{\epsilon_{j}^2}{2}))t}\left[\frac{\ln (4u_{j}| C_{j}|t)}
{2\sqrt{4\pi |C_{j}|t}}\right]^{d}\right)$.
By contrast, when $\Theta_{j}=d|\gamma_{j}|>0$, the density behaves as
$\langle n_{m}^{j}\rangle (t) \sim \rho_{j}(\infty) + \psi_{m}^{j} 
e^{-d|\gamma_{j}|t} $. Here we have restricted our attention to the case where  $0<\alpha_{i}<1$, 
$\forall i, 0<\alpha_{i}<1$, while in general we could consider different
regimes in the different directions (for example, $0<\alpha_{1}<1$, 
$\alpha_{1}=1$ and 
$\alpha_{1}>1$ ). The corresponding asymptotic behavior follows as above. 

To conclude this section let us focus on the manifold $V_2$, where the density
is translationally invariant.
On account of (\ref{eq.2.6}), we obtain two coupled linear differential
equations which are easily integrated. The result is
\begin{eqnarray}
\label{eq.2.16}
\rho_{j}(t)=\rho_{j}(\infty)+(\rho_{j}(0)- \rho_{j}(\infty))e^{-d|\gamma_j|t}
\end{eqnarray}
The above solution allows us to solve for the correlation functions on $V_{2}(d)$. 
These in turn will be useful to solve perturbatively the problem on the manifold
$V_{transl-inv}(d)$. 
\\

Let us now discuss the relationship between our results and the solution of some models  solved exactly in $d\geq 1$ \cite{Clement,Fichthorn,Toussaint,Bramson,Belitsky} earlier.

In reference \cite{Clement}, Cl{\'e}ment et al., solved exactly (the fast adsorption rates version) of  the Fichthorn, Gulari, Ziff (FGZ) model \cite{Fichthorn} introduced to describe the conversion of $CO$ and $O$ to $CO_{2}$ on platinum substrates. The model solved in \cite{Clement} describes the dynamics of classical stochastic particles of two-species (called $A$ and $B$) {\it with hard-core constraint}, and so our approach applies to this model. However, as the system solved in \cite{Clement} is a {\it two-states model} (in \cite{Clement},
 the stochastic variable is a ``spin variable'' $z_{j}=\pm 1$, $+1$ corresponding to an $A$ particle and $-1$ to a $B$ particle at site $j$. In this model there are no vacancies), it is not in the class of ({\it three-states}) models which we specifically study there. It is however possible to recover previous results \cite{Clement} considering the system \cite{Clement} in the framework of the {\it two-states} analog of our approach \cite{Schutz,Fujii}. To do so we relabel Cl\'ement et al. $B$ particles  by vacancies symbolized by $0$ (the A particles are symbolized by $1$), thus reactions occurring in \cite{Clement} are described by the rates:
$\Gamma_{1 1}^{1 0}=\Gamma_{1 1}^{0 1}=\Gamma_{0 0}^{1 0}= \Gamma_{0 0}^{0 1}=p/4$; 
$\Gamma_{1 0}^{0 1}=\Gamma_{0 1}^{1 0}=1/4d$; $\Gamma_{1 0}^{0 0}=\Gamma_{1 0}^{1 1}=
 \Gamma_{0 1}^{1 1}=\Gamma_{0 1}^{0 0}=p/4 + 1/4d $; $\Gamma_{1 1}^{1 1}=\Gamma_{0 0}^{0 0}=
-p/2$ and $\Gamma_{1 0}^{1 0}=\Gamma_{0 1}^{0 1}=-(p/2+3/4d)  $. For such systems (where $s$=1), the $2s^{3}=2$ {\it solubility constraints}, which are the analogs of (\ref{eq.1.9}), read \cite{Schutz,Fujii}:
$\Gamma_{1 0}^{0 0}+\Gamma_{0 0}^{1 0}+\Gamma_{1 0}^{0 1}+\Gamma_{0 0}^{1 1}
= \Gamma_{1 1}^{0 0}+\Gamma_{1 1}^{0 1}+\Gamma_{0 1}^{1 0}+\Gamma_{0 1}^{1 1} $
and $\Gamma_{0 1}^{0 0}+\Gamma_{0 0}^{0 1}+\Gamma_{0 1}^{1 0}+\Gamma_{0 0}^{1 1}
= \Gamma_{1 1}^{0 0}+\Gamma_{1 1}^{1 0}+\Gamma_{1 0}^{0 1}+\Gamma_{1 0}^{1 1} $.
These constraints are fulfilled for the previous choice of rates (similar to the choice of \cite{Clement}) and thus the dynamics of the system is soluble in arbitrary  dimensions, i.e.  the equations of motion of the correlation functions are closed and obtained as in (\ref{eq.1.4},\ref{eq.1.7},\ref{eq.1.8}). As an example, for the density we have the following equation of motion \cite{Schutz}: $\frac{d}{dt}\langle\widetilde n_{j}\rangle=B_{1}\sum_{\pm \alpha}\left(\langle\widetilde n_{j+\alpha}\rangle -\langle\widetilde n_{j}\rangle  \right) -p\langle\widetilde n_{j}\rangle $, where $\widetilde{n_{j}}\equiv n_{j}-\frac{1}{2}$ and $B_{1}=\Gamma_{1 0}^{0 1}+
\Gamma_{1 0}^{1 1}-\Gamma_{0 0}^{0 1}-\Gamma_{0 0}^{1 1}
=1/2d$ are the same quantities defined in \cite{Schutz}. Noting that in language of \cite{Clement} $\gamma_{j}\equiv\langle z_{j}\rangle=2\langle\widetilde n_{j}\rangle$, we recover the result of \cite{Clement}:
$\frac{d}{dt}\gamma_{j}=\frac{1}{2d}\triangle \gamma_{j}-p\gamma_{j}$, where $\triangle \gamma_{j}\equiv \sum_{\pm \alpha}(\gamma_{j+\alpha}-\gamma_{j})$. Similarly we can reproduce the (closed) equations of motion for higher correlation functions. Saturation phenomena as in \cite{Clement} should also occur in the class of three states models. However the analytical treatment would be more complex than in \cite{Clement} two-states models.

 Let us sketch the strategy which one should follow to treat  saturation in the models considered here. For translationally invariant systems one should solve a (linear) differential-difference systems of coupled equations describing equations of motion of correlation functions $\langle n_{m}^{i}n_{l}^{j}\rangle (t), j\in(A,B)$ paying due attention to the boundary terms $m=l$ and $|m-l|=1$ (see also section V). This system is solved in Fourier space and involves  a general $3\times 3$ matrix with non-constant entries. One should, as it has been done for the density,  carefully discuss the properties of this matrix, which is  a technical matter. In fact such a study should be carried out for some specific model.  

A further two-states model which can be solved exactly in $d\geq 1$ is the Voter model (see e.g. \cite{Bennaim}) described by the reactions-rates: $\Gamma=\Gamma_{0 1}^{0 0}=\Gamma_{1 0}^{0 0}=
\Gamma_{0 1}^{1 1}=\Gamma_{0 1}^{1 1} >0$. Since this model fulfill the previous ``two-states''  {\it solubility constraints}, it is soluble in arbitrary dimensions. With the {\it two-states} analog of (\ref{eq.1.4},\ref{eq.1.7},\ref{eq.1.8}) we obtain the (closed) equations of motion of the correlation functions. As an example, for the density, we have \cite{Bennaim}:
$\frac{d}{dt}\langle n_{j}\rangle=2\Gamma \sum_{\alpha}\left(\langle n_{j+\alpha}\rangle
+\langle n_{j-\alpha}\rangle -2\langle n_{j}\rangle \right)$.

Another important model which has been studied rigorously in dimensions $d\geq 1$ is the (irreversible) reaction $A+B\rightarrow \emptyset+\emptyset$. For this model, Bramson and Lebowitz \cite{Bramson} obtained, rigorously,  upper and lower bounds for the long time behavior of densities. However, systems considered in these works, \cite{Bramson}, allow the multiple occupancy of a site by particles of the same species. Later,  Belitsky \cite{Belitsky} generalized the study of \cite{Bramson} to the case of hard-core particles reacting according to $A+B\rightarrow \emptyset+\emptyset$ (with rates $\Gamma_{1 2}^{0 0}=\Gamma_{2 1}^{0 0}=\Gamma(=1)$) and $A+\emptyset \leftrightarrow\emptyset +A $; $B+\emptyset \leftrightarrow\emptyset +B $ (with rates $\Gamma_{1 0}^{0 1}=\Gamma_{0 1}^{1 0}=
\Gamma_{2 0}^{0 2}=\Gamma_{0 2}^{2 0}=1$). He obtained rigorously an upper bound for long-time behavior of the density [for an uncorrelated initial state with equal species densities: $\rho_{A}(0)=\rho_{B}(0)\leq 1/2$]: $\forall \epsilon>0, \exists T(\epsilon)<\infty$, so that for $t>T(\epsilon)$, $\rho_{t\rightarrow \infty}(t)\leq t^{-d/4 + \epsilon}$, for $d\leq 4$ and
  $\rho_{t\rightarrow \infty}(t)\leq C^{\star}t^{-1}$, for $d>4$; where $C^{\star}$ is a positive constant. We can now wonder whether such a model can be 
dealt with in our approach. As the model considered by Belitsky is a {\it  three-states} model, the equations of motion of correlation functions are given by (\ref{eq.1.4},\ref{eq.1.7},\ref{eq.1.8}) with $A_{2}^{a}=C_{1}^{a}=B_{2}^{b}=D_{1}^{b}=-A_{1}^{a}=-B_{1}^{b}=-C_{2}^{a}=-D_{2}^{b}=1$ and $A_{0}^{a}=B_{1}^{a}=B_{2}^{a}=D_{1}^{a}=D_{2}^{a}=A_{0}^{b}=A_{1}^{b}= A_{2}^{b}=
C_{0}^{b}=C_{1}^{b}=C_{2}^{b}=0$. Unfortunately the solubility constraints (\ref{eq.1.9}) are not fulfilled for such a model (consider e.g. the fourth constraint: $ A_{2}^{a}+B_{1}^{a}+A_{0}^{a}= \Gamma_{2 1}^{1 0}+\Gamma_{2 1}^{1 1}+\Gamma_{2 1}^{1 2}\Rightarrow \Gamma =0 $, but in this model $\Gamma=1$ ) and the equations of motion give rise to an open hierarchy which cannot be solved.

Results in $d\geq 1$ have also been obtained by approximate methods (mean-field theories), and/or by scaling and heuristic arguments (see e.g. \cite{Privman} and references therein).
As an illustration of these studies let us  consider the work of Toussaint and Wilczek. In \cite{Toussaint}, a system of two species  $A$ and $B$ reacting  according to $A+B\rightarrow \emptyset + \emptyset$, in addition to their diffusive  motion is  studied numerically and an approximate method for calculating the densities at long-time  is proposed (for system with equal densities : $\rho_{A}(0)=\rho_{B}(0)$). Approximate results \cite{Toussaint} predict $\rho(t) \sim t^{-d/4}$, in agreement with Bramson, Lebowitz \cite{Bramson}  and Belitsky \cite{Belitsky} rigorous results in the one-dimensional case, but in disagreement in higher dimensions $d>1$. The approach in \cite{Toussaint} is a continuum  macroscopic approach and cannot take into account the hard-core  constraint of the particles . In addition, it
 takes into account of fluctuations in an approximate and uncontrolled way. It is therefore difficult to compare their method  with the microscopic exact results presented here. 

In sum  we have seen that the two-states formulation \cite{Schutz} of the method dicussed here allows to recover some previous exact results in arbitrary dimensions \cite{Clement,Bennaim} for the stochastic models of hard-core particles. Our approach applied on  three-states models (of hard-core particles) is in a sense complementary to the rigorous results of \cite{Belitsky}  and is  useful to describe exactly  physical models such as a {\it three-states} growth model \cite{Mobar2}. 
\section{Non-instantaneous Correlation functions}
To our knowledge, the only exact computations of two-time correlation functions
in non-equilibrium statistical mechanics that are available are those for
single species models, in particular for one-dimensional models which can be 
mapped onto free fermion
\cite{Schutz2,Grynberg2,BM1} and other related \cite{Grynberg,Bennaim} models .
These exact results are useful, as starting points for perturbative calculations
or for checking numerical computations.  
This section is devoted to the study of the $50$-parametric 
manifold $V_{1}$. We are interested in the density-density 
correlation functions
\begin{eqnarray}
\label{eq.6.1}
\langle n_{m}^{i}(t) n_{l}^{j}(0)\rangle \equiv 
\langle \widetilde \chi | n_{m}^{i}e^{-Ht}n_{l}^{j}|P(0)\rangle=
\langle \widetilde \chi | n_{m}^{i}e^{-Ht}|P'(0)\rangle , \;\; i,j\in (A,B)
\end{eqnarray}
%
%
%
where $|P(0)\rangle$ denotes the initial state of the system.
From the above we see that the evaluation for the correlation function 
with respect to the initial
state $|P(0)\rangle$ is equivalent
to computation of the density of particles of species $i$ at site  $m$ for a 
sytem in an initial state described by 
$|P'(0)\rangle\equiv n_{l}^{j}|P(0)\rangle$.

We now distinguish the case where correlations are absent in the initial 
(with broken translation invariance) state
from that where they are present.

\subsection{Non-instantaneous two-point correlation functions for
uncorrelated initial states }

In this subsection we assume uncorrelated initial states with
a random and translationally invariant distribution of particles of type 
$A$ (density
$\rho_{A}(0)$) and of type $B$ (density   
$\rho_{B}(0)$).
Therefore in our notations $|P(0)\rangle$ becomes
\begin{eqnarray}
\label{eq.6.7}
|P(0)\rangle =
\left(
 \begin{array}{c}
 1-\rho_{A}(0)-\rho_{B}(0)\\
\rho_{A}(0)\\
\rho_{B}(0)
 \end{array}\right)^{\otimes L^d}
\end{eqnarray}
So that we have

\begin{eqnarray}
\label{eq.6.8}
\langle n_{m}^{A}(0) n_{l}^{A}(0)\rangle &=& \rho_{A}(0)\delta_{m, l} +
\rho_{A}(0)^{2}(1-\delta_{m, l})\nonumber\\
\langle n_{m}^{B}(0) n_{l}^{B}(0)\rangle &=&\rho_{B}(0)\delta_{m, l} +
\rho_{B}(0)^{2}(1-\delta_{m, l}) \nonumber\\
\langle n_{m}^{A}(0) n_{l}^{B}(0) \rangle &=&\langle n_{m}^{B}(0) n_{l}^{A}(0) \rangle=
 \rho_{A}(0)\rho_{B}(0) (1-\delta_{m, l})
\end{eqnarray}
We begin this section by computing explicitely the Fourier-Laplace transform of    
the two-points non-instantaneous correlation functions, i.e., the dynamic
form factors measured in the light scattering experiments \cite{Kroon,Grynberg2}
\begin{eqnarray}
\label{eq.6.8.0}
{\cal S}_{1}^{ij}(\vec{p},\omega)&\equiv& \frac{1}{L^d}\sum_{m'\equiv m-l}\int_{0}^{\infty} dt e^{-i\vec{p}.m'-\omega t}\langle n_{m}^{i}(t)n_{l}^{j}(0)\rangle \nonumber\\
&=&  \frac{1}{L^d}\sum_{m'\equiv m-l}\int_{0}^{\infty} dt e^{-i\vec{p}.m'-\omega t }\langle n_{m'=m-l}^{i}(t)n_{0}^{j}(0)\rangle, (i,j)\in(A,B)
\end{eqnarray}
Using the results of the previous section and
assuming regularity and diagonalizability of 
${\cal M}(p)$, we find for the dynamic form factors
\begin{eqnarray}
\label{eq.6.8.5}
&{\cal S}_{1}^{AA}(\vec{p},\omega)&=
\frac{1}{\lambda_{-}(p)-\lambda_{+}(p)}\left[\frac{\lambda_{-}(p)-
{\cal M}_{1,1}(p)}{\omega -\lambda_{+}(p)} -\frac{\lambda_{+}(p)-
{\cal M}_{1,1}(p)}{\omega -\lambda_{-}(p)}  \right] \left((\rho_{A}^2(0) +
\rho_{A}(0))\delta_{p,0} -\rho_{A}(0)\right) \nonumber\\
 &+& \frac{{\cal M}_{1,2}(p) \left(\rho_{A}(0) \rho_{B}(0)\right)
 (\delta_{p,0}-1)  }{(\omega -\lambda_{+}(p))(\omega -\lambda_{-}(p))} 
 \nonumber\\
&+& \frac{\delta_{p,0}}{\gamma_{-}-\gamma_{+}}\left[(\gamma_{-}-
{\cal M}_{1,1}(0))(A_{0}^{a}+ C_{0}^{a})d -{\cal M}_{1,2}(0)
(A_{0}^{b}+ C_{0}^{b})d\right] \frac{1}{\omega(\omega -\gamma_{+})} 
\nonumber\\
&-& \frac{\delta_{p,0}}{\gamma_{-}-\gamma_{+}}\left[(\gamma_{-}-
{\cal M}_{1,1}(0))(A_{0}^{a}+ C_{0}^{a})d -{\cal M}_{1,2}(0)(A_{0}^{b}+ 
C_{0}^{b})d\right] \frac{1}{\omega(\omega -\gamma_{-})}
\end{eqnarray}
\begin{eqnarray}
\label{eq.6.8.6}
&{\cal S}_{1}^{AB}(\vec{p},\omega)&=
\frac{1}{\lambda_{-}(p)-\lambda_{+}(p)}\left[\frac{\lambda_{-}(p)-
{\cal M}_{1,1}(p)}{\omega -\lambda_{+}(p)} -\frac{\lambda_{+}(p)-
{\cal M}_{1,1}(p)}{\omega -\lambda_{-}(p)}  \right]\rho_{A}(0)\rho_{B}(0)
(\delta_{p,0}-1) \nonumber\\
 &+& \frac{{\cal M}_{1,2}(p) \left((\rho_{B}^2(0) +\rho_{B}(0))\delta_{p,0} -
 \rho_{B}(0) \right) }{(\omega -\lambda_{+}(p))(\omega -\lambda_{-}(p))} 
 \nonumber\\
&+& \frac{\delta_{p,0}}{\gamma_{-}-\gamma_{+}}\left[(\gamma_{-}-
{\cal M}_{1,1}(0))(A_{0}^{a}+ C_{0}^{a})d -{\cal M}_{1,2}(0)
(A_{0}^{b}+ C_{0}^{b})d\right] \frac{1}{\omega(\omega -\gamma_{+})} 
\nonumber\\
&-& \frac{\delta_{p,0}}{\gamma_{-}-\gamma_{+}}\left[(\gamma_{-}-
{\cal M}_{1,1}(0))(A_{0}^{a}+ C_{0}^{a})d -{\cal M}_{1,2}(0)
(A_{0}^{b}+ C_{0}^{b})d\right] \frac{1}{\omega(\omega -\gamma_{-})}
\end{eqnarray}
and
\begin{eqnarray}
\label{eq.6.8.7}
&{\cal S}_{1}^{BB}(\vec{p},\omega)&=
\left((\rho_{B}^2(0) +\rho_{B}(0))\delta_{p,0} -\rho_{B}(0) \right) 
 \left[\frac{\lambda_{-}(p) - {\cal M}_{1,1}(p)}{ \omega-\lambda_{-}(p)}-
 \frac{\lambda_{+}(p) - {\cal M}_{1,1}(p)}{\omega-\lambda_{+}(p)}\right]
 \nonumber\\&-&\left[\frac{(\lambda_{+}(p) - {\cal M}_{1,1}(p) )
 (\lambda_{-}(p) - {\cal M}_{1,1}(p) )}{{\cal M}_{1,2} (p) 
 (\omega-\lambda_{+}(p))(\omega-\lambda_{-}(p))}\right] \rho_{A}(0)
 \rho_{B}(0)(\delta_{p,0}-1) \nonumber\\
&+& \frac{\delta_{p,0}}{\gamma_{-}-\gamma_{+}}\left[\frac{(\gamma_{-}-
{\cal M}_{1,1}(0))}{{\cal M}_{1,2}(0)}(A_{0}^{a}+ C_{0}^{a})d -
(A_{0}^{b}+ C_{0}^{b})d\right] \frac{\gamma_{+}-{\cal M}_{1,1}(0) }
{\omega(\omega -\gamma_{+})} \nonumber\\
&-& \frac{\delta_{p,0}}{\gamma_{-}-\gamma_{+}}\left[\frac{(\gamma_{+}-
{\cal M}_{1,1}(0))}{{\cal M}_{1,2}(0)}(A_{0}^{a}+ C_{0}^{a})d -
(A_{0}^{b}+ C_{0}^{b})d\right] \frac{\gamma_{-}-{\cal M}_{1,1}(0)}
{\omega(\omega -\gamma_{-})}
\end{eqnarray}
\begin{eqnarray}
\label{eq.6.8.8}
&{\cal S}_{1}^{BA}(\vec{p},\omega)&=
\rho_{A}(0)\rho_{B}(0)(\delta_{p,0}-1) \left[\frac{\lambda_{-}(p) - 
{\cal M}_{1,1}(p)}{ \omega-\lambda_{-}(p)}-\frac{\lambda_{+}(p) - 
{\cal M}_{1,1}(p)}{\omega-\lambda_{+}(p)}\right]\nonumber\\
&-&
\left[\frac{(\lambda_{+}(p) - {\cal M}_{1,1}(p) )(\lambda_{-}(p) - 
{\cal M}_{1,1}(p) )}{{\cal M}_{1,2} (p) (\omega-\lambda_{+}(p))
(\omega-\lambda_{-}(p))}\right] \left((\rho_{A}^2(0)+\rho_{A}(0) )
\delta_{p,0}-\rho_{A}(0)  \right) \nonumber\\
&+& \frac{\delta_{p,0}}{\gamma_{-}-\gamma_{+}}\left[\frac{(\gamma_{-}-
{\cal M}_{1,1}(0))}{{\cal M}_{1,2}(0)}(A_{0}^{a}+ C_{0}^{a})d -
(A_{0}^{b}+ C_{0}^{b})d\right] \frac{\gamma_{+}-{\cal M}_{1,1}(0) }
{\omega(\omega -\gamma_{+})} \nonumber\\
&-& \frac{\delta_{p,0}}{\gamma_{-}-\gamma_{+}}
\left[\frac{(\gamma_{+}-{\cal M}_{1,1}(0))}{{\cal M}_{1,2}(0)}
(A_{0}^{a}+ C_{0}^{a})d -(A_{0}^{b}+ C_{0}^{b})d\right] \frac{\gamma_{-}-
{\cal M}_{1,1}(0)}{\omega(\omega -\gamma_{-})}
\end{eqnarray}
Again the poles of the dynamic form factors give the relaxational eigenvalues.
As in the previous section, we could also compute 
the correlation functions in 
the case where ${\cal M}(p)$ is non-diagonalizable, triangular or already 
diagonal, but for brevity's sake we prefer to 
focus here on the  non-instantaneous correlation functions on the 
$50-$parameterss 
manifold $V_{1}$.

With help of equations (\ref{eq.2.5}-\ref{eq.2.7}), we obtain the  
non-instantaneous two-point 
correlation
functions on the manifold $V_{1}$ as
\begin{eqnarray}
\label{eq.6.9}
\langle n_{m}^{A}(t) n_{l}^{A}(0)\rangle =
\rho_{A}(\infty) (1-e^{-d|\gamma_{A}| t})+
\rho_{A}^{2}(0)e^{-d|\gamma_{A}|t}
+  (\rho_{A}(0)-\rho_{A}^2(0))\prod_{\alpha=1\dots d}
\mu_{A}^{m_{\alpha}-l_{\alpha}}e^{-\frac{|B_{A}|t}{2}}I_{m_{\alpha}-l_{\alpha}}(2C_{A}t)
\end{eqnarray}
\begin{eqnarray}
\label{eq.6.10}
\langle n_{m}^{A}(t) n_{l}^{B}(0)\rangle= 
\rho_{A}(\infty) (1-e^{-d|\gamma_{A}| t})+
\rho_{A}(0)\rho_{B}(0)\left[e^{-d|\gamma_{A}| t}
-  \prod_{\alpha=1\dots d}
\mu_{A}^{m_{\alpha}-l_{\alpha}}
e^{-\frac{|B_{A}|t}{2}}I_{m_{\alpha}-l_{\alpha}}(2C_{A}t) \right]
\end{eqnarray}
\begin{eqnarray}
\label{eq.6.11}
\langle n_{m}^{B}(t) n_{l}^{B}(0)\rangle =
\rho_{B}(\infty) (1-e^{-d|\gamma_{B}| t})+
\rho_{B}^2(0)e^{-d|\gamma_{B}|t} + (\rho_{B}(0)-\rho_{B}^2(0))\prod_{\alpha=1\dots d}
\mu_{B}^{m_{\alpha}-l_{\alpha}}e^{-\frac{|B_{B}|t}{2}}I_{m_{\alpha}-l_{\alpha}}(2C_{B}t)
\end{eqnarray}
\begin{eqnarray}
\label{eq.6.12}
\langle n_{m}^{B}(t) n_{l}^{A}(0)\rangle =
\rho_{B}(\infty) (1-e^{-d|\gamma_{B}|t})+
\rho_{A}(0)\rho_{B}(0)\left[e^{-d|\gamma_{B}|t}-\prod_{\alpha=1\dots d}
\mu_{B}^{m_{\alpha}-l_{\alpha}}e^{-\frac{|B_{B}|t}{2}}
I_{m_{\alpha}-l_{\alpha}}(2C_{B}t)\right]
\end{eqnarray}

with the notations, $\vec{r}\equiv \sum_{\alpha}r_{\alpha} e_{\alpha}$, 
$\vec{\epsilon_{i}}\equiv \epsilon_{i}\sum_{\alpha}e_{\alpha}$ and 
$r_{\alpha}=m_{\alpha}-l_{\alpha} =\sigma L$. We are interested in the asymptotic behavior 
($|C_{j}|t \gg 1, j\in (A,B)$ and  $u_{j}=L^2/4|C_{j}|t <\infty$) of the 
above correlation functions in two regimes:
\\
i) when $|m-l|\equiv |r|\equiv\left(\sum_{\alpha=1\dots d} r_{\alpha}^2\right)^{1/2}\sim L \gg 1$, in this case $\sigma=r/L={\cal O}(1)$.\\
ii)  when $|m-l|\equiv |r|\ll 1$, in this case $\sigma=r/L={\cal O}(1/L)$.

It is worth noting that the 
autocorrelation functions (where $|m-l|=0$) are obtained in the second 
regimes (ii).
With the above, we finally arrive at
\begin{eqnarray}
\label{eq.6.13}
\langle n_{m}^{A}(t) n_{l}^{A}(0)\rangle &=& 
\rho_{A}(\infty) (1-e^{-d|\gamma_{A}| t})+\rho_{A}^{2}(0)e^{-d|\gamma_{A}|t} +exp\left(2d\sigma^2 u_{A}-\frac{(\vec{r}-
\vec{\epsilon}_{A}|C_{A}|t)^2}{2|C_{A}|t}\right)e^{-d(\frac{|B_{A}|}{2}-|C_{A}|(2+\epsilon_{A}^2/2))t}\nonumber\\&\times&
\left[ \frac{(\rho_{A}(0)-\rho_{A}^2(0))e^{-d\sigma^{2}u_{A}}}{(4\pi |C_{A}|t)^{d/2}} +
 {\cal O}(1/t^d)\right]
\end{eqnarray}
\begin{eqnarray}
\label{eq.6.14}
\langle n_{m}^{A}(t) n_{l}^{B}(0)\rangle &=& 
\rho_{A}(\infty) (1-e^{-d|\gamma_{A}| t})+ \rho_{A}(0)\rho_{B}(0)e^{-d|\gamma_{A}|t}+ exp\left(2d\sigma^2 u_{A}-\frac{(\vec{r}-
\vec{\epsilon}_{A}|C_{A}|t)^2}{2|C_{A}|t}\right)e^{-d(\frac{|B_{A}|}{2}-|C_{A}|(2+\epsilon_{A}^2/2))t}\nonumber\\&\times&
\rho_{A}(0)\rho_{B}(0)\left[1-\frac{e^{-d\sigma^{2}u_{A}}}{(4\pi |C_{A}|t)^{d/2}} + {\cal O}(1/t^d)  \right]
\end{eqnarray}
\begin{eqnarray}
\label{eq.6.15}
\langle n_{m}^{B}(t) n_{l}^{B}(0)\rangle &=& 
\rho_{B}(\infty) (1-e^{-d|\gamma_{B}| t})+\rho_{B}^{2}(0)e^{-d|\gamma_{B}|t}+exp\left(2d \sigma^2 u_{B}-\frac{(\vec{r}-
\vec{\epsilon}_{B}|C_{B}|t)^2}{2|C_{B}|t}\right)e^{-d(\frac{|B_{B}|}{2}-|C_{B}|(2+\epsilon_{B}^2/2))t}\nonumber\\&\times&
\left[ \frac{(\rho_{B}(0)-\rho_{B}^2(0))e^{-d\sigma^2 u_{B}}  }{(4\pi |C_{A}|t)^{d/2}} + {\cal O}(1/t^d) \right]
\end{eqnarray}
\begin{eqnarray}
\label{eq.6.16}
\langle n_{m}^{B}(t) n_{l}^{A}(0)\rangle &=& 
\rho_{B}(\infty) (1-e^{-d|\gamma_{B}| t})+ \rho_{A}(0)\rho_{B}(0)e^{-d|\gamma_{B}|t}+ exp\left(2d\sigma^2 u_{B}-\frac{(\vec{r}-
\vec{\epsilon}_{B}|C_{B}|t)^2}{2|C_{B}|t}\right)e^{-d(\frac{|B_{B}|}{2}-|C_{B}|(2+\epsilon_{B}^2/2))t}\nonumber\\&\times&
\rho_{A}(0)\rho_{B}(0)\left[1-\frac{e^{-d\sigma^2 u_{B}} }{(4\pi |C_{B}|t)^{d/2}}+ {\cal O}(1/t^d)  \right]
\end{eqnarray}
In the regime $|B_{i}|=|C_{i}|(2+\epsilon_{i}^2/2d), 
\; i\in (A,B) $, the two-point correlation functions decay as
\begin{eqnarray}
\label{eq.6.17}
\langle n_{m}^{i}(t) n_{l}^{j}(0)\rangle\sim 
\frac{exp\left(2d \sigma^2 u_{i}-\frac{(\vec{r}-
\vec{\epsilon}_{i}|C_{i}|t)^2}{2|C_{i}|t}\right) }{|C_{i}|t^{d/2}}   , (i,j)\in (A,B)
\end{eqnarray}
Note the non-trivial dependence on dimensionality and the drift term in $exp\left(2d \sigma^2 u_{i}-\frac{(\vec{r}-
\vec{\epsilon}_{i}|C_{i}|t)^2}{2|C_{i}|t}\right)$,  ($\epsilon_{i}\neq 0$). 
We remark that this is consistent with the result obtained in one
dimension for free fermions\cite{Grynberg2}. If $\epsilon_{i}=0$, than there is no drift:
$exp\left(2d \sigma^2 u_{i}-\frac{(\vec{r}-
\vec{\epsilon}_{i}|C_{i}|t)^2}{2|C_{i}|t}\right)=1$.
\subsection{ Non-instantaneous two-point correlation functions on $V_{1}$: 
correlated initial states}
Let us consider correlated initial states described by a distribution
having the following properties:

i) when $dist(m-l)>0$:
\begin{eqnarray}
\label{eq.6.18}
\langle n_{m}^{i}(0) n_{l}^{j}(0)\rangle ={\cal K}_{ij}\prod_{\alpha=1,\dots, d}(1-\delta_{r_{\alpha},0})|r_{\alpha}|^{-\Delta_{ij}^{\alpha}},\; \Delta_{ij}>0, {\cal K}_{ij}>0
\;,r_{\alpha}\equiv |m_{\alpha}-l_{\alpha}|,\; (i,j)\in (A,B),
\end{eqnarray}

ii) when  $m=l$, 
\begin{eqnarray}
\label{eq.6.19}
\langle n_{m}^{i}(0) n_{l}^{j}(0)\rangle = \langle n_{m}^{i}(0)\rangle \delta_{i,j}=\langle n_{l}^{j}(0)\rangle \delta_{i,j} =\rho_{i}(0)\delta_{i,j}
\end{eqnarray}
The initial distribution in this subsection has been chosen in a special form,
namely in such a way that computations can be carried out explicitly to the end.
There is a priori no physical justification for such a choice, which has already been considered in \cite{Grynberg} for a  single-species reaction-diffusion system. However and most
importantly, 
our goal here is to investigate the dependence of the asymptotics on the initial correlations. 

We remark that in one dimension the initial state (\ref{eq.6.18},\ref{eq.6.19}) 
is translationally invariant, 
\begin{eqnarray}
\label{eq.6.19.1}
\langle n_{m}^{i}(0) n_{l}^{j}(0)\rangle=
\langle n_{|r|=|m-l|}^{i}(0) n_{0}^{j}(0)\rangle=
{\cal K}_{ij}(1-\delta_{|r|,0})|r|^{-\Delta_{ij}}+
\rho_{i}(0)\delta_{i,j}\delta_{|r|,0}
\end{eqnarray}
while in higher dimensions ($d\geq 2$, see \ref{eq.6.18}) 
translational invariance is broken.
This state of affairs has lead us to distinguish in the discussions the one
one dimensional case from its higher dimensional counterparts.

i) We begin with one dimension ($d=1$), where $r=r_{\alpha}\equiv m-l$. 
Because of the translational invariance in the initial state, 
we expect the non-instantaneous correlation function to depend 
on $r=m-l$, indeed,
\begin{eqnarray}
\label{eq.6.20.1}
\langle n_{m}^{A}(t) n_{l}^{A}(0)\rangle=\langle n_{r}^{A}(t) n_{0}^{A}(0)\rangle  &=& 
\rho_{A}(\infty) (1-e^{-|\gamma_{A}| t})+
\rho_{A}(0) e^{-\frac{|B_{A}|t}{2}}\mu_{A}^{r}I_{r}(2 C_{A}t) \nonumber\\&+&
{\cal K}_{AA}\sum_{r'\neq 0}\mu_{A}^{r-r'} |r'|^{-\Delta_{AA}} e^{-\frac{|B_{A}|t}{2}} 
I_{r-r'}(2C_{A}t)
\end{eqnarray}
\begin{eqnarray}
\label{eq.6.21.1}
\langle n_{m}^{A}(t) n_{l}^{B}(0)\rangle =
\rho_{A}(\infty) (1-e^{-|\gamma_{A}| t}) +
{\cal K}_{AB}\sum_{r'\neq 0} 
\mu_{A}^{r-r'} |r'|^{-\Delta_{AB} } e^{-\frac{|B_{A}|t}{2}} I_{r-r'}(2 C_{A}t)
\end{eqnarray}
\begin{eqnarray}
\label{eq.6.22.1}
\langle n_{m}^{B}(t) n_{l}^{B}(0)\rangle=\langle n_{r}^{B}(t) n_{0}^{B}(0)\rangle  &=& 
\rho_{B}(\infty) (1-e^{-|\gamma_{B}| t})+
\rho_{B}(0) e^{-\frac{|B_{B}|t}{2}}\mu_{B}^{r}I_{r}(2 C_{B}t) \nonumber\\&+&
{\cal K}_{BB}\sum_{r'\neq 0}\mu_{B}^{r-r'} |r'|^{-\Delta_{BB}} e^{-\frac{|B_{B}|t}{2}} 
I_{r-r'}(2C_{B}t)
\end{eqnarray}
\begin{eqnarray}
\label{eq.6.23.1}
\langle n_{m}^{B}(t) n_{l}^{A}(0)\rangle =
\rho_{B}(\infty) (1-e^{-|\gamma_{B}| t}) +
{\cal K}_{BA}\sum_{r'\neq 0} 
\mu_{B}^{r-r'} |r'|^{-\Delta_{BA}} e^{-\frac{|B_{B}|t}{2}} I_{r-r'}(2 C_{B}t)
\end{eqnarray}
Notice that when $\mu_{A,B}\neq 0$, then 
$\langle n_{r}^{A,B} (t)n_{0}^{A,B} (0)\rangle \neq \langle n_{-r}^{A,B}(t)
n_{0}^{A,B} (0)\rangle$ because of the drift which is 
due to an asymmetric Markov generator $H$. 
Such a behaviour has been observed in single-species one-dimensional 
free-fermionic models \cite{Grynberg2,Schutz2}.

ii) In higher dimensions ($d\geq 2$), the initial state is no longer 
translationally invariant and
with (\ref{eq.6.18},\ref{eq.6.19}) and (\ref{eq.2.7}), we  find 
\begin{eqnarray}
\label{eq.6.19.1}
\langle n_{m}^{A}(t) n_{l}^{A}(0)\rangle &=& 
\rho_{A}(\infty) (1-e^{-d|\gamma_{A}| t})+
\rho_{A}(0) e^{-\frac{d|B_{A}|t}{2}}\prod_{\alpha=1\dots d}\mu_{A}^{m_{\alpha}-m'_{\alpha}}I_{m_{\alpha}-l_{\alpha}}(2 C_{A}t) \nonumber\\&+&
{\cal K}_{AA}\sum_{(m_{1}'\neq l_{1},\dots, m_{d}'\neq l_{d})} \prod_{\alpha=1\dots d}\mu_{A}^{m_{\alpha}-m'_{\alpha}} |m'_{\alpha}-l_{\alpha}|^{-\Delta_{AA}^{\alpha}} e^{-\frac{|B_{A}|t}{2}} 
I_{m_{\alpha} - m'_{\alpha}}(2C_{A}t)
\end{eqnarray}
\begin{eqnarray}
\label{eq.6.20}
\langle n_{m}^{A}(t) n_{l}^{B}(0)\rangle &=& 
\rho_{A}(\infty) (1-e^{-d|\gamma_{A}| t}) \nonumber\\&+&
{\cal K}_{AB}\sum_{(m_{1}'\neq l_{1},\dots, m_{d}'\neq l_{d})} \prod_{\alpha=1\dots d}
\mu_{A}^{m_{\alpha}-m'_{\alpha}} |m'_{\alpha}-l_{\alpha}|^{-\Delta_{AB}^{\alpha} } e^{-\frac{|B_{A}|t}{2}} I_{m_{\alpha}-m'_{\alpha}}(2 C_{A}t)
\end{eqnarray}
\begin{eqnarray}
\label{eq.6.21}
\langle n_{m}^{B}(t) n_{l}^{B}(0)\rangle &=& 
\rho_{B}(\infty) (1-e^{-d|\gamma_{B}| t})+
\rho_{B}(0) e^{-\frac{d|B_{B}|t}{2}}\prod_{\alpha=1\dots d}\mu_{B}^{m_{\alpha}-m'_{\alpha}}I_{m_{\alpha}-m'_{\alpha}}(2 C_{B}t) \nonumber\\&+&
{\cal K}_{BB}\sum_{(m_{1}'\neq l_{1},\dots, m_{d}'\neq l_{d})} \prod_{\alpha=1\dots d}\mu_{B}^{m_{\alpha}-m'_{\alpha}} |m'_{\alpha}-l_{\alpha}|^{-\Delta_{BB}^{\alpha}} e^{-\frac{| B_{B}|t}{2}} I_{m_{\alpha}-m'_{\alpha}}(2 C_{B}t)
\end{eqnarray}
\begin{eqnarray}
\label{eq.6.22}
\langle n_{m}^{B}(t) n_{l}^{A}(0)\rangle &=& 
\rho_{B}(\infty) (1-e^{-d|\gamma_{B}| t}) \nonumber\\&+&
{\cal K}_{BA}\sum_{(m_{1}'\neq l_{1},\dots, m_{d}'\neq l_{d})} \prod_{\alpha=1\dots d}\mu_{B}^{m_{\alpha}-m'_{\alpha}} |m'_{\alpha}-l_{\alpha}|^{-\Delta_{BA}^{\alpha}} e^{-\frac{|B_{B}|t}{2}} I_{m_{\alpha}-m'_{\alpha}}(2 C_{B}t)
\end{eqnarray}
We see that in higher dimensions, because of the broken symmetry of the initial 
state, 
the non-instantaneous  correlation functions no longer depends on $r_{\alpha}=m_{\alpha}-l_{\alpha}$.
We can study the asymptotic behavior of these non-instantaneous correlation 
functions in an 
unified way (including both $d=1$ and $d\geq 2$), namely 
for $r_{\alpha}=m_{\alpha}-l_{\alpha}$, with 
$|r_{\alpha}|=|m_{\alpha}-l_{\alpha}|\sim|m_{\alpha}|\gg 1$, 
with $r_{\alpha}=\sigma_{\alpha}L$ and $u_{i}=L^2/4|C_{i}|t <\infty$, 
in the regime where $|C_{i}|t, r \gg1$, $(i,j) \in (A,B)$, the correlation
functions are given by
\begin{eqnarray}
\label{eq.6.23}
\langle n_{m}^{i}(t) n_{l}^{j}(0)\rangle &=& 
\rho_{i}(\infty) (1-e^{-d|\gamma_{i}| t})+
 exp\left(2 \sum_{\alpha} \sigma_{\alpha}^{2} u_{i}-(\frac{\vec{r} - 
\vec{\epsilon_{i}}|C_{i}|t}{2|C_{i}|t})^2\right) 
e^{-d(\frac{|B_{i}|}{2}-|C_{i}|(2+\epsilon_{i}^{2}/2))t} \nonumber\\
&\times&
\left(\frac{\rho_{i}(0) e^{-\sum_{\alpha}\sigma_{\alpha}^2 u_{i}}\delta_{i,j}}{(4\pi|C_{i}|t)^{d/2}}  
+{\cal K}_{ij} \prod_{\alpha=1\dots d}
 \left[\frac{e^{-\sigma_{\alpha}^{2} u_{i}}}{1-\Delta_{ij}^{\alpha}}\sqrt{\frac{u_{i}\sigma_{\alpha}^{2}}{\pi}}\frac{1}{4u_{i}|C_{i}|\sigma_{\alpha}^{2}t^{\Delta_{ij}^{\alpha}/2}}\right] + {\cal O}(t^{-2d})\right)\;,\;
 0<\Delta_{ij}^{\alpha}<1
\end{eqnarray}
where we use $\vec{r}\equiv \sum_{\alpha}r_{\alpha} e_{\alpha}$ and
$\vec{\epsilon_{i}}\equiv\epsilon_{i}\sum_{\alpha}e_{\alpha} $.
Moreover, 
\begin{eqnarray}
\label{eq.6.24}
\langle n_{m}^{i}(t) n_{l}^{j}(0)\rangle &=& 
\rho_{i}(\infty) (1-e^{-d|\gamma_{i}| t})+
 exp\left(2\sum_{\alpha} \sigma_{\alpha}^{2} u_{i}-(\frac{\vec{r} - 
\vec{\epsilon_{i}}|C_{i}|t}{2|C_{i}|t})^2\right) 
e^{-d(\frac{|B_{i}|}{2}-|C_{i}|(2+\epsilon_{i}^{2}/2))t} \nonumber\\
&\times&
\frac{1}{(4\pi|C_{i}|t)^{d/2}} \left(\rho_{i}(0) e^{-\sum_{\alpha}\sigma_{\alpha}^2 u_{i}}  \delta_{i,j} + {\cal K}_{ij}
 \prod_{\alpha=1\dots d} \zeta(\Delta_{ij}^{\alpha} ) + {\cal O}(t^{-2d})\right)\;,\;
 \;\Delta_{ij}^{\alpha}>1
\end{eqnarray}
When $\Delta_{ij}^{\alpha}=1$, a cross-over takes place and logarithmic
corrections arise, namely, 
\begin{eqnarray}
\label{eq.6.25}
\langle n_{m}^{i}(t) n_{l}^{j}(0)\rangle &=& 
\rho_{i}(\infty) (1-e^{-d|\gamma_{i}| t})+
 exp\left(2\sum_{\alpha} \sigma_{\alpha}^{2} u_{i}-(\frac{\vec{r} - 
\vec{\epsilon_{i}}|C_{i}|t}{2|C_{i}|t})^2\right) 
e^{-d(\frac{|B_{i}|}{2}-|C_{i}|(2+\epsilon_{i}^{2}/2))t} \nonumber\\
&\times&
\frac{1}{(4\pi|C_{i}|t)^{d/2}} \left(\rho_{i}(0)  e^{-\sum_{\alpha}\sigma_{\alpha}^2 u_{i}}  \delta_{i,j} + {\cal K}_{ij}
 \prod_{\alpha=1\dots d}\ln(4u_{i}\sigma_{\alpha}|C_{i}|t) + {\cal O}(t^{-2d}) 
 \right)\;,\;\;
 \Delta_{ij}^{\alpha}=1
\end{eqnarray}
Therefore, when the spatial correlations are important, i.e.,
$\Delta_{ij}^{\alpha}<1 $, correlation functions
(at $\frac{|B_{i}|}{2}-|C_{i}|(2+\epsilon_{i}^{2}/2d))=0$) decay as
\begin{eqnarray}
\label{eq.6.26}
\langle n_{m}^{i}(t) n_{l}^{j}(0)\rangle \sim
 \frac{ exp\left(2\sum_{\alpha} \sigma_{\alpha}^{2} u_{i}-(\frac{\vec{r} - 
\vec{\epsilon_{i}}|C_{i}|t}{2|C_{i}|t})^2\right)  }{(4\pi|C_{i}|t)^{\sum_{\alpha}\Delta_{ij}^{\alpha} /2}}, \Delta_{ij}^{\alpha}<1 
\end{eqnarray}
On the contrary, weak spatial initial correlations do not affect the long time
behavior of correlation functions, since 
\begin{eqnarray}
\label{eq.6.27}
\langle n_{m}^{i}(t) n_{l}^{j}(0)\rangle \sim
 \frac{ exp\left(2\sum_{\alpha} \sigma_{\alpha}^{2} u_{i}-(\frac{\vec{r} - 
\vec{\epsilon_{i}}|C_{i}|t}{2|C_{i}|t})^2\right)  }
{(4\pi|C_{i}|t)^{d/2}} \;, \; \Delta_{ij}^{\alpha}>1 
\end{eqnarray}
The marginal case $\Delta_{ij}^{\alpha}=1$ has logarithmic corrections
\begin{eqnarray}
\label{eq.6.28}
\langle n_{m}^{i}(t) n_{l}^{j}(0)\rangle \sim
 \frac{  exp\left(2\sum_{\alpha} \sigma_{\alpha}^{2} u_{i}-(\frac{\vec{r} - 
\vec{\epsilon_{i}}|C_{i}|t}{2|C_{i}|t})^2\right)\left(\ln{4|C_{i}|t}\right)^d}{(4\pi|C_{i}|t)^{d/2}} \;, \; \Delta_{ij}^{\alpha}=1 
\end{eqnarray}
Notice the drift which occurs for  $\epsilon \neq 0$ and 
the effect of dimensionality. 
The fact that initially the state is translationally-invariant ($d=1$) 
gives rise to the same  asymptotic behavior $(t^{-d/2})$ 
as for the non-translationally-invariant system in higher dimensions ($d\geq 2$).

\section{Instantaneous two-point correlation functions on the manifold $V_{2}$}
We now pass to the computation of the two-point correlation function
on the translation invariant manifold $V_2(d)$ (\ref{eq.1.13}). 
From (\ref{eq.1.7},\ref{eq.1.8}), the evolution equations of correlation functions follow.
We shall discuss both cases, when the initial state is correlated and when 
it is uncorrelated.
We shall evaluate
\begin{eqnarray}
\label{eq.3.1}
{\cal G}_{|r|=|n-m|}^{AA} (t)&\equiv&\langle n_{n}^{A}n_{m}^{A}\rangle(t)
\equiv\langle n_{|m-n|}^{A}n_{0}^{A}\rangle(t)\equiv {\cal G}_{r}^{AA} (t)\nonumber\\
{\cal G}_{|r|=|n-m|}^{BB} (t)&\equiv&\langle n_{n}^{B}n_{m}^{B}\rangle(t)
\equiv\langle n_{|m-n|}^{B}n_{0}^{B}\rangle(t)\equiv {\cal G}_{r}^{BB} (t)\nonumber\\
{\cal G}_{|r|=|n-m|}^{AB} (t)&\equiv&\langle n_{n}^{A}n_{m}^{B}\rangle(t)\equiv\langle n_{n}^{B}n_{m}^{A}\rangle(t)\equiv\langle n_{|m-n|}^{A}n_{0}^{B}\rangle(t)\equiv\langle n_{|m-n|}^{B}n_{0}^{A}\rangle(t)\equiv {\cal G}_{r}^{AB} (t)\equiv {\cal G}_{r}^{BA} (t)
\end{eqnarray}
with the boundary conditions at $r=0$ (for the densities see (\ref{eq.2.16})):
\begin{eqnarray}
\label{eq.3.1}
{\cal G}_{r=0}^{AA}(t)\equiv\rho_{A}(t) \; ;\;
{\cal G}_{r=0}^{BB}(t)\equiv\rho_{B}(t) \; ;\;
{\cal G}_{r=0}^{AB}(t)\equiv 0
\end{eqnarray}
Notice that the point $|r|=1$ must be dealt with care.
Further, it will be convenient to distinguish the one dimensional problem from
that in $d\geq 2$. In this section, in addition to the definition of the Appendix A, we also introduce some specific notations and abbreviations (see Appendix B, (\ref{eq.B.1}), (\ref{eq.B.2}) and (\ref{eq.B.3})).
Let us start the discussion with one spatial dimension.
\subsection{One-dimensional instantaneous two-point correlation functions on $V_{2}(d=1)$}
On account of the above remarks and (\ref{eq.1.7}), the equations of motion 
for the correlation functions read ($|r|>1$):
\begin{eqnarray}
\label{eq.3.2}
\frac{d}{dt}{\cal G}_{r}^{AA}(t)= B_{A} {\cal G}_{r}^{AA}(t)+
{\cal C}_{A}\left({\cal G}_{r+1}^{AA}(t)+{\cal G}_{r-1}^{AA}(t)\right) + {\cal A}_{A} \rho_{A}(t)
\end{eqnarray}
while for $|r|=1$, we have   (\ref{eq.1.8}),
\begin{eqnarray}
\label{eq.3.3}
\frac{d}{dt}{\cal G}_{1}^{AA}(t)&=& (G_{1}^{a}+B_{A}/2){\cal G}_{1}^{AA}(t)
+{\cal C}_{A} {\cal G}_{2}(t) +E_{0}^{a}+(F_{1}^{a}+
F_{2}^{a}+{\cal A}_{A}/2)\rho_{A}(t)+(F_{3}^{a}+F_{4}^{a})\rho_{B}(t)
\end{eqnarray}
For $r=0$, we recover 
\begin{eqnarray}
\label{eq.3.4}
\frac{d}{dt}{\cal G}_{0}^{AA}(t)=\frac{d}{dt}\rho_{A}(t)=\frac{{\cal A}_{A}}{2}+\left(\frac{ B_A+2{\cal C}_A}{2}\right) \rho_{A}(t)
\end{eqnarray}
The solution of the above set of coupled differential equations
(\ref{eq.3.2}-\ref{eq.3.4})  
can be expressed in terms of the modified Bessel functions
$I_{\nu}(z)$ (See appendix C), and for $|r|\equiv|n-m|>0$ we have:
\begin{eqnarray}
\label{eq.3.10}
{\cal G}_{r}^{AA}(t)&-&(\rho_{A}(t))^2 = -(\rho_{A}(0) e^{-|\gamma_{A}|t})^2 +\left(\rho_{A}(0)+\frac{{\cal D}_{2,0}^{A} +{\cal D}_{2,1}^{A}+{\cal D}_{2,2}^{A} }{{\cal C}_{A}}\right) e^{-|B_{A}|t}I_{r}(2{\cal C}_{A}t)+\sum_{r'\neq 0}{\cal G}_{r'}^{AA}(0)
 e^{-|B_{A}|t}I_{r-r'}(2{\cal C}_{A}t) \nonumber\\
&+& \left({\cal D}_{1,0}^{A}+ \frac{{\cal D}_{2,0}^{A}|B_{A}|}{{\cal C}_{A}}\right) \int_{0}^{t}d\tau e^{-|B_{A}|\tau}I_{r}(2{\cal C}_{A}\tau)+ \left({\cal D}_{1,1}^{A}+ \frac{{\cal D}_{2,1}^{A}(|B_{A}|-|\gamma_{A}|)}{{\cal C}_{A}} \right) e^{-|\gamma_{A}|t} \int_{0}^{t}d\tau e^{-(|B|-|\gamma_{A}|)\tau}I_{r}(2{\cal C}_{A}\tau)\nonumber\\
&+& \left( \frac{{\cal D}_{2,2}^{A}(|B_{A}|-|\gamma_{B}|)}{{\cal C}_{A}} \right) e^{-|\gamma_{B}|t} \int_{0}^{t}d\tau e^{-(|B_{A}|-|\gamma_{B}|)\tau}I_{r}(2{\cal C}_{A}\tau)\nonumber\\
&-& \int_{0}^{t} dt'  e^{-|B_{A}|(t-t')}  {\cal G}_{1}^{AA}(t')\left[\frac{(G_{1}^{a}-B_{A}/2)}{{\cal C}_{A} } \frac{\partial}{\partial t'}I_{r}(2{\cal C}_{A}(t-t')) +2{\cal C}_{A} I_{r}(2{\cal C}_{A}(t-t'))\right],
\end{eqnarray}
similarly, we find
\begin{eqnarray}
\label{eq.3.11}
{\cal G}_{r}^{BB}(t)&-&(\rho_{B}(t))^2 = -(\rho_{B}(0) e^{-|\gamma_{B}|t})^2 +\left(\rho_{B}(0)+\frac{{\cal D}_{2,0}^{B} +{\cal D}_{2,1}^{B}+{\cal D}_{2,2}^{B} }{{\cal C}_{B}}\right) e^{-|B_{B}|t}I_{r}(2{\cal C}_{B}t)+\sum_{r'\neq 0}{\cal G}_{r'}^{BB}(0)
 e^{-|B_{A}|t}I_{r-r'}(2{\cal C}_{B}t) \nonumber\\
&+& \left({\cal D}_{1,0}^{B} + \frac{{\cal D}_{2,0}^{B}|B_{B}|}{{\cal C}_{B}}\right) \int_{0}^{t}d\tau e^{-|B_{B}|\tau}I_{r}(2{\cal C}_{B}\tau)+ \left({\cal D}_{1,1}^{B}+ \frac{{\cal D}_{2,1}^{B}(|B_{B}|-|\gamma_{B}|)}{{\cal C}_{B}} \right) e^{-|\gamma_{B}|t} \int_{0}^{t}d\tau e^{-(|B_{B}|-|\gamma_{B}|)\tau}I_{r}(2{\cal C}_{B}\tau)\nonumber\\
&+& \left( \frac{{\cal D}_{2,2}^{B}(|B_{B}|-|\gamma_{A}|)}{{\cal C}_{B}} \right) e^{-|\gamma_{A}|t} \int_{0}^{t}d\tau e^{-(|B_{B}|-|\gamma_{A}|)\tau}I_{r}(2{\cal C}_{B}\tau)\nonumber\\
&-& \int_{0}^{t} dt'  e^{-|B_{B}|(t-t')}  {\cal G}_{1}^{BB}(t')\left[\frac{(G_{2}^{b}-B_{B}/2)}{{\cal C}_{B}} \frac{\partial}{\partial t'}I_{r}(2{\cal C}_{B}(t-t')) +2{\cal C}_{B} I_{r}(2{\cal C}_{B}(t-t'))\right],
\end{eqnarray}
and we also obtain
\begin{eqnarray}
\label{eq.3.12}
{\cal G}_{r}^{AB}(t)&-&(\rho_{A}(t)\rho_{B}(t))=\left(\rho_{A}(\infty)\rho_{B}(0) + \rho_{A}(0)\rho_{B}(\infty) - \rho_{A}(\infty)\rho_{B}(\infty)\right)e^{-|\gamma_{A}+\gamma_{B}|t}\nonumber\\
&+&\left(\frac{{\cal D}_{2,0}^{AB} +{\cal D}_{2,1}^{AB}+{\cal D}_{2,2}^{AB} }{{\cal C}_{AB}}\right) e^{-|B_{AB}|t}I_{r}(2{\cal C}_{AB}t)
+\sum_{r'\neq 0} {\cal G}_{r'}^{AB}(0) e^{-|B_{AB}|t} I_{r-r'}(2{\cal C}_{AB}t)\nonumber\\
&+& \left({\cal D}_{1,0}^{AB}+ \frac{{\cal D}_{2,0}^{AB}|B_{AB}|}{{\cal C}_{AB}}\right) \int_{0}^{t}d\tau e^{-|B_{AB}|\tau} I_{r}(2{\cal C}_{AB}\tau)
\nonumber\\ &+& \left({\cal D}_{1,1}^{A}+ \frac{{\cal D}_{2,1}^{AB}(|B_{AB}|-|\gamma_{A}|)}{{\cal C}_{AB}} \right) e^{-|\gamma_{A}|t} \int_{0}^{t}d\tau e^{-(|B_{AB}|-|\gamma_{A}|)\tau}I_{r}(2{\cal C}_{AB}\tau)\nonumber\\
&+& \left({\cal D}_{1,2}^{AB} +  \frac{{\cal D}_{2,2}^{AB}(|B_{AB}|-|\gamma_{B}|)}{{\cal C}_{AB}} \right) e^{-|\gamma_{B}|t} \int_{0}^{t}d\tau e^{-(|B_{AB}|-|\gamma_{B}|)\tau}I_{r}(2{\cal C}_{AB}\tau)\nonumber\\
&-&\int_{0}^{t} dt'e^{-|B_{AB}|(t-t')} {\cal G}_{1}^{AB}(t')
\left(\frac{H_{1}^{ab}+ H_{2}^{ab} -A_{1}^{a}-D_{2}^{b}}{{\cal C}_{AB} }\right)\frac{\partial}{\partial t'}I_{r}(2{\cal C}_{AB}(t-t'))
\nonumber\\
&-& 2{\cal C}_{AB} \int_{0}^{t} dt'e^{-| B_{AB}|(t-t')} {\cal G}_{1}^{AB}(t')
I_{r}(2{\cal C}_{AB}(t-t'))
\end{eqnarray}
To study the asymptotic behavior of these expressions, we shall distinguish 
two regimes:

 i) long time , i.e., $|{\cal C}_{j} |t\gg 1$ and large distances
, i.e., $r\sim L\gg 1$ with the ratios
$\frac{r^2}{|{\cal C}_{j}|t}<\infty$ and 
$u_{l}\equiv L^2/4|{\cal C}_{l}|t <\infty$ hold finite. 

ii) long time , i.e., $|{\cal C}_{j}|t\gg 1$ and finite distances, 
$r\ll L \rightarrow \infty$ with $\frac{r^2}{|{\cal C}_{j}|t}\ll 1$ and
$u_{l}\equiv L^2/4|{\cal C}_{l}|t <\infty$.

In order to investigate the effect of initial correlations on the dynamics, 
we consider
\begin{eqnarray}
\label{eq.3.17.1}
 {\cal G}_{|r|>0}^{l}(0)
=\rho_{i}(0)\rho_{j}(0)(1 - |\kappa_{l}|(sign({\cal C}_{l}))^{r} r^{-\nu_{l}}),
\;\nu_{l}\geq 0 ,
\end{eqnarray}
where ($l \in (AA, BB, AB), \; i,j \in {A,B}$).

Such a choice has been made for the one-dimensional single-species symmetric 
$A+A\leftrightarrow \emptyset+\emptyset$ process \cite{Grynberg}.
In this subsection, we want to proceed with a systematic study of instantaneous 
correlations, on the manifold $V_{2}(d=1)$, with the choice 
(\ref{eq.3.17.1}) for the initial state. 

Below we shall use the incomplete gamma function, $\Gamma(\nu,u)\equiv 
\int_{u}^{\infty} dx e^{-x} x^{\nu-1}$, as well as the Riemann zeta functions
$\zeta(\nu)=\sum_{k\geq 1} k^{-\nu}, \nu>1$. 
For notational simplicity, we write $\nu$ instead of $\nu_{l}$.
The results for the asymptotics are summarized as follows 

1. For $|B_{l}|>2|{\cal C}_{l}|$, the decay of correlations is exponential.
With the definition
\begin{eqnarray}
\label{eq.3.18}
\varphi_{jj}\equiv min(|\gamma_{j}|, |B_{j}|-2|{\cal C}_{j}|)
\end{eqnarray}
we have\\
\vskip 0.5cm 
1.a For $\varphi_{jj}=|\gamma_{A}|\neq|B_{j}|-2|{\cal C}_{j}|$, 
\begin{eqnarray}
\label{eq.3.18.1}
{\cal G}_{r}^{jj}(t)-{\cal G}_{r}^{jj}(\infty)\sim e^{-|\gamma_{j}|t}
\end{eqnarray}
\vskip 0.5cm
1.b For $\varphi_{jj}=|B_{j}|-2|{\cal C}_{j}|>0$, and 
${\cal G}_{|r|>0}(0)\neq 0$
\begin{eqnarray}
\label{eq.3.18.2}
{\cal G}_{r}^{jj}(t)-{\cal G}_{r}^{jj}(\infty)\sim 
 \rho_{j}(0)^2 e^{-(|B_{j}|-2|{\cal C}_{j}|)t}
\end{eqnarray}
Note that in the case where ${\cal G}_{r}(0)=0, \forall r$, we have
\begin{eqnarray}
\label{eq.3.18.4}
{\cal G}_{r}^{jj}(t)-{\cal G}_{r}^{jj}(\infty)\sim 
\left\{
  \begin{array}{l l}
   \frac{Q_{1}(u_{j}, \sigma)e^{-(|B_{j}|-2|{\cal C}_{j}|)t} }{\sqrt{4\pi|{\cal C}_{j}|t}}  &\mbox{if $r\ll L$}\\
 \frac{Q_{1}(u_{j}, \sigma)e^{-\sigma^{2}u_{j}-(|B_{j}|-2|{\cal C}_{j}|)t}}{\sqrt{4\pi|{\cal C}_{j}|t}} &\mbox{si $r\gg 1$}\\
\end{array}
\right.
\end{eqnarray}
where  $Q_{1}(u_{j}, \sigma)$ is a function explicitely determined
by the processes which occur in the system under consideration.
\vskip 0.5cm
1.c For $|B_{j}|\neq 2|{\cal C}_{j}|$, $|\gamma_{j'\neq\j}|=
|B_{j}|- 2|{\cal C}_{j}|$, $|\gamma_{j}|>0$, and  ${\cal D}_{1,2}^{j}+
2(sign {\cal C}_{j}){\cal D}_{2,2}^{j}\neq 0 $, we have
\begin{eqnarray}
\label{eq.3.18.5}
{\cal G}_{r}^{jj}(t)-{\cal G}_{r}^{jj}(\infty)\sim 
\left\{
  \begin{array}{l l}
 e^{-|\gamma_{j}|t} &\mbox{if $\varphi_{jj}=|\gamma_{j}|$}\\
 e^{-|\gamma_{j'\neq j}|t}\sqrt{\frac{t}{\pi|{\cal C}_{j}|}}
 &\mbox{if $\varphi_{jj}=|\gamma_{j'\neq j}|$}\\
\end{array} 
\right.
\end{eqnarray}
\vskip 0.5cm
1.d  For $|B_{j}|\neq 2|{\cal C}_{j}|$, $|\gamma_{j}|=
|B_{j}|- 2|{\cal C}_{j}|$, $|\gamma_{j'\neq j}|>0$, and  ${\cal D}_{1,1}^{j}+
2(sign {\cal C}_{j}){\cal D}_{2,1}^{j}\neq 0 $, we have
\begin{eqnarray}
\label{eq.3.18.6}
{\cal G}_{r}^{jj}(t)-{\cal G}_{r}^{jj}(\infty)\sim  e^{-|\gamma_{j}|t}\sqrt{\frac{t}{\pi|{\cal C}_{j}|}}
\end{eqnarray}
\vskip 0.5cm
The correlation functions ${\cal G}_{r}^{AB}(t)$ 
have to be discussed separately. With the definition 
\begin{eqnarray}
\label{eq.3.18.7}
\varphi_{AB}\equiv min(|\gamma_{A}|,|\gamma_{B}|, |B_{AB}|-2|{\cal C}_{AB}|)
\end{eqnarray}

1.e  For $\varphi_{AB}=|\gamma_{A}|\neq|B_{j}|-2|{\cal C}_{j}|$, 
we have
\begin{eqnarray}
\label{eq.3.18.8}
{\cal G}_{r}^{AB}(t)-{\cal G}_{r}^{AB}(\infty)\sim e^{-|\gamma_{A}|t}
\end{eqnarray}
\vskip 0.5cm
1.f  For $\varphi_{AB}=|\gamma_{B}|\neq|B_{AB}|-2|{\cal C}_{AB}|$, we have
\begin{eqnarray}
\label{eq.3.18.9}
{\cal G}_{r}^{AB}(t)-{\cal G}_{r}^{AB}(\infty)\sim e^{-|\gamma_{B}|t}
\end{eqnarray}
\vskip 0.5cm
1.g  For $\varphi_{AB}=|B_{AB}|-2|{\cal C}_{AB}|>0$ and ${\cal G}_{|r|>0}^{AB}
\neq 0$
\begin{eqnarray}
\label{eq.3.18.10}
{\cal G}_{r}^{AB}(t)-{\cal G}_{r}^{AB}(\infty)
\sim \rho_{A}(0)\rho_{B}(0) e^{-(|B_{AB}|-2|{\cal C}_{AB}|)t}
\end{eqnarray}
Notice again that if ${\cal G}_{r}^{AB}(0)=0, \forall r$, we arrive at
\begin{eqnarray}
\label{eq.3.18.12}
{\cal G}_{r}^{AB}(t)-{\cal G}_{r}^{AB}(\infty)\sim 
\left\{
  \begin{array}{l l}
   \frac{Q_{1}(u_{j}, \sigma)e^{-(|B_{AB}|-2|{\cal C}_{AB}|)t} }{\sqrt{4\pi|{\cal C}_{AB}|t}}  &\mbox{if $r\ll L$}\\
 \frac{Q_{1}(u_{AB}, \sigma)e^{-\sigma^{2}u_{AB}-(|B_{AB}|-2|{\cal C}_{AB}|)t}}{\sqrt{4\pi|{\cal C}_{AB}|t}} &\mbox{if $r\gg 1$}
\end{array}
\right.
\end{eqnarray}
Where $Q_{1}(u_{j}, \sigma)$ is the same quantity as above.
\vskip 0.5cm
1.h For $|B_{AB}|\neq 2|{\cal C}_{AB}|$, $|\gamma_{A}|=
|B_{AB}|- 2|{\cal C}_{AB}|$, and ${\cal D}_{1,1}^{AB}+
2(sign {\cal C}_{AB}){\cal D}_{2,1}^{AB}\neq 0 $, we have
\begin{eqnarray}
\label{eq.3.18.13}
{\cal G}_{r}^{AB}(t)-{\cal G}_{r}^{AB}(\infty)\sim 
\left\{
  \begin{array}{l l}
 e^{-|\gamma_{B}|t} &\mbox{if $\varphi_{AB}=|\gamma_{B}|$}\\
 e^{-|\gamma_{A}|t}\sqrt{\frac{t}{\pi|{\cal C}_{AB}|}}
 &\mbox{if $\varphi_{AB}=|\gamma_{A}|$}\\
\end{array} 
\right.
\end{eqnarray}
\vskip 0.5cm
1.i  For $|B_{AB}|\neq 2|{\cal C}_{AB}|$, $|\gamma_{B}|=
|B_{AB}|- 2|{\cal C}_{AB}|$, and ${\cal D}_{1,2}^{AB}+
2(sign {\cal C}_{AB}){\cal D}_{2,2}^{AB}\neq 0 $, we have
\begin{eqnarray}
\label{eq.3.18.14}
{\cal G}_{r}^{AB}(t)-{\cal G}_{r}^{AB}(\infty)\sim 
\left\{
  \begin{array}{l l}
 e^{-|\gamma_{A}|t} &\mbox{if $\varphi_{AB}=|\gamma_{A}|$}\\
 e^{-|\gamma_{B}|t}\sqrt{\frac{t}{\pi|{\cal C}_{AB}|}}
 &\mbox{if $\varphi_{AB}=|\gamma_{B}|$}\\
\end{array}
\right. 
\end{eqnarray}
\vskip 0.5cm

2. For $|B_{l}|=2|{\cal C}_{l}|$, the correlation functions decay algebraically.
It is appropriate to distinguish, $r\ll L$ and $r\sim L$. 
Again, $\sigma=r/L$, $l\in(AA, BB, AB)$.

In the regime $r\ll L$,
\begin{eqnarray}
\label{eq.3.19.1}
{\cal G}_{r}^{l}(t)-{\cal G}_{r}^{l}(\infty)\sim 
\left\{
  \begin{array}{l l l}
     \frac{{\cal F}_{1}(u_{l},\sigma,\nu)  }{(4|{\cal C}_{l}|t)^{\nu/2}}
&\mbox{if $0<\nu<1$}\\    
 \frac{ \left[2\zeta(\nu) + (4\sigma^{2}u_{l}|{\cal C}_{l}|t)^{(1-\nu)/2} \right]}{(4\pi|{\cal C}_{l}|t)^{1/2}} &\mbox{if $\nu >1$}\\
 \frac{\ln{[4|{\cal C}_{l}|u_{l}(1-\sigma)t]} }{(4\pi|{\cal C}_{l}|t)^{1/2}}  &\mbox{if $\nu =1$},
\end{array}
\right.
\end{eqnarray}
where the following auxiliary function  has been used
\begin{eqnarray}
\label{eq.3.19.2.1}
{\cal F}_{1}(u,\sigma, \nu)\equiv \frac{ \left(\Gamma(\frac{1-\nu}{2}) +\Gamma(\frac{1-\nu}{2},\sigma^2 u_{l})
-\Gamma(\frac{1-\nu}{2},u_{l}(1-\sigma)^2)-
 \Gamma(\frac{1-\nu}{2},u_{l}(1+\sigma)^2)\right) }{\sqrt{4\pi}}
\end{eqnarray}
while for $r\gg 1, r\equiv \sigma L \sim L$, we find
\begin{eqnarray}
\label{eq.3.19.2}
{\cal G}_{r}^{l}(t)-{\cal G}_{r}^{l}(\infty)\sim 
\left\{
  \begin{array}{l l l}
    \frac{{\cal F}_{2}(u_{l},\sigma, \nu) }{(4|{\cal C}_{l}|t)^{\nu/2}}
&\mbox{if $0<\nu<1$}\\
    \frac{\left[(1+e^{-\sigma^{2}u_{l}})\zeta(\nu) +  ((1-\sigma)/\sigma) (4\sigma^{2}u_{l}|{\cal C}_{l}|t)^{(1-\nu)/2} \right]  }{(4 \pi|{\cal C}_{l}|t)^{1/2}} &\mbox{if $\nu >1$}\\
 \frac{\ln{(4|{\cal C}_{l}| u_{l} t})}{(4 \pi|{\cal C}_{l}|t)^{1/2}} e^{-\sigma^{2}u_{l}} &\mbox{if $\nu =1$},
\end{array}
\right.
\end{eqnarray}
with the auxiliary function
\begin{eqnarray}
\label{eq.3.19.2.1}
{\cal F}_{2}(u_{l},\sigma, \nu)\equiv \frac{e^{-\sigma^2u_{l} }}{(1-\nu)}\sqrt{\frac{u_{l}\sigma^2}{\pi}}
\end{eqnarray}
Notice that when the initial correlations are absent ($\nu =0$),
\begin{eqnarray}
\label{eq.3.19.3}
{\cal G}_{r}^{l}(t)-{\cal G}_{r}^{l}(\infty)\sim 
\left\{
  \begin{array}{l l l}
\frac{Q_{1}(u,\sigma) +{\cal O}(1)}{\sqrt{4\pi|{\cal C}_{l}|t}},  &\mbox{if ${\cal C}_{l}< 0$}\\
   \frac{Q_{1}(u,\sigma)}{\sqrt{4\pi|{\cal C}_{l}|t}}+ {\cal F}_{1}(u,\sigma, \nu)\frac{\rho_{i}(0)\rho_{j}(0)(1-|\kappa_{l}|)}{8|{\cal C}_{l}|t}  &\mbox{if $r\ll L$ and ${\cal C}_{l}>0$}\\
 \frac{e^{-\sigma^{2}u_{l}}Q_{1}(u,\sigma) }{\sqrt{4\pi|{\cal C}_{l}|t}}+ {\cal F}_{2}(u,\sigma, \nu)\frac{\rho_{i}(0)\rho_{j}(0)(1-|\kappa_{l}|)}{8|{\cal C}_{l}|t}  &\mbox{if $r\gg 1$ and ${\cal C}_{l}>0$ }\\
\end{array}
\right.
\end{eqnarray}
where $Q_{1}(u,\sigma)$ has been defined previously.

From the above we infer that the initial conditions affect the asymptotic
behavior of correlation functions only when
the latter decay algebraically 
(\ref{eq.3.19.1}, \ref{eq.3.19.2}, \ref{eq.3.19.3}). 
Provided the initial correlations are dominant,
($0<\nu_{l}<1$), the critical exponent is renormalized, while for weak initial 
correlations,
($\nu_{l}>1$), the exponent is independent of initial correlations, i.e., 
$1/2$. The intermediate regime, $\nu_{l}=1$, exhibits logarithmic dependence 
consistent with a marginal behavior.

\subsection{Two-point instantaneous correlation functions 
on $V_{2}(d)$ in arbitrary dimension}
This section is devoted to the computation of correlation functions
in arbitrary dimensions ($d\geq 2$) on the manifold $V_{2}(d)$.

With the notations
$r=(r_1, \dots, r_{\alpha}, \dots, r_{d})$,  
$|r|=\sqrt{\sum_{\alpha} r_{\alpha}^{2}}$ (sometimes denoted by $r$, 
for notational simplicity) and $r_{\alpha}^{\pm}\equiv 
\sqrt{(r_{\alpha}\pm 1)^2+\sum_{\alpha\neq \alpha'} r_{\alpha'}^2}$, solving the multidimensional equations of motion of the correlation functions  
(see appendix D), we arrive at the following explicit forms
\begin{eqnarray}
\label{eq.3.31}
&&{\cal G}_{|r|=|(r_{1},\dots,r_{d})|>0}^{AA}(t)-(\rho_{A}(t))^{2}=
-(\rho_{A}(0))^2e^{-2|\gamma_{A}|dt} + 
\left(\frac{{\cal C}_{A} \rho_{A}(0) + {\cal D}_{2,0}^{A}+ {\cal D}_{2,1}^{A}+  {\cal D}_{2,2}^{A} }{{\cal C}_{A}}\right)\prod_{\alpha=1\dots d}\left(e^{-|B_{A}|t} I_{r_{\alpha}}(2{\cal C}_{A}t) \right)\nonumber\\&+& \sum_{r'\neq 0}{\cal G}_{|r'|}^{AA}(0)\prod_{\alpha = 1 \dots d}\left(e^{-|B_{A}|t} I_{r_{\alpha}-r'_{\alpha}}(2{\cal C}_{A}t)\right)
+ \left({\cal D}_{1,0}^{A}+\frac{{\cal D}_{2,0}^{A}|B_{A}|d}{{\cal C}_{A}}\right) \int_{0}^{t} d\tau \prod_{\alpha=1 \dots d }\left(e^{-|B_{A}|t}I_{r_{\alpha}}(2{\cal C}\tau)\right)
\nonumber\\
&+& \left({\cal D}_{1,1}^{A}+d \frac{{\cal D}_{2,1}^{A}(|B_{A}|-|\gamma_{A}|)}{{\cal C}_{A}}\right)
e^{-d|\gamma_{A}|t} \int_{0}^{t}d\tau e^{-d(|B_{A}|-|\gamma_{A}|)\tau}
\prod_{\alpha=1 \dots d }I_{r_{\alpha}}(2{\cal C}_{A}\tau) \nonumber\\
&+& d {\cal D}_{2,2}^{A} e^{-d|\gamma_{B}|t}\left(\frac{|B_{A}|-|\gamma_{B}|}{{\cal C}_{A}}\right) \int_{0}^{t}d\tau e^{-d(|B_{A}|-|\gamma_{B}|)\tau}\prod_{\alpha=1\dots d}I_{r_{\alpha}}(2{\cal C}_{A}\tau)
 \nonumber\\
&-& \int_{0}^{t}dt' {\cal G}_{1}(t')e^{-d|B_{A}|(t-t')}\left(2{\cal C}_{A}d \prod_{\alpha=1\dots d} I_{r_{\alpha}}(2{\cal C}_{A}(t-t')) + \frac{G_{1}^{a}-B_{A}/2}{{\cal C}_{A}} \frac{\partial}{\partial t'}\prod_{\alpha=1\dots d} I_{r_{\alpha}}(2{\cal C}_{A}(t-t'))\right),
\end{eqnarray}
we also have,
\begin{eqnarray}
\label{eq.3.32}
&&{\cal G}_{|r|=|(r_{1},\dots,r_{d})|>0}^{BB}(t)-(\rho_{B}(t))^{2}=
-(\rho_{B}(0))^2e^{-2|\gamma_{B}|dt} + 
\left(\frac{{\cal C}_{B} \rho_{A}(0) + {\cal D}_{2,0}^{B}+ {\cal D}_{2,1}^{B}+  {\cal D}_{2,2}^{B} }{{\cal C}_{B}}\right)\prod_{\alpha=1\dots d}\left(e^{-|B_{B}|t} I_{r_{\alpha}}(2{\cal C}_{B}t) \right)\nonumber\\&+& \sum_{r'\neq 0}{\cal G}_{|r'|}^{BB}(0)\prod_{\alpha = 1 \dots d}\left(e^{-|B_{B}|t} I_{r_{\alpha}-r'_{\alpha}}(2{\cal C}_{B}t)\right)
+\left({\cal D}_{1,0}^{B}+ \frac{{\cal D}_{2,0}^{B}|B_{B}|d}{{\cal C}_{B}}\right) \int_{0}^{t} d\tau \prod_{\alpha=1\dots d}\left(e^{-|B_{B}|t}I_{r_{\alpha}}(2{\cal C}_{B}\tau)\right)
\nonumber\\
&+& \left({\cal D}_{1,1}^{B}+d \frac{{\cal D}_{2,1}^{B}(|B_{B}|-|\gamma_{B}|)}{{\cal C}_{B}}\right)
e^{-d|\gamma_{B}|t} \int_{0}^{t}d\tau e^{-d(|B_{B}|-|\gamma_{B}|)\tau}\prod_{\alpha=1\dots d}I_{r_{\alpha}}(2{\cal C}_{B}\tau) \nonumber\\
&+& d {\cal D}_{2,2}^{B}e^{-d|\gamma_{A}|t}\left(\frac{|B_{B}|-|\gamma_{A}|}{{\cal C}_{B}}\right) \int_{0}^{t}d\tau e^{-d(|B_{B}|-|\gamma_{A}|)\tau}\prod_{\alpha=1\dots d}I_{r_{\alpha}}(2{\cal C}_{B}\tau)
 \nonumber\\
&-& \int_{0}^{t}dt' {\cal G}_{1}(t')e^{-d|B_{B}|(t-t')}\left(2{\cal C}_{B}d \prod_{\alpha=1\dots d} I_{r_{\alpha}}(2{\cal C}_{B}(t-t')) + \frac{G_{2}^{b}-B_{B}/2}{{\cal C}_{B}} \frac{\partial}{\partial t'}\prod_{\alpha=1\dots d} I_{r_{\alpha}}(2{\cal C}_{B}(t-t'))\right),
\end{eqnarray}
and
\begin{eqnarray}
\label{eq.3.33}
&&{\cal G}_{|r|=|(r_{1},\dots,r_{d})|>0}^{AB}(t)-(\rho_{A}(t)\rho_{B}(t))=
-(\rho_{A}(\infty)\rho_{B}(0) + \rho_{A}(0)\rho_{B}(\infty) -  \rho_{A}(\infty)\rho_{B}(\infty)) e^{-|\gamma_{A}+\gamma_{B}|dt} \nonumber\\ &-&(\rho_{A}(0) \rho_{B}(0))e^{-d(|\gamma_{A}|+|\gamma_{B}|)t} + 
\left(\frac{ {\cal D}_{2,0}^{AB}+ {\cal D}_{2,1}^{AB}+  {\cal D}_{2,2}^{AB} }{{\cal C}_{AB}}\right)\prod_{\alpha=1\dots d}\left(e^{-|B_{AB}|t} I_{r_{\alpha}}(2{\cal C}_{AB}t) \right)\nonumber\\&+& \sum_{r'\neq 0}{\cal G}_{|r'|}^{AB}(0)\prod_{\alpha = 1 \dots d}\left(e^{-|B_{AB}|t} I_{r_{\alpha}-r'_{\alpha}}(2{\cal C}_{AB}t)\right)
+\left({\cal D}_{1,0}^{AB}+  \frac{{\cal D}_{2,0}^{AB}|B_{AB}|d}{{\cal C}_{AB}}\right) \int_{0}^{t} d\tau \prod_{\alpha=1\dots d}\left(e^{-|B_{AB}|t}I_{r_{\alpha}}(2{\cal C}_{AB}\tau)\right)
\nonumber\\
&+& \left({\cal D}_{1,1}^{AB}+d \frac{{\cal D}_{2,1}^{AB}(|B_{AB}|-|\gamma_{A}|)}{{\cal C}_{AB}}\right)
e^{-d|\gamma_{A}|t} \int_{0}^{t}d\tau e^{-d(|B_{AB}|-|\gamma_{A}|)\tau}\prod_{\alpha=1\dots d}I_{r_{\alpha}}(2{\cal C}_{AB}\tau) \nonumber\\
&+& \left({\cal D}_{1,2}^{AB}+d {\cal D}_{2,2}^{AB}\left(\frac{|B_{AB}|-|\gamma_{B}|}{{\cal C}_{AB}}\right)\right)e^{-d|\gamma_{B}|t} \int_{0}^{t}d\tau e^{-d(|B_{AB}|-|\gamma_{B}|)\tau}\prod_{\alpha=1\dots d}I_{r_{\alpha}}(2{\cal C}_{AB}\tau)
 \nonumber\\
&-& \int_{0}^{t}dt' {\cal G}_{1}(t')e^{-d|B_{AB}|(t-t')}\left(2{\cal C}_{AB}d \prod_{\alpha=1\dots d} I_{r_{\alpha}}(2{\cal C}_{AB}(t-t')) + \frac{H_{1}^{ab}+H_{2}^{ab}-A_{1}^{a}-D_{2}^{b}}{{\cal C}_{AB}} \frac{\partial}{\partial t'}\prod_{\alpha=1\dots d} I_{r_{\alpha}}(2{\cal C}_{AB}(t-t'))\right)
\end{eqnarray}
Notice that these expression are valid, on $V_{2}(d)$, in arbitrary dimension 
and setting $d=1$ we recover the one-dimensional expressions of the previous 
section.

We assume here that the initial state is characterized by
a random, translationally invariant, but uncorrelated initial 
distribution, i.e., ${\cal G}_{|r|>0}^{l}(0)
=\rho_{i}(0)\rho_{j}(0) $.

The asymptotic behavior is obtained similarly to
the one dimensional case when $|{\cal C}_{l}|t\gg 1$ with $u_{l}=L^2/4|
{\cal C}_{l}|t$ \\
1. For $|B_{l}|>2|{\cal C}_{l}|$, the decay of the correlation function
is exponential, 
\begin{eqnarray}
\label{eq.3.34}
\varphi_{jj}\equiv min(|\gamma_{j}|, |B_{j}|-2|{\cal C}_{j}|)
\end{eqnarray}
\hskip 0.5cm 1.a  If $\varphi_{jj}=|\gamma_{A}|\neq|B_{j}|-2|{\cal C}_{j}|$, 
we have
\begin{eqnarray}
\label{eq.3.34.1}
{\cal G}_{r}^{jj}(t)-{\cal G}_{r}^{jj}(\infty)\sim e^{-d|\gamma_{j}|t}
\end{eqnarray}
\vskip 0.5cm
1.b If $\varphi_{jj}=|B_{j}|-2|{\cal C}_{j}|>|\gamma_{j}|$ and ${\cal G}_{r\neq 0}^{jj}(0)\neq 0$, we have
\begin{eqnarray}
\label{eq.3.34.2}
{\cal G}_{r}^{jj}(t)-{\cal G}_{r}^{jj}(\infty)\sim 
e^{-d(|B_{j}|-2|{\cal C}_{j}|)t}
\end{eqnarray}
Finally, note that provided ${\cal G}_{r\neq 0}(0)=0$, we find
\begin{eqnarray}
\label{eq.3.34.4}
{\cal G}_{r}^{jj}(t)-{\cal G}_{r}^{jj}(\infty)\sim 
\left\{
  \begin{array}{l l}
   \frac{ Q_{2}(u_{j}, \sigma_{\alpha}) e^{-(|B_{j}|-2|{\cal C}_{j}|)t}}{(4\pi|{\cal C}_{j}|t)^{d/2}}  &\mbox{if $r\ll L$}\\
 \frac{ Q_{2}(u_{j}, \sigma_{\alpha})exp(-\sum_{\alpha} \sigma_{\alpha}^2 u_{j}-(|B_{j}|-2|{\cal C}_{j}|)t)}
{(4\pi|{\cal C}_{j}|t)^{d/2}} &\mbox{if $r\gg 1$}\\
\end{array}
\right.
\end{eqnarray}
where $ Q_{2}(u_{j}, \sigma_{\alpha})$ is a known function determined by
the processes occuring in the model under consideration.
\vskip 0.5cm
1.c If $|B_{j}|\neq 2|{\cal C}_{j}|$, $|\gamma_{j'\neq\j}|=
|B_{j}|- 2|{\cal C}_{j}|$, $|\gamma_{j}|>0$, and ${\cal D}_{1,2}^{j}+
2(sign {\cal C}_{j}){\cal D}_{2,2}^{j}\neq 0 $, we have
\begin{eqnarray}
\label{eq.3.34.5}
{\cal G}_{r}^{jj}(t)-{\cal G}_{r}^{jj}(\infty)\sim 
\left\{
  \begin{array}{l l l l}
 e^{-d|\gamma_{j}|t} &\mbox{, if $\varphi_{jj}=|\gamma_{j}|$}\\
 e^{-|\gamma_{j'\neq j}|t}\sqrt{\frac{t}{\pi|{\cal C}_{j}|}}
 &\mbox{, if $\varphi_{jj}=|\gamma_{j'\neq j}|$}\\
e^{-2|\gamma_{j'\neq j}|t} \ln{t}  &\mbox{, if $\varphi_{jj}=|\gamma_{j'\neq j}|$ 
and $d=2$}\\
e^{-d|\gamma_{j'\neq j}|t} (4\pi|{\cal C}_{j}|t)^{1-d/2} 
&\mbox{, if $\varphi_{jj}=|\gamma_{j'\neq j}|$ and $d\geq 3$}
\end{array}
\right. 
\end{eqnarray}
\vskip 0.5cm
1.d  If $|B_{j}|\neq 2|{\cal C}_{j}|$, $|\gamma_{j}|=
|B_{j}|- 2|{\cal C}_{j}|$, $|\gamma_{j'\neq j}|>0$, and ${\cal D}_{1,1}^{j}+
2(sign {\cal C}_{j}){\cal D}_{2,1}^{j}\neq 0 $, we have
\begin{eqnarray}
\label{eq.3.34.6}
{\cal G}_{r}^{jj}(t)-{\cal G}_{r}^{jj}(\infty)\sim 
\left\{
  \begin{array}{l l l}
 e^{-|\gamma_{j}|t}\sqrt{\frac{t}{\pi|{\cal C}_{j}|}}
 &\mbox{, if $d=1$}\\
e^{-2|\gamma_{j}|t} \ln{t}  &\mbox{, if $d=2$}\\
e^{-d|\gamma_{j}|t} (4\pi|{\cal C}_{j}|t)^{1-d/2} 
&\mbox{, if $d\geq 3$}
\end{array} 
\right.
\end{eqnarray}
\vskip 0.5cm
The functions ${\cal G}_{r}^{AB}(t)$ requires a separate discussion.
With the definition of $\varphi_{AB}$
\begin{eqnarray*}
\label{eq.3.34.7}
\varphi_{AB}\equiv min(|\gamma_{A}|,|\gamma_{B}|, |B_{AB}|-2|{\cal C}_{AB}|)
\end{eqnarray*}
\hskip0.5cm 1.e  If $\varphi_{AB}=|\gamma_{A}|\neq|B_{j}|-2|{\cal C}_{j}|$, 
we have
\begin{eqnarray}
\label{eq.3.34.8}
{\cal G}_{r}^{AB}(t)-{\cal G}_{r}^{AB}(\infty)\sim e^{-d|\gamma_{A}|t}
\end{eqnarray}
1.f  If $\varphi_{AB}=|\gamma_{B}|\neq|B_{AB}|-2|{\cal C}_{AB}|$, we have
\begin{eqnarray}
\label{eq.3.34.9}
{\cal G}_{r}^{AB}(t)-{\cal G}_{r}^{AB}(\infty)\sim e^{-d|\gamma_{B}|t}
\end{eqnarray}
1.g  If $\varphi_{AB}=|B_{AB}|-2|{\cal C}_{AB}|\neq |\gamma_{A,B}|>0$ and ${\cal G}_{r\neq 0}(0)\neq 0$ 
, we have
\begin{eqnarray}
\label{eq.3.34.10}
{\cal G}_{r}^{AB}(t)-{\cal G}_{r}^{AB}(\infty)\sim e^{-d(|B_{AB}|-2|{\cal C}_{AB}|)t}
\end{eqnarray}
Again, when ${\cal G}_{r\neq 0}^{AB}(0)=0, \forall r$, we arrive at
\begin{eqnarray}
\label{eq.3.34.12}
{\cal G}_{r}^{AB}(t)-{\cal G}_{r}^{AB}(\infty)\sim 
\left\{
  \begin{array}{l l l}
   \frac{Q_{2}(u_{AB}, \sigma_{\alpha})e^{-d(|B_{AB}|-2|{\cal C}_{AB}|)t}}{(4\pi|{\cal C}_{AB}|t)^{d/2}}  &\mbox{if $r\ll L$}\\
 \frac{Q_{2}(u_{AB}, \sigma_{\alpha})e^{-du_{AB}-d(|B_{AB}|-2|{\cal C}_{AB}|)t}}{(4\pi|{\cal C}_{AB}|t)^{d/2}} &\mbox{if $r\gg 1$}
\end{array}
\right.
\end{eqnarray}
Where $Q_{2}(u_{AB}, \sigma_{\alpha})$ has been defined previously.

1.h If $|B_{AB}|\neq 2|{\cal C}_{AB}|$, $|\gamma_{A}|=
|B_{AB}|- 2|{\cal C}_{AB}|$, and ${\cal D}_{1,1}^{AB}+
2(sign {\cal C}_{AB}){\cal D}_{2,1}^{AB}\neq 0 $, we have
\begin{eqnarray}
\label{eq.3.34.13}
{\cal G}_{r}^{AB}(t)-{\cal G}_{r}^{AB}(\infty)\sim 
\left\{
  \begin{array}{l l l l}
 e^{-d|\gamma_{B}|t} &\mbox{if $\varphi_{AB}=|\gamma_{B}|$}\\
 e^{-|\gamma_{A}|t}\sqrt{\frac{t}{\pi|{\cal C}_{AB}|}}
 &\mbox{if $\varphi_{AB}=|\gamma_{A}|$ and $d=1$}\\
e^{-2|\gamma_{A}|t}\ln{t}
 &\mbox{if $\varphi_{AB}=|\gamma_{A}|$ and  $d=2$}\\
e^{-d|\gamma_{A}|t}(4\pi|{\cal C}_{AB}|t)^{1-d/2}
 &\mbox{if $\varphi_{AB}=|\gamma_{A}|$ and  $d\geq 3$}
\end{array} 
\right.
\end{eqnarray}
\vskip 0.5cm
1.i  If $|B_{AB}|\neq 2|{\cal C}_{AB}|$, $|\gamma_{B}|=
|B_{AB}|- 2|{\cal C}_{AB}|$, and ${\cal D}_{1,2}^{AB}+
2(sign {\cal C}_{AB}){\cal D}_{2,2}^{AB}\neq 0 $, we have
\begin{eqnarray}
\label{eq.3.34.14}
{\cal G}_{r}^{AB}(t)-{\cal G}_{r}^{AB}(\infty)\sim 
\left\{
  \begin{array}{l l l}
 e^{-d|\gamma_{A}|t} &\mbox{if $\varphi_{AB}=|\gamma_{A}|$}\\
 e^{-|\gamma_{B}|t}\sqrt{\frac{t}{\pi|{\cal C}_{AB}|}}
&\mbox{if $\varphi_{AB}=|\gamma_{B}|$ and  $d=1$}\\
e^{-|\gamma_{B}|t}\ln{t}
&\mbox{if $\varphi_{AB}=|\gamma_{B}|$ and $d=2$}\\
e^{-|\gamma_{B}|t}(4\pi|{\cal C}_{AB}t|^{1-d/2})
&\mbox{if $\varphi_{AB}=|\gamma_{B}|$ and $d\geq 3$}\\
\end{array} 
\right.
\end{eqnarray}
\vskip 0.5cm

2. For $|B_{l}|=2|{\cal C}_{l}|$,  the decay of correlation functions is
algebraic.
We distinguish the regime $r\ll L$, from that where $r\sim L$
($\sigma=r/L$, $l\in (AA,BB, AB)$.

In the limit $r\ll L$,
\begin{eqnarray}
\label{eq.3.35.1}
{\cal G}_{r}^{l}(t)-{\cal G}_{r}^{l}(\infty)\sim 
\left\{
  \begin{array}{l l}
   \frac{e^{\sum_{\alpha}\sigma_{\alpha}^2 u} \rho_{i}(0)\rho_{j}(0)}{(4\pi|{\cal C}_{l}|t)^{d/2}} &\mbox{if ${\cal C}_{l} < 0$}\\
 \frac{Q_{2}(u_{l},\sigma_{\alpha})}{(4\pi|{\cal C}_{l}|t)^{d/2}}
+\frac{\rho_{i}(0)\rho_{j}(0)\prod_{\alpha=1\dots d}{\cal F}_{1,\alpha}(u_{l},\sigma_{\alpha},\nu=0)}{8|{\cal C}_{l}|t}
 &\mbox{if ${\cal C}_{l} > 0$}  ,
\end{array}
\right.
\end{eqnarray}
with the definition of the following auxiliary functions:
\begin{eqnarray}
\label{eq.3.35.2}
{\cal F}_{1,\alpha}(u,\sigma_\alpha, \nu)\equiv \frac{ \left(\Gamma
(\frac{1-\nu}{2}) +\Gamma(\frac{1-\nu}{2},\sigma_{\alpha}^2 u_{l})
-\Gamma(\frac{1-\nu}{2},u_{l}(1-\sigma_{\alpha})^2)-
 \Gamma(\frac{1-\nu}{2},u_{l}(1+\sigma_{\alpha})^2)\right) }{\sqrt{4\pi}} 
\end{eqnarray}
In the regime where $r\gg 1, r_{\alpha}\equiv \sigma_{\alpha} 
L \sim L$, we find
\begin{eqnarray}
\label{eq.3.35.2}
{\cal G}_{r}^{l}(t)-{\cal G}_{r}^{l}(\infty)\sim 
\left\{
  \begin{array}{l l}
   \frac{e^{\sum_{\alpha}\sigma_{\alpha}^2 u} \rho_{i}(0)\rho_{j}(0)}{(4\pi|{\cal C}_{l}|t)^{d/2}} &\mbox{if ${\cal C}_{l} < 0$}\\
 \frac{Q_{2}(u_{l},\sigma_{\alpha})}{(4\pi|{\cal C}_{l}|t)^{d/2}}
+\frac{\rho_{i}(0)\rho_{j}(0)\prod_{\alpha=1\dots d}{\cal F}_{2,\alpha}(u_{l},\sigma_{\alpha},\nu=0)}{8|{\cal C}_{l}|t}
 &\mbox{if ${\cal C}_{l} > 0$}  ,
\end{array}
\right.
\end{eqnarray}
with the auxiliary functions defined as:
\begin{eqnarray}
\label{eq.3.35.3}
{\cal F}_{2,\alpha}(u_{l},\sigma_\alpha, \nu)\equiv 
\frac{e^{-\sigma_{\alpha}^2 u_{l} }}{1-\nu}
\sqrt{\frac{u_{l}\sigma_{\alpha}^2}{\pi}} 
\end{eqnarray}
Notice that setting ${\cal G}_{r}(0)=0, \forall r$, leads to
\begin{eqnarray}
\label{eq.3.35.4}
{\cal G}_{r}^{l}(t)-{\cal G}_{r}^{l}(\infty)\sim 
\left\{
  \begin{array}{l l}
   \frac{Q_2(u_{l},\sigma_{\alpha})}{(4\pi|{\cal C}_{j}t|)^{d/2}}  &\mbox{if $r\ll L$}\\
   \frac{Q_2(u_{l},\sigma_{\alpha} )e^{-\sum_{\alpha} \sigma_{\alpha}^2 u_{l}}}{(4\pi|{\cal C}_{j}t|)^{d/2}}  &\mbox{if $r\gg 1$}
\end{array}
\right.
\end{eqnarray}
where the function $Q_2(u,\sigma)$ is as above.

We see that for the uncorrelated initial state under consideration, 
the dimensionality of the problem has a non-trivial effect on the dynamics.  
In fact, in the critical regime (\ref{eq.3.35.2}, \ref{eq.3.35.4}), 
when $d>2$ the correlation functions decay as $t^{-1}$ instead of $t^{-1/2}$, 
as in lower dimensions. We further remark that also in the massive case, 
the dimensionality of the model can have particular non-trivial effects 
on the  asymptotic regime (see e.g. \ref{eq.3.34.5}, \ref{eq.3.34.6}, 
\ref{eq.3.34.13}, \ref{eq.3.34.14} ).
Let us conclude by noting that all results obtained in this subsection are 
compatible and in agreement with the previous one-dimensional results.

\section{Conclusions}

In this technical paper, we have classified the  solutions of the two-species
bimolecular diffusion-limited reaction models and have been able to obtain exact and explicit results, namely :
\\

-  The Fourier-Laplace transform of the density and of the non-instanateous two-point correlation functions (dynamic form factors)  on a $56-$parameters space, in arbitrary dimensions.

- Exact computation, in arbitrary dimensions, of the density on a  $50-$parameters manifold
for various initial conditions.

- Exact computation of the non-instantaneous two-point correlation functions on a  $50-$parameters manifold  for uncorrelated homogeneous, but random, initial states as well as for initially correlated states, in arbitrary dimensions.

- Exact results for the instantaneous two-point correlation functions on a translationally-invariant  $31-$parameters space manifold in arbitrary dimensions.
\\
 
Exploring the various classes of solutions for the one- and 
two-point correlation functions, we have seen in real space and time that there are essentially 
two regimes, a massive one and an algebraic one for the density and two-point 
correlation functions. For non-instantaneous correlation function we have
 pointed
out that a drift can occur due to an asymmetry of  the reaction-rates characterizing the  stochastic Hamiltonian.
We have also shown that when initial correlations are strong
enough, the critical exponents in the asymptotic regime are renormalized while
for weak initial correlations, the long-time behavior is insensitive to and
independent of the initial state.

Our approach applied on three-states models (of hard-core particles) is in a sense complementary to previous rigorous results \cite{Belitsky} (see the end of section III) and  allows to study exactly  physical models such as a {\it three-states} growth model \cite{Mobar2}. From our analysis, it follows (alternatively, using  the Rauth-Hurwitz conditions and simple algebra) that in arbitrary dimensions, two-species models belonging to the class of soluble models discussed here, do not exhibit phase-transition, nor pattern formation. This has lead us to  {\it conjecture} that such a property holds true, in  {\it soluble} models (in the sense discussed here) for  an  arbitrary number of species $s$  and in arbitrary dimensions. 
\section{Acknowledgments}
The support of Swiss National Fonds is gratefully acknowledged.
\section{Appendix A : Definitions and abreviations}
In this appendix we introduce the definitions  which are adopted throughout the paper.
In (\ref{eq.1.9}) and in the following, we have used these notations
\begin{eqnarray}
\label{eq.A.1}
A_{0}^{a}&\equiv& \Gamma_{0 0}^{1 0}+\Gamma_{0 0}^{1 1}+\Gamma_{0 0}^{1 2} \;;\;
A_{1}^{a} \equiv \Gamma_{1 0}^{1 0}+\Gamma_{1 0}^{1 1}+\Gamma_{1 0}^{1 2}-A_{0}^{a} \;;\;
A_{2}^{a} \equiv \Gamma_{0 1}^{1 0}+\Gamma^{1 1}_{0 1}+\Gamma^{1 2}_{0 1}-A_{0}^{a} \nonumber\\
B_{1}^{a} &\equiv& \Gamma_{2 0}^{1 0}+\Gamma_{2 0}^{1 1}+\Gamma_{2 0}^{1 2}-A_{0}^{a}  \;;\;
B_{2}^{a} \equiv \Gamma_{0 2}^{1 0}+\Gamma_{0 2}^{1 1}+\Gamma_{0 2}^{1 2}-A_{0}^{a} \;;\;
C_{0}^{a} \equiv \Gamma_{0 0}^{0 1}+\Gamma_{0 0}^{1 1}+\Gamma_{0 0}^{2 1} \nonumber\\
C_{1}^{a} &\equiv& \Gamma_{1 0}^{0 1}+\Gamma_{1 0}^{1 1}+\Gamma_{1 0}^{2 1}-C_{0}^{a} \;;\;
C_{2}^{a} \equiv \Gamma_{0 1}^{0 1}+\Gamma_{0 1}^{1 1}+\Gamma_{0 1}^{2 1}-C_{0}^{a}  \;;\;
D_{1}^{a} \equiv \Gamma_{2 0}^{0 1}+\Gamma_{2 0}^{1 1}+\Gamma_{2 0}^{2 1}-C_{0}^{a} \nonumber\\
D_{2}^{a} &\equiv& \Gamma_{0 2}^{0 1}+\Gamma_{0 2}^{1 1}+\Gamma_{0 2}^{2 1}-C_{0}^{a} \;;\;
A_{0}^{b}\equiv \Gamma_{0 0}^{2 0}+\Gamma_{0 0}^{2 1}+\Gamma_{0 0}^{2 2} \;;\;
A_{1}^{b} \equiv \Gamma_{1 0}^{2 0}+\Gamma_{1 0}^{2 1}+\Gamma_{1 0}^{2 2}-A_{0}^{b} \nonumber\\
A_{2}^{b} &\equiv& \Gamma_{0 1}^{2 0}+\Gamma_{0 1}^{2 1}+\Gamma_{0 1}^{2 2}-A_{0}^{b} \;;\;
B_{1}^{b} \equiv \Gamma_{2 0}^{2 0}+\Gamma_{2 0}^{2 1}+\Gamma_{2 0}^{2 2}-A_{0}^{b} \;;\;
B_{2}^{b} \equiv \Gamma_{0 2}^{2 0}+\Gamma_{0 2}^{2 1}+\Gamma_{0 2}^{2 2}-A_{0}^{b} \nonumber\\
C_{0}^{b} &\equiv& \Gamma_{0 0}^{0 2}+\Gamma_{0 0}^{1 2}+\Gamma_{0 0}^{2 2} \;;\;
C_{1}^{b} \equiv \Gamma_{1 0}^{0 2}+\Gamma_{1 0}^{1 2}+\Gamma_{1 0}^{2 2}-C_{0}^{b} \;;\;
C_{2}^{b} \equiv \Gamma_{0 1}^{0 2}+\Gamma_{0 1}^{1 2}+\Gamma_{0 1}^{2 2}-C_{0}^{b} \nonumber\\
D_{1}^{b} &\equiv& \Gamma_{2 0}^{0 2}+\Gamma_{2 0}^{1 2}+\Gamma_{2 0}^{2 2}-C_{0}^{b} \;;\;
D_{2}^{b} \equiv \Gamma_{0 2}^{0 2}+\Gamma_{0 2}^{1 2}+\Gamma_{0 2}^{2 2}-C_{0}^{b}
\end{eqnarray}
Further we also use the following notations:
\begin{eqnarray}
\label{eq.A.2}
E_{0}^{a}&\equiv& \Gamma_{0 0}^{1 1} \;;\;
E_{0}^{b}\equiv \Gamma_{0 0}^{2 2} \;;\;
E_{0}^{ab}\equiv \Gamma_{0 0}^{1 2} \nonumber\\
E_{0}^{ba}&\equiv& \Gamma_{0 0}^{2 1} \;;\;
F_{1}^{a}\equiv \Gamma_{1 0}^{1 1}- \Gamma_{0 0}^{1 1}\;;\;
F_{1}^{b}\equiv \Gamma_{1 0}^{2 2}- \Gamma_{0 0}^{2 2}  \nonumber\\
F_{1}^{ab}&\equiv& \Gamma_{1 0}^{1 2}- \Gamma_{0 0}^{1 2} \;;\;
F_{1}^{ba}\equiv \Gamma_{1 0}^{2 1}- \Gamma_{0 0}^{2 1} \;;\;
F_{2}^{a}\equiv \Gamma_{0 1}^{1 1}- \Gamma_{0 0}^{1 1}\nonumber\\
F_{2}^{b}&\equiv& \Gamma_{0 1}^{2 2}- \Gamma_{0 0}^{2 2}  \;;\;
F_{2}^{ab}\equiv \Gamma_{0 1}^{1 2}- \Gamma_{0 0}^{1 2} \;;\;
F_{2}^{ba}\equiv \Gamma_{0 1}^{2 1}- \Gamma_{0 0}^{2 1} \nonumber\\
F_{3}^{a} &\equiv& \Gamma_{2 0}^{1 1}-\Gamma_{0 0}^{1 1}  \;;\;
F_{3}^{b}\equiv \Gamma_{2 0}^{2 2}- \Gamma_{0 0}^{2 2}  \;;\;
F_{3}^{ab} \equiv \Gamma_{2 0}^{1 2}- \Gamma_{0 0}^{1 2} \nonumber\\
F_{3}^{ba}&\equiv& \Gamma_{2 0}^{2 1}- \Gamma_{0 0}^{2 1} \;;\;
F_{4}^{a}\equiv \Gamma_{0 2}^{1 1}- \Gamma_{0 0}^{1 1}\;;\;
F_{4}^{b}\equiv \Gamma_{0 2}^{2 2}- \Gamma_{0 0}^{2 2}  \nonumber\\
F_{4}^{ab}&\equiv& \Gamma_{0 2}^{1 2}- \Gamma_{0 0}^{1 2} \;;\;
F_{4}^{ba}\equiv \Gamma_{0 2}^{2 1}- \Gamma_{0 0}^{2 1} \;;\;
G_{1}^{a}\equiv \Gamma_{0 0}^{1 1}+ \Gamma_{1 1}^{1 1}- \Gamma_{0 1}^{1 1} - \Gamma_{1 0}^{1 1} \nonumber\\
G_{1}^{b}&\equiv& \Gamma_{0 0}^{2 2}+ \Gamma_{1 1}^{2 2}- \Gamma_{0 1}^{2 2} - \Gamma_{1 0}^{2 2}  \;;\;
G_{1}^{ab}\equiv \Gamma_{0 0}^{1 2} + \Gamma_{1 1}^{1 2}- \Gamma_{0 1}^{1 2} - \Gamma_{1 0}^{1 2} \;;\;
G_{1}^{ba}\equiv  \Gamma_{0 0}^{2 1} + \Gamma_{1 1}^{2 1}- \Gamma_{0 1}^{2 1} - \Gamma_{1 0}^{2 1}  \nonumber\\
G_{2}^{a}&\equiv& \Gamma_{0 0}^{1 1}+ \Gamma_{2 2}^{1 1}- \Gamma_{0 2}^{1 1} - \Gamma_{2 0}^{1 1} \;;\;
G_{2}^{b}\equiv \Gamma_{0 0}^{2 2}- \Gamma_{0 2}^{2 2}- \Gamma_{2 0}^{2 2} - \Gamma_{2 2}^{2 2} \;;\;
G_{2}^{ab}\equiv \Gamma_{0 0}^{1 2} + \Gamma_{2 2}^{1 2}- \Gamma_{0 2}^{1 2} - \Gamma_{2 0}^{1 2} \nonumber\\
G_{2}^{ba}&\equiv&  \Gamma_{0 0}^{2 1} + \Gamma_{2 2}^{2 1}- \Gamma_{0 2}^{2 1} - \Gamma_{2 0}^{2 1} \;;\;
H_{1}^{a} \equiv \Gamma_{2 1}^{1 1}+ \Gamma_{0 0}^{1 1}-  \Gamma_{0 1}^{1 1} - \Gamma_{2 0}^{1 1}   \;;\;
H_{2}^{a} \equiv  \Gamma_{0 0}^{1 1}+ \Gamma_{1 2}^{1 1}-  \Gamma_{1 0}^{1 1} - \Gamma_{0 2}^{1 1}   \nonumber\\
H_{1}^{b} &\equiv & \Gamma_{2 1}^{2 2}+ \Gamma_{0 0}^{2 2}-  \Gamma_{0 1}^{2 2} - \Gamma_{2 0}^{2 2}   \;;\;
H_{2}^{b} \equiv  \Gamma_{0 0}^{2 2}+ \Gamma_{1 2}^{2 2}-  \Gamma_{1 0}^{2 2} - \Gamma_{0 2}^{2 2}   \;;\;
H_{2}^{ab} \equiv  \Gamma_{0 0}^{1 2}+ \Gamma_{1 2}^{1 2}-  \Gamma_{0 2}^{1 2} - \Gamma_{1 0}^{1 2}   \nonumber\\
H_{2}^{ba} &\equiv & \Gamma_{0 0}^{2 1}+ \Gamma_{1 2}^{2 1}-  \Gamma_{0 2}^{2 1} - \Gamma_{1 0}^{2 1} \;;\;
H_{1}^{ab} \equiv\Gamma_{0 0}^{1 2}+ \Gamma_{2 1}^{1 2}-  \Gamma_{2 0}^{1 2} - \Gamma_{0 1}^{1 2}  \;;\;
H_{1}^{ba} \equiv  \Gamma_{0 0}^{2 1}+ \Gamma_{2 1}^{2 1}-  \Gamma_{2 0}^{2 1} - \Gamma_{0 1}^{2 1}
\end{eqnarray}
Notice that in the expressions  above, (\ref{eq.A.1}) and (\ref{eq.A.2}), 
the rates
$\Gamma_{\alpha \beta}^{\alpha \beta}$ have not been made explicit for brevity.
\section{Appendix B: definitions for section V}
%
In section $V$, in addition to (\ref{eq.A.1}) and (\ref{eq.A.2}), we will also use the following additional definitions :
\begin{eqnarray}
\label{eq.B.1}
{\cal A}_{A}&\equiv& 2(A_{0}^{a}+C_{0}^{a}), \nonumber\\
B_{A}&\equiv& 2(A_{1}^{a}+C_{2}^{a}), \nonumber\\
{\cal C}_{A}&\equiv& A_{2}^{a}+C_{1}^{a}, \nonumber\\
{\cal D}_{1,0}^{A} &\equiv&  \frac{{\cal A}_{A}d}{2}-\left(\frac{B_{A}}{2}-{\cal C}_{A}+ {\cal A}_{A} \right)d\rho_{A}(\infty), \nonumber\\
{\cal D}_{1,1}^{A} &\equiv& \left(\frac{B_{A}}{2}-{\cal C}_{A}+ {\cal A}_{A} \right)d(\rho_{A}(\infty) - \rho_{A}(0)), \nonumber\\
{\cal D}_{2,0}^{A} &\equiv& E_{0}^{a}+(F_{1}^{a}+ F_{2}^{a} -\frac{{\cal A}_{A}}{2} -{\cal C}_{A})\rho_{A}(\infty)+(F_{3}^{a}+ F_{4}^{a})\rho_{B}(\infty), \nonumber\\
{\cal D}_{2,1}^{A} &\equiv& (F_{1}^{a}+ F_{2}^{a} -\frac{{\cal A}_{A}}{2} -{\cal C}_{A}  )(\rho_{A}(0)-\rho_{A}(\infty)), \nonumber\\
{\cal D}_{2,2}^{A} &\equiv& (F_{3}^{a}+ F_{4}^{a})(\rho_{B}(0)-\rho_{B}(\infty)),
\end{eqnarray}
\begin{eqnarray}
\label{eq.B.2}
{\cal A}_{B} &\equiv&  2(A_{0}^{b}+C_{0}^{b}), \nonumber\\
B_{B} &\equiv& 2(B_{1}^{b}+D_{2}^{b}), \nonumber\\
{\cal C}_{B} &\equiv& B_{2}^{b}+D_{1}^{b}, \nonumber\\
{\cal D}_{1,0}^{B}&\equiv&  \frac{{\cal A}_{B}d}{2}-\left(\frac{B_{B}}{2}-{\cal C}_{B}+ {\cal A}_{B} \right)d\rho_{B}(\infty), \nonumber\\
{\cal D}_{1,1}^{B} &\equiv& \left(\frac{B_{B}}{2}-{\cal C}_{B}+ {\cal A}_{B} \right)d(\rho_{A}(\infty) - \rho_{A}(0)), \nonumber\\
{\cal D}_{2,0}^{B} &\equiv&  E_{0}^{b}+(F_{1}^{b}+ F_{2}^{b} )\rho_{A}(\infty)+(F_{3}^{a}+ F_{4}^{a}  -\frac{{\cal A}_{B}}{2} -{\cal C}_{B} )\rho_{B}(\infty), \nonumber\\
{\cal D}_{2,1}^{B} &\equiv&  (F_{3}^{b}+ F_{4}^{b} -\frac{{\cal A}_{B}}{2} -{\cal C}_{B} )(\rho_{B}(0)-\rho_{B}(\infty)), \nonumber\\
{\cal D}_{2,2}^{B} &\equiv&  (F_{1}^{b}+ F_{2}^{b})(\rho_{A}(0)-\rho_{A}(\infty)),
\end{eqnarray}
and
\begin{eqnarray}
\label{eq.B.3}
{\cal A}_{AB,1} &\equiv&  {\cal A}_{B}/2, \; {\cal A}_{AB,2}\equiv {\cal A}_{A}/2  \nonumber\\
B_{AB}&\equiv& \frac{(B_{A}+B_{B})}{2}, \nonumber\\
{\cal C}_{AB} &\equiv&  A_{2}^{a}+D_{1}^{b}=B_{2}^{b}+D_{1}^{b}, \nonumber\\
{\cal D}_{1,0}^{AB} &\equiv&  -\left({\cal A}^{AB}_{1} \rho_{A}(\infty)+ {\cal A}^{AB}_{2} \rho_{B}(\infty)\right)d \nonumber\\
{\cal D}_{1,1}^{AB} &\equiv&  -{\cal A}^{AB}_{1}\left(\rho_{A}(0)-\rho_{A}(\infty)\right)d \nonumber\\
{\cal D}_{1,2}^{AB} &\equiv&  -{\cal A}^{AB}_{2}\left(\rho_{B}(0)-\rho_{B}(\infty)\right)d \nonumber\\
{\cal D}_{2,0}^{AB} &\equiv& E_{0}^{ab}+(F_{1}^{ab}+ F_{2}^{ab}-C_{0}^{b})\rho_{A}(\infty)+(F_{3}^{ab} + F_{4}^{ab}-A_{0}^{a})\rho_{B}(\infty),\nonumber\\ 
{\cal D}_{2,1}^{AB} &\equiv& (F_{1}^{ab}+F_{2}^{ab}-C_{0}^{b})(\rho_{A}(0)- \rho_{A}(\infty))\nonumber\\
{\cal D}_{2,2}^{AB} &\equiv& (F_{3}^{ab}+F_{4}^{ab}-C_{0}^{b})(\rho_{B}(0)- \rho_{B}(\infty)) ,
\end{eqnarray}
Where $\rho_{A}(\infty)=\frac{A_{0}^{a}+C_{0}^{a}}{2|\gamma_{A}|}$ and 
$\rho_{B}(\infty)=\frac{A_{0}^{b}+C_{0}^{b}}{2|\gamma_{B}|} $, as in section III.
\section{ Appendix C: One dimensional two-point correlation function on $V_{2}$, the equations of motion and their solutions}
In one dimension the equations of motion (\ref{eq.3.2}-\ref{eq.3.4}) of the correlation functions can be
written as an unique difference equation:
\begin{eqnarray}
\label{eq.C.1}
\frac{d}{dt}{\cal G}_{r}^{AA}(t)&=& B_{A} {\cal G}_{r}^{AA}(t)+{\cal C}_{A}\left( {\cal G}_{r+1}^{AA}(t) + {\cal G}_{r-1}^{AA}(t) \right)+ {\cal A}_{A}\rho_{A}(t)\nonumber\\ &+& \left({\cal D}_{1,0}^{A} +{\cal D}_{1,1}^{A}e^{\gamma_{A}t} \right)\delta_{r,0} + \left({\cal D}_{2,0}^{A}+ {\cal D}_{2,1}^{A}e^{\gamma_{A}t} +  {\cal D}_{2,2}^{A} e^{\gamma_{B}t} \right)(\delta_{r,1}+\delta_{r,-1}) \nonumber\\
&+& \left[(G_{1}^{a}-B_{A}/2)(\delta_{r,1}+ \delta_{r,-1})-2{\cal C}_{A}\delta_{r,0}\right]{\cal G}_{1}^{AA}(t)
\end{eqnarray}
The solution is 
\begin{eqnarray}
\label{eq.C.2}
&&{\cal G}_{r}^{AA}(t)-(\rho_{A}(t))^2= -(\rho_{A}(0) e^{-|\gamma_{A}|t})^2 +\rho_{A}(0) e^{-|B_{A}|t}I_{r}(2{\cal C}_{A}t)+\sum_{r'\neq 0}{\cal G}_{r'}^{AA}(0)
 e^{-|B_{A}|t}I_{r-r'}(2{\cal C}_{A}t) \nonumber\\
 &+& \int_{0}^{t} dt'  e^{-|B_{A}|(t-t')}\left({\cal D}_{1,0}^{A} I_{r}(2{\cal C}_{A}(t-t')) +{\cal D}_{2,0}^{A}\left[I_{r+1}(2{\cal C}_{A}(t-t'))+ I_{r-1}(2{\cal C}_{A}(t-t'))\right] \right) \nonumber\\&+& {\cal D}_{1,1}^{A} \int_{0}^{t} dt'  e^{-| B_{A}|(t-t')} e^{-|\gamma_{A}|t'} I_{r}(2{\cal C}_{A}(t-t')) \nonumber\\
&+& \int_{0}^{t} dt'  e^{-| B_{A}|(t-t')}  ({\cal D}_{2,1}^{A} e^{-|\gamma_{A}|t'} + {\cal D}_{2,2}^{A} e^{-|\gamma_{B}|t'})\left[I_{r+1}(2{\cal C}_{A}(t-t'))+I_{r-1}(2{\cal C}_{A}(t-t'))\right]\nonumber\\
&+& \int_{0}^{t} dt'  e^{-|B_{A}|(t-t')}  {\cal G}_{1}^{AA}(t')\left[(G_{1}^{a}-B_{A}/2)\left(I_{r+1}(2{\cal C}_{A}(t-t'))+I_{r-1}(2{\cal C}_{A}(t-t')) \right) -2{\cal C}_{A} I_{r}(2{\cal C}_{A}(t-t'))\right]
\end{eqnarray}
Similarly for the $B-B$ correlation functions we get 
\begin{eqnarray}
\label{eq.C.3}
\frac{d}{dt}{\cal G}_{r}^{BB}(t)&=& B_{B} {\cal G}_{r}^{BB}(t)+{\cal C}_{B}\left( {\cal G}_{r+1}^{BB}(t) + {\cal G}_{r-1}^{BB}(t) \right)+ {\cal A}_{B}\rho_{B}(t)\nonumber\\ &+& \left({\cal D}_{1,0}^{B} +{\cal D}_{1,1}^{B}e^{\gamma_{B}t} \right)\delta_{r,0} + \left({\cal D}_{2,0}^{B}+ {\cal D}_{2,1}^{B}e^{\gamma_{B}t} +  {\cal D}_{2,2}^{B}e^{\gamma_{A}t} \right)(\delta_{r,1}+\delta_{r,-1}) \nonumber\\
&+& \left[(G_{2}^{b}-B_{B}/2)(\delta_{r,1}+ \delta_{r,-1})-2{\cal C}_{B}\delta_{r,0}\right]{\cal G}_{1}^{BB}(t)
\end{eqnarray}
and for the $A-B$ correlation function, we have
\begin{eqnarray}
\label{eq.C.5}
\frac{d}{dt}{\cal G}_{r}^{AB}(t)&=& B_{AB} {\cal G}_{r}^{AB}(t)+{\cal C}_{AB}\left( {\cal G}_{r+1}^{AB}(t) + {\cal G}_{r-1}^{AB}(t) \right)+ {\cal A}_{AB,1}\rho_{A}(t)+ {\cal A}_{AB,2}\rho_{B}(t) \nonumber\\ &+& \left({\cal D}_{1,0}^{AB} + {\cal D}_{1,1}^{AB}e^{\gamma_{A}t}+ {\cal D}_{1,2}^{AB}e^{\gamma_{B}t} \right)\delta_{r,0} + \left({\cal D}_{2,0}^{AB}+ {\cal D}_{2,1}^{AB}e^{\gamma_{A}t} +  {\cal D}_{2,2}^{AB}e^{\gamma_{B}t} \right)(\delta_{r,1}+\delta_{r,-1}) \nonumber\\
&+& \left[(H_{1}^{ab}+H_{2}^{ab}-(A_{1}^{a}+D_{2}^{b}))(\delta_{r,1}+ \delta_{r,-1})-2{\cal C}_{AB}\delta_{r,0}\right]{\cal G}_{1}^{AB}(t)
\end{eqnarray}
Eqs. (\ref{eq.C.3}) and (\ref{eq.C.5}) are solved in asimilar way as (\ref{eq.C.2}).
The above expressions for ${\cal G}_{r}^{ij}(t),\; (i,j)
\in (A, B)$ can be rewritten in a more compact form (\ref{eq.3.10}-\ref{eq.3.12}) using the properties of the
modified Bessel functions $I_{n}(z)$.
\section{Appendix D: Correlation functions on $V_{2}$ in arbitrary dimension. the equations of motion and their solutions }
The equations of motion are the higher dimensional counterparts of the previous
equations (\ref{eq.C.1},\ref{eq.C.3},\ref{eq.C.5}), i.e., 
\begin{eqnarray}
\label{eq.D.7}
&&\frac{d}{dt}{\cal G}_{|r|=|(r_{1},\dots,r_{d})|}^{AA}(t)=  B_{A}d {\cal G}_{r}^{AA}(t)+{\cal C}_{A}\sum_{\alpha}\left({\cal G}_{r^{+}_{\alpha}}^{AA}(t) + {\cal G}_{r^{-}_{\alpha}}^{AA}(t) \right)+ d{\cal A}_{A}\rho_{A}(t)\nonumber\\ &+& \left({\cal D}_{1,0}^{A} +{\cal D}_{1,1}^{A}e^{d\gamma_{A}t} \right)\prod_{\alpha=1\dots d}\delta_{r_{\alpha},0} + \left({\cal D}_{2,0}^{A}+ {\cal D}_{2,1}^{A}e^{d\gamma_{A}t} +  {\cal D}_{2,2}^{A}e^{d\gamma_{B}t} \right)\sum_{\alpha}(\delta_{r_{\alpha},e^{\alpha}} +\delta_{r_{\alpha},-e^{\alpha}})\prod_{\alpha'\neq \alpha}\delta_{r_{\alpha '},0} \nonumber\\
&+& \left[(G_{1}^{a}-B_{A}/2)\sum_{\alpha}(\delta_{r_{\alpha},e_{\alpha}} +\delta_{r_{\alpha},-e_{\alpha}} )\prod_{\alpha'\neq \alpha}\delta_{r_{\alpha '},0}-2{\cal C}_{A}d\prod_{\alpha=1\dots d}\delta_{r_{\alpha},0}\right]{\cal G}_{|r|=1}^{AA}(t),
\end{eqnarray}
and
\begin{eqnarray}
\label{eq.D.8}
&&\frac{d}{dt}{\cal G}_{|r|=|(r_{1},\dots,r_{d})|}^{BB}(t)=  B_{B}d {\cal G}_{r}^{BB}(t)+{\cal C}_{B}\sum_{\alpha} \left({\cal G}_{r^{+}_{\alpha}}^{BB}(t) + {\cal G}_{r^{-}_{\alpha}}^{BB}(t) \right)+ d{\cal A}_{B}\rho_{B}(t)\nonumber\\ &+& \left({\cal D}_{1,0}^{B} +{\cal D}_{1,1}^{B}e^{d\gamma_{B}t} \right)\prod_{\alpha=1\dots d}\delta_{r_{\alpha},0} + \left({\cal D}_{2,0}^{B}+ {\cal D}_{2,1}^{B}e^{d\gamma_{B}t} +  {\cal D}_{2,2}^{B}e^{d\gamma_{A}t} \right)\sum_{\alpha}(\delta_{r_{\alpha},e^{\alpha}}+\delta_{r_{\alpha},-e^{\alpha}})\prod_{\alpha\neq\alpha'}\delta_{r_{\alpha'},0} \nonumber\\
&+& \left[(G_{2}^{b}-B_{B}/2)\sum_{\alpha}(\delta_{r_{\alpha},e_{\alpha}} +\delta_{r_{\alpha},-e_{\alpha}} )\prod_{\alpha\neq\alpha'}\delta_{r_{\alpha'},0}-2{\cal C}_{B}d\prod_{\alpha=1\dots d}\delta_{r_{\alpha},0}\right]{\cal G}_{|r|=1}^{AA}(t),
\end{eqnarray}
and also
\begin{eqnarray}
\label{eq.D.9}
&&\frac{d}{dt}{\cal G}_{|r|=|(r_{1},\dots,r_{d})|}^{AB}(t)=  B_{AB}d {\cal G}_{r}^{AB}(t)+{\cal C}_{AB}\sum_{\alpha} \left({\cal G}_{r^{+}_{\alpha}}^{AB}(t) + {\cal G}_{r^{-}_{\alpha}}^{AB}(t) \right)+ d{\cal A}_{AB,1}\rho_{A}(t) + d{\cal A}_{AB,2}\rho_{B}(t) 
\nonumber\\ &+& \left({\cal D}_{1,0}^{AB} +{\cal D}_{1,1}^{AB}e^{d\gamma_{A}t} + {\cal D}_{1,2}^{AB}e^{d\gamma_{B}t} \right)\prod_{\alpha=1\dots d}\delta_{r_{\alpha},0} \nonumber\\ 
&+& \left({\cal D}_{2,0}^{AB}+ {\cal D}_{2,1}^{AB}e^{d\gamma_{A}t} +  {\cal D}_{2,2}^{AB}e^{d\gamma_{B}t} \right)\sum_{\alpha}(\delta_{r_{\alpha},e^{\alpha}}+\delta_{r_{\alpha},-e^{\alpha}})\prod_{\alpha\neq\alpha'}\delta_{r_{\alpha'},0} \nonumber\\
&+& \left[(H_{1}^{ab}+H_{2}^{ab}-A_{1}^{a} - D_{2}^{b})\sum_{\alpha}(\delta_{r_{\alpha},e_{\alpha}} +\delta_{r_{\alpha},-e_{\alpha}} )\prod_{\alpha\neq\alpha'}\delta_{r_{\alpha'},0}-2{\cal C}_{AB}d\prod_{\alpha=1\dots d}\delta_{r_{\alpha},0}\right]{\cal G}_{|r|=1}^{AB}(t)
\end{eqnarray}
The solution of (\ref{eq.D.7}) is :
\begin{eqnarray}
\label{eq.D.10}
&&{\cal G}_{|r|=|(r_{1},\dots,r_{d})|}^{AA}(t)-(\rho_{A}(t))^{2}=
-(\rho_{A}(0))^2e^{-2|\gamma_{A}|dt} +\rho_{A}(0)e^{-|B_{A}|dt}\prod_{\alpha=1\dots d} I_{r_{\alpha}}(2{\cal C}_{A}t)\nonumber\\ 
&+&\sum_{r'\neq 0} {\cal G}_{|r'|}^{AA}(0) e^{-|B_{A}|dt}\prod_{\alpha=1\dots d}I_{r_{\alpha}-r'_{\alpha}}(2{\cal C_{A}}t) \nonumber\\
&+& \int_{0}^{t} dt' e^{-|B_{A}|d(t-t')}{\cal D}_{1,0}^{A} \prod_{\alpha=1\dots d}
I_{r_{\alpha}}(2{\cal C}_{A}(t-t')) \nonumber\\
&+& {\cal D}_{2,0}^{A}  \int_{0}^{t} dt' e^{-|B_{A}|d(t-t')}  \sum_{\alpha}\left[\prod_{\alpha'\neq \alpha}
I_{r_{\alpha'}}(2{\cal C}_{A}(t-t'))  \right](I_{r_{\alpha}+1}(2{\cal C}_{A}(t-t')) + I_{r_{\alpha}-1}(2{\cal C}_{A}(t-t'))) \nonumber\\
&+& \int_{0}^{t} dt' e^{-|B_{A}|d(t-t')}{\cal D}_{1,1}^{A} e^{-|\gamma_{A}|dt'}\prod_{\alpha=1\dots d} I_{r_{\alpha}}(2{\cal C}_{A}(t-t')) \nonumber\\
&+&  \int_{0}^{t} dt' e^{-|B_{A}|d(t-t')}({\cal D}_{2,1}^{A} e^{-|\gamma_{A}|dt'}  + {\cal D}_{2,2}^{A} e^{-|\gamma_{B}|dt'} ) \sum_{\alpha}\left[\prod_{\alpha'\neq \alpha}
I_{r_{\alpha'}}(2{\cal C}_{A}(t-t'))  \right](I_{r_{\alpha}+1}(\dots) + I_{r_{\alpha}-1}(\dots)) ) \nonumber\\
&+&  \int_{0}^{t} dt' e^{-|B_{A}|d(t-t')}{\cal G}_{|r|=1}^{AA}(t')(G_{1}^{a}-B_{A}/2) \sum_{\alpha}\left[\prod_{\alpha'\neq \alpha}
I_{r_{\alpha'}}(2{\cal C}_{A}(t-t'))  \right]( I_{r_{\alpha}+1}(\dots) +I_{r_{\alpha}-1}(\dots)) ) \nonumber\\
&-& 2{\cal C}_{A}d \int_{0}^{t} dt' e^{-|B_{A}|d(t-t')}{\cal G}_{|r|=1}^{AA}(t')
\prod_{\alpha=1\dots d}  e^{-|B_{A}|d(t-t')}  I_{r_{\alpha}}(2{\cal C}_{A}(t-t')) 
\end{eqnarray}
where the abreviated notation $(\dots)$ instead of $(2{\cal C}_{A}(t-t') )$ has
been used.
Other correlation functions ${\cal G}^{BB}_{|r|}(t)$ and  ${\cal G}^{AB}_{|r|}(t)$ are obtained in a similar way. Properties of the Bessel functions and elementary manipulations
lead to the more compact forms (\ref{eq.3.31}-\ref{eq.3.33}). 
\end{document}